\documentclass{jnmp}

\usepackage[dvips]{graphics,graphicx}
\usepackage{amsmath}
\usepackage{amssymb}                 
\usepackage{wasysym}
\usepackage{cite}
\usepackage{psfrag}
\DeclareGraphicsExtensions{.eps}

\JNMPnumberwithin{equation}{section}
\resetfootnoterule

\DeclareGraphicsExtensions{.eps}

\newcommand{\ri}{{ \rm i }}
\newcommand{\rmi}{{ \rm i }}
\newcommand{\re}{{ \rm e }}
\newcommand{\rme}{{ \rm e }}
\newcommand{\rd}{{ \rm d }}
\newcommand{\cn}{{\rm cn}}
\newcommand{\sn}{{\rm sn}}
\newcommand{\dn}{{\rm dn}}
\newcommand{\nn}{\nonumber}
\newcommand{\abs}[1]{\vert#1\vert}

\begin{document}

\renewcommand{\evenhead}{D Witthaut, K Rapedius and H J Korsch}
\renewcommand{\oddhead}{The nonlinear Schr\"odinger equation for the delta-comb potential}

\thispagestyle{empty}


\copyrightnote{2008}{D Witthaut, K Rapedius and H J Korsch}

\Name{The nonlinear Schr\"odinger equation for the delta-comb potential:
quasi-classical chaos and bifurcations of periodic stationary solutions}

\label{firstpage}

\Author{D WITTHAUT~$^a$, K RAPEDIUS~$^b$ and H J KORSCH~$^c$}

\Address{$^a$ QUANTOP, The Niels Bohr Institute, University of Copenhagen, 2100 
Copenhagen, Denmark \\
~~E-mail: dirk.witthaut@nbi.dk\\[10pt]
$^b$ FB Physik, Technical University of Kaiserslautern,
 D-67653 Kaiserslautern, Germany  \\
~~E-mail: rapedius@physik.uni-kl.de \\[10pt]
$^c$ FB Physik, Technical University of Kaiserslautern,
 D-67653 Kaiserslautern, Germany\\
~~E-mail: korsch@physik.uni-kl.de}


\begin{abstract}
The nonlinear Schr\"odinger equation is studied for a periodic sequence of
delta-potentials (a delta-comb) or narrow Gaussian potentials.
For the delta-comb the time-independent nonlinear Schr\"odinger equation can be
solved analytically in terms of Jacobi elliptic functions and thus provides useful
insight into the features of nonlinear stationary states of periodic potentials.
Phenomena well-known from classical chaos are found, such as a bifurcation of
periodic stationary states and a transition to spatial chaos.
The relation of new features of nonlinear Bloch bands, such as looped and period
doubled bands, are analyzed in detail. An analytic expression for the critical
nonlinearity for the emergence of looped bands is derived.
The results for the delta-comb are generalized to a more realistic potential
consisting of a periodic sequence of narrow Gaussian peaks and the dynamical
stability of periodic solutions in a Gaussian comb is discussed.

\end{abstract}

\section{Introduction}

In the case of low temperatures, the dynamics of a Bose-Einstein
condensate (BEC) can be described in a mean--field approach by the
nonlinear Schr\"odinger equation (NLSE) or Gross--Pitaevskii equation
(see, e.g., \cite{Pita03,Peth02})
\be
  \left( - \frac{\hbar^2}{2M} \nabla^2 + V(x) + g |\psi(x,t)|^2 \right)
  \psi(x,t) = \ri \hbar \frac{\partial \psi(x,t)}{\partial t} \, ,
  \label{eqn-NLSE-timedep}
\ee
where $g$ is the nonlinear interaction strength. Another important
application of the nonlinear Schr\"odinger equation is the propagation 
of electromagnetic waves in nonlinear media (see, e.g., \cite{Dodd82},
chapter 8).

Due to the possibility to control all experimental parameters accurately 
over wide ranges and monitor the dynamics of a BEC in situ, these have 
become one of the most prominent models for the study of nonlinear 
dynamical systems. An increasing number of important nonlinear phenomena 
has been demonstrated experimentally in the last decade, such as the 
motion of dark \cite{Burg99a} and bright solitons \cite{Khay02} and the
self-trapping effect \cite{Albi05}. Furthermore, it is possible to
reduce the dimensionality of the NLSE in confined geometries (see, 
e.g., \cite{Grei01} and references therein). An excellent review of
nonlinear phenomena in BECs can be found in the recent article by
Carretero-Gonz\'alez {\it et al.\/} \cite{Carr08}.
A particular important subdiscipline is the study of the nonlinear 
dynamics of ultracold atoms and BECs in periodic potentials since 
these systems provide excellent model systems for fundamental problems 
in condensed matter physics (see \cite{Mors06} and \cite{Bloc08} for
recent reviews). Among other phenomena, recent experiments have
demonstrated the excistence of gap solitons \cite{Eier04a} and
of looped Bloch bands and the corresponding dynamical instability
\cite{Fall04,Jona03} as well as nonlinear de- and rephasing \cite{Naeg08,04bloch_bec}.

In this paper paper we will present an analytic study of stationary 
states of the nonlinear Schr\"odinger equation, i.e. states of the 
form $\psi(x,t) = \exp(-\ri \mu t/\hbar) \psi(x)$, in a one-dimensional 
periodic potential. These states satisfy the time-independent NLSE
\be
  \left( - \frac{\hbar^2}{2M} \frac{\rd^2}{\rd x^2} 
     + V(x) + g |\psi(x)|^2 \right) \psi(x) = \mu \psi(x) 
  \label{eqn-NLSE-general}
\ee
with $V(x+d)=V(x)$ and $\mu \in \mathbb{R}$. Here we adopt a novel
 approach to nonlinear stationary states by exploiting the 
analogy to nonlinear dynamical systems in considering the wave function 
$\psi(x)$ as a dynamical variable.
Although a bit unusual, this approach leads to a variety of important results, 
explaining several phenomena umambigiously in the language of nonlinear dynamical 
systems. For instance, the appearance of period-doubled Bloch bands, as
described previously in \cite{Mach04}, is readily understood as a familiar 
period-doubling bifurcation. Similarly, the bifurcation leading to the 
appearance of looped Bloch bands is easily understood so that it is possible
to calculate the critical nonlinearity for this process {\it analytically}.
From a fundamental point of view it becomes clear that Bloch states, i.e.
solutions which are periodic up to a phase factor $e^{i \kappa d}$, are 
extremely fragile objects. While in the linear case $g=0$ all states are Bloch states 
or linear combinations of degenerate Bloch states
according to Bloch's theorem, these states are of measure
zero in the nonlinear case $g \neq 0$. The reason for this difference is 
that almost all solutions are only quasi-periodic for $g \neq 0$, approaching
a true periodicity smoothly in the limit $g \rightarrow 0$. With increasing 
strength of the nonlinearity, more and more (quasi)-periodic orbits are lost 
and chaos sets in in the wake of a period doubling process. Spatial chaoticity 
usually implies a divergence of the wave function. Thus, well-behaved nonlinear 
stationary states are very rare for certain parameter values, in fact large 
interaction constants.

As a convenient model for a space-periodic structure, we consider a particular
simple model system, the delta-comb potential
\be
  V(x) = \lambda \sum_n \delta(x- d n) = \lambda \delta_d(x).
  \label{eqn-delta-pot}
\ee 
The potential (\ref{eqn-delta-pot}) has the great advantage that analytical solutions
of the NLSE can be constructed in terms of Jacobi elliptic functions \cite{Carr00}.
Analytic solutions of the nonlinear Schr\"odinger equation for a non-vanishing
potential $V(x)$ are very rare and hence it will be of interest to study such a
potential in some detail. Its linear counterpart represents a basic model for a
quantum mechanical Bloch system (see, e.g. \cite{Grif92}). A generalization to
a more realistic setup consisting of a periodic series of Gaussian potential 
barriers will be discussed at the end of this paper. Narrow Gaussian potential 
barriers can be realized in a good approximation in current atom-chip experiments 
(see \cite{Fort07} for a recent review).
For example the resonant transport of BECs through two such peaks is analyzed
in \cite{Paul05}.
We generally restrict ourselves to the case of a repulsive interaction, $g>0$
and $\mu > 0$. Some differences that arise for an attractive interaction will 
be discussed briefly in Sec.~\ref{sec-attractive-nonlin}. For convenience we 
set $\hbar = M = 1$ and fix the period of the potential as $d = 2 \pi$.
If the normalization of the wave function is arbitrary, one can rescale the
wave function as $\psi \rightarrow \psi/\sqrt{g}$, leaving two free parameters
($\lambda$, $\mu$), otherwise three parameters must be considered.

Up now, quite a number of articles were devoted to the important
problem of nonlinear stationary states in a periodic potential 
and we refer to the review articles \cite{Mors06,Carr08} for an
overview. There also have been some papers devoted to the NLSE 
with a delta-comb potential, however focusing on completly 
different aspects. Bound states were studied in \cite{Theo97} and 
the nonlinear transport problem through such a series of barriers 
was discussed in \cite{Tara99}. Bloch bands were studied in
\cite{Li04,Seam05,Seam05b,Dans07a}, however relying mostly on
numerical approaches. Recently, Paul {\it et al.\/} used 
a randomized delta-comb potential to study  the crossover to
Anderson localization in \cite{Paul07c} in ultracold gases.
A related dynamical approach to nonlinear eigenstates can be found
in \cite{Port04}, however without taking much benefit from the
theory of nonlinear Bloch bands. It should also be noted that bound
states of nonlinear Schr\"odinger equations with a nonlinear periodic 
microstructure where $g=g(x)$ is space-periodic have been explored 
\cite{Wan90,Fibi06}.

In particular this paper is organized as follows. The stationary 
solutions of the free NLSE (i.e. $V(x) = 0$) are reviewed in 
Sec.~\ref{eqn-sol-general} to lay the foundation for our further 
work. With these results in mind we then construct the solutions 
for the delta-comb potential in terms of Jacobi elliptic function 
in Sec.~\ref{sec-dcomb-real-states}, thus reducing the NLSE to a 
discrete mapping for their parameters. In Sec.~\ref{sec-periodic-states} 
we focus on periodic solutions and derive conditions for their 
existence. The stability of these solutions is discussed in 
detail in Sec.~\ref{sec-periodic-stability}. We then apply our
results to the important problem of nonlinear Bloch bands in
Sec.~\ref{sec-bloch-bands}, deriving the critical nonlinearity 
for their emergence analytically. Finally, we generalize our 
approach to a more realistic setup consisting of a periodic 
series of Gaussian potential barriers in Sec.~\ref{sec-gauss}.
In addition we discuss the {\it temporal} stability of 
nonlinear periodic solutions, showing that the stability 
properties are fundamentally different for different types of 
nonlinear Bloch states. A summary and discussion of our results 
is presented in Sec.~\ref{sec-summary}

\section{Solutions of the nonlinear Schr\"odinger equation}
\label{eqn-sol-general}

Before discussing particular solutions, we start with some general remarks
on the time-independent nonlinear Schr\"odinger equation (NLSE)

First, we note that the real-valued time-independent NLSE is mathematically
equivalent to a classical nonlinear oscillator described by the nonlinear
Hill equation
\be
  \ddot y + f(t) y + \beta y^3 = 0, \quad f(t+T) = f(t).
  \label{eqn-hill-nonlin}
\ee
Replacing the amplitude $y(t)$ by the wave function $\psi(x)$ and the time $t$
by the position variable $x$ and identifying $\beta= -2g$ and $f(t) = 2(\mu - V(x))$
one recovers the time-independent NLSE (\ref{eqn-NLSE-general}).
For $V(x) = 0$ one recovers the undamped Ueda oscillator (see, e.g.,
\cite{Nayf79,Moon87,08chaos} and references therein). 
Within the framework of nonlinear eigenstates of BECs, this equation has
been analyzed in \cite{Carr00}.
For a cosine-potential $V(x) = V_0 \cos(x)$, Eqn.~(\ref{eqn-hill-nonlin})
is just the nonlinear Matthieu equation, which is a popular example in classical
nonlinear dynamics (see, e.g., \cite{Nayf79,Chir79,Bowm81,Mont82}).
The NLSE with a delta-comb potential $V(x) = \lambda \delta_d(x)$, or analogously
the kicked nonlinear Hill equation, studied in the present paper has the advantage that
it allows analytical solutions.
Further exactly solvable examples can be found when the potential is given
in terms of Jacobi elliptic functions \cite{Hsu74,Bron01a}.

In quantum systems, however, we also encounter complex solutions of the NLSE
discussed in sections \ref{sec-bloch-bands} and \ref{sec-gauss} in the context
of nonlinear Bloch bands and the temporal evolution of a wave function, which
is governed by the {\it time-dependent} NLSE.

Furthermore, the time-independent NLSE (\ref{eqn-NLSE-general}) can be written
as a Hamiltonian system with the conjugate variables $(\psi^*,\psi')$ and
$(\psi,\psi'^*)$, where $\psi' = \rd \psi / \rd x$.
Introducing the Hamiltonian function
\be
  {\cal H} = |\psi'|^2 - 2 (V(x)-\mu) |\psi|^2 - g |\psi|^4,
\ee
Hamilton's equations read
\begin{eqnarray}
  \frac{\rd}{\rd x} \psi' &=& - \frac{\partial {\cal H}}{\partial \psi^*} =
  2(V(x) - \mu) \psi + 2 g |\psi|^2 \psi = \psi''  \\
  \frac{\rd}{\rd x} \psi^* &=& \frac{\partial {\cal H}}{\partial \psi'} = \psi'^*
\end{eqnarray}
and analogously for $(\psi,\psi'^*)$. 
Considering only real-valued solutions, one is left with a two-dimensional
phase space $(\psi,\psi')$ where the flow is is area-preserving.

For the special case of a delta-comb,
\be
  V(x) = \lambda \delta_{2\pi}(x) =  \lambda \sum_n \delta(x-2\pi n),
\ee
the solutions are essentially the ones of the free NLSE. Hence we give a brief
review of the real solutions of the free nonlinear Schr\"odinger equation
in the following section. The first derivative of the wave function $\psi'$ is
discontinuous at $x = 2 \pi n$ (cf.~the studies 
\cite{04nls_delta,Adam05,Sacc07,LeCo08,09ddshell} of a NLSE for a single delta-potential)
\be
   \lim_{\epsilon \to 0+}\big( \psi'(2\pi n+\epsilon) - \psi'(2 \pi n-\epsilon)\big)
   = 2 \lambda \psi(2 \pi n),
  \label{eqn-delta-condition}
\ee
whereas the wave function itself is continuous.
The discontinuity of the delta-comb potential does not affect the area-preserving
quality of the flow $(\psi(x_0),\psi'(x_0)) \rightarrow (\psi(x),\psi'(x))$.
To clarify this issue, we linearize the flow in the vicinity of a delta peak at
$x= 2 \pi n$\,:
\be
  \left(\begin{array}{*{1}{c}} \psi(2\pi n+\epsilon) \\ \psi'(2\pi n+\epsilon) \end{array} \right)
  = \left(\begin{array}{*{2}{c}} 1 & 0 \\ 2\lambda & 1 \end{array} \right)
  \left(\begin{array}{*{1}{c}} \psi(2\pi n-\epsilon) \\ \psi'(2\pi n-\epsilon) \end{array} \right) \, .
\ee
As the determinant of the matrix equals unity, the flow is clearly area-preserving.

\subsection{Real solutions of the free nonlinear Schr\"odinger equation}
\label{ss-free}

The free nonlinear Schr\"odinger equation has well known real solutions in
terms of Jacobi elliptic functions \cite{Carr00,Dago00}
(see, e.g., \cite{Abra72,Lawd89} for an introduction).

\begin{figure}[htb]
\centering
\includegraphics[width=6cm,  angle=0]{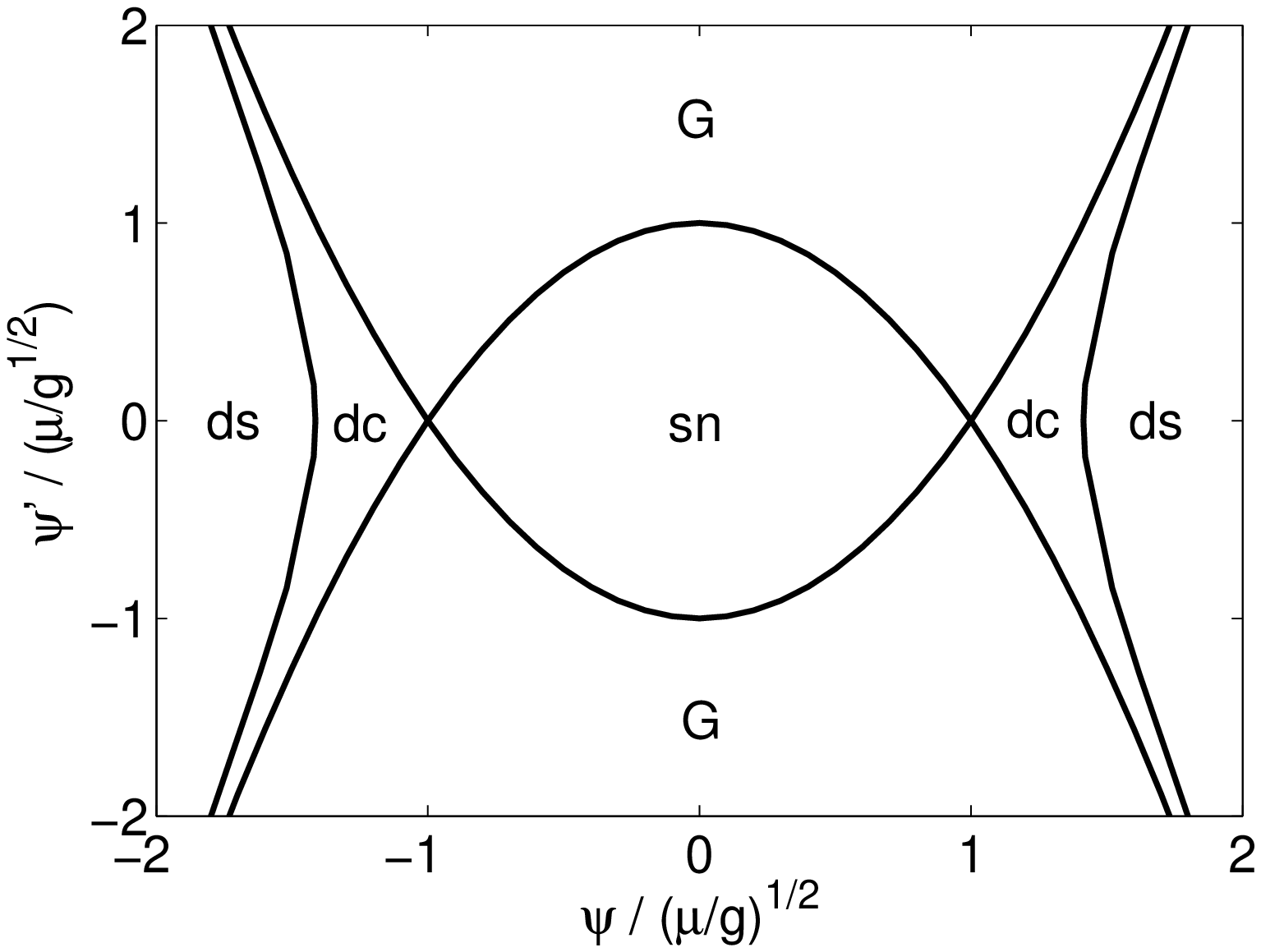}
\includegraphics[width=6cm,  angle=0]{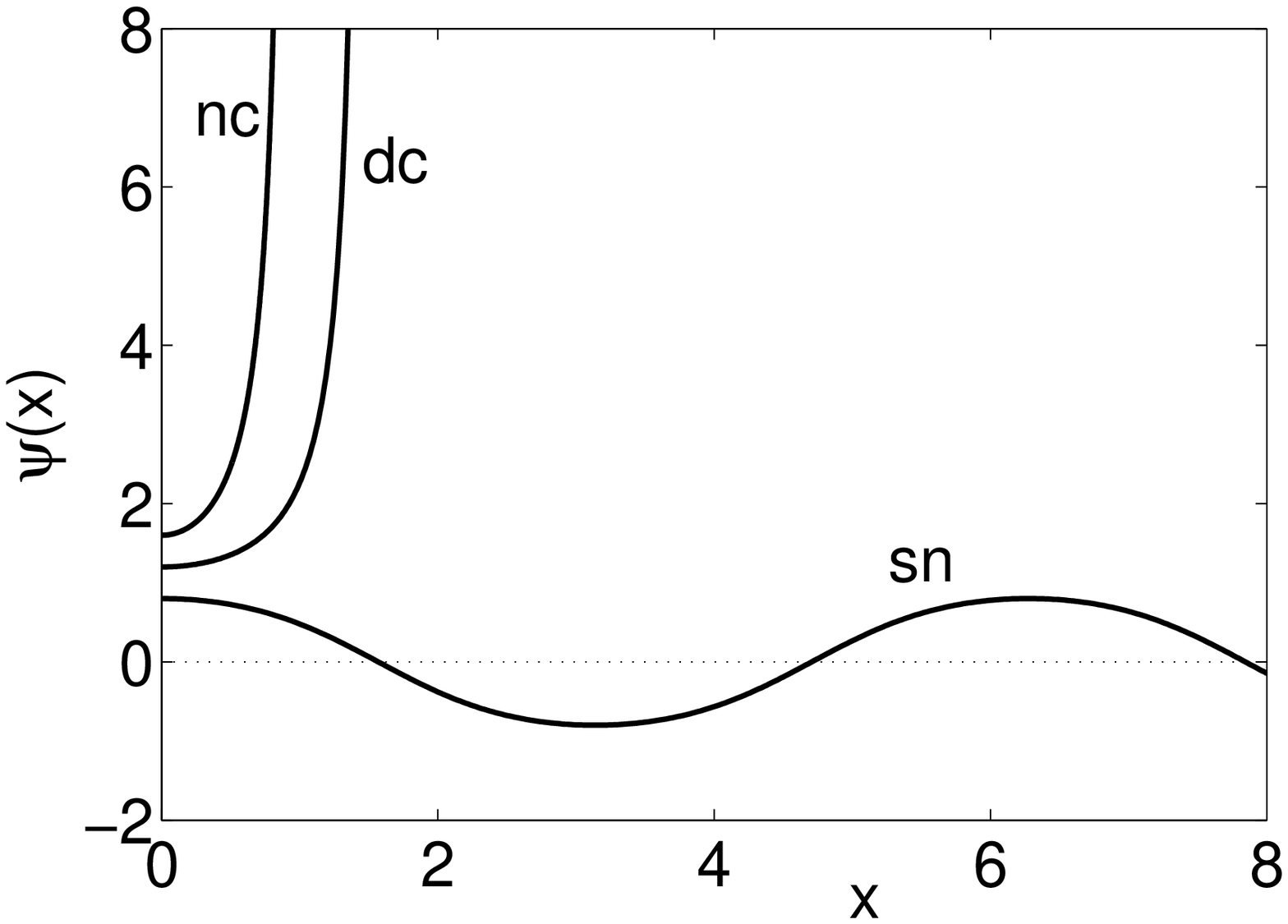}
\caption{\label{fig-rep_soltypes1}
Solution types of the free nonlinear Schr\"odinger equation with a repulsive 
nonlinearity $g$. Left panel: Dependence of the solution type on the initial condition
($\psi(x_0),\psi'(x_0)$). Right panel: Examples of the solution types for
$\psi'(x_0) = 0$ and $\psi(x_0) = 0.8$ (sn), $1.2$ (dc), $1.8$ (nc) and
$\mu = g = 1$.}
\end{figure}

It will be instructive to calculate the general solution of the free NLSE explicitly.
Multiplying Eqn.~(\ref{eqn-NLSE-general}) for $V(x) = 0$ by $\psi'$ and integrating
once yields the general solution of the free NLSE in the form
\be
  x - x_0 = \int_{\psi(x_0)}^{\psi(x)} \frac{\rd s}{\sqrt{g s^4 - 2 \mu s^2 + 2 E}} \, .
  \label{eqn-elliptic_int1}
\ee
The type of solution of Eqn.~(\ref{eqn-elliptic_int1}) strongly depends on the
initial values $\psi(x_0)$ and $\psi'(x_0)$, which also determine the constant 
of integration $E$ via
\be
   E =  \frac{1}{2} \psi'(x_0)^2  - \frac{g}{2} \psi(x_0)^4 + \mu \psi(x_0)^2.
\ee
This dependence on the initial values is illustrated in Fig.~\ref{fig-rep_soltypes1}.
For initial values of $\psi(x_0)$ and $\psi'(x_0)$ inside the regions marked by
sn, dc and nc, the integral in Eqn.~(\ref{eqn-elliptic_int1}) can be reduced
to the canonical form of the Jacobi elliptic functions sn, dc and nc, respectively.
For, e.g., $\psi'(x_0) = 0$ a simple scaling of the wave function $\tilde \psi =
\sqrt{g/\mu} \, \psi$ and the position $\tilde x = \sqrt{\mu} \, x$ leads directly
to the standard forms given in \cite{Lawd89}, Chapter 3.
Examples of these solutions types are shown on the right-hand side of
Fig.~\ref{fig-rep_soltypes1}.  Note that only the sn-type solution is periodic
while all other types diverge at a finite value of $x$.  

The situation is more involved for initial values in the regions marked by
G in Fig.~\ref{fig-rep_soltypes1}.
A simple scaling is not sufficient any longer, but a transformation
$t = (s+q)/(s-q)$ with $q = (2E/g)^{1/4}$ brings the integral
(\ref{eqn-elliptic_int1}) to the standard form of the sc function
\cite{Lawd89}.

\begin{figure}[htb]
\centering
\includegraphics[width=7cm,  angle=0]{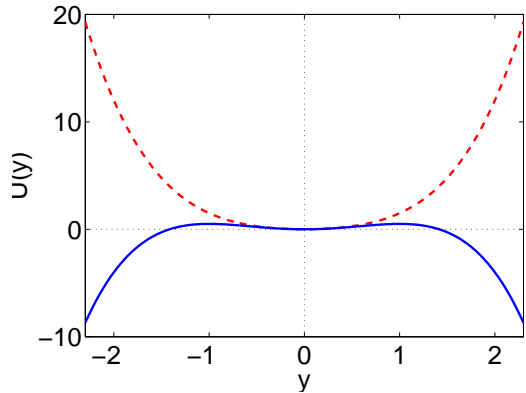}
\caption{\label{fig-rep_cass_pot}
Classical potential $U(y)$ describing the nonlinear Hill equation
resp. undamped Ueda equation (\ref{eqn-hill-nonlin}) for $\mu = 1$
and $g =1$ (---) resp. $g =-1$ ($---$).}
\end{figure}

Note that the sn-type solutions are periodic, while all other solutions
(nc, dc and G) diverge at a finite value of $x$. This is easily understood
in terms of the classical analog, the nonlinear Hill equation
(\ref{eqn-hill-nonlin}). This equation describes a classical particle with
coordinate $y$ 
and kinetic energy $T=\dot y^2/2$ 
moving in a potential $U(y) = \mu y^2 + \beta y^4/4$ with
$\beta = -2 g$, that is illustrated in Fig.~\ref{fig-rep_cass_pot}.
For $g > 0$ this potential has a minimum at $y = 0$ and diverges to $- \infty$
for $|y| \rightarrow \infty$. Thus $y$ will diverge as soon as the particle can
leave the potential minimum around $y = 0$. In contrast, 
the motion is always bounded for $g < 0$. 
The different regions in Fig.~\ref{fig-rep_soltypes1} are directly related
to the turning points for a system with energy $E=T+U=\dot y^2/2+\mu y^2 + \beta y^4/4$.
For $g>0$ and energies in the interval $0 < E < U_{\rm max}=\mu^2/2g$, 
we have four real valued turning points with bounded trajectories in
the sn-region and unbounded ones in the dc-region. For $E<0$, we have two real and
two complex turning points and in the G-region for $E>U_{\rm max}$ all turning
points are complex. The curves in Fig.~\ref{fig-rep_soltypes1} are the
boundaries of these regions, i.e.~$E=U_{\rm max}$ and $E=0$.

In the following we focus on the sn-type solutions since they do not diverge. For fixed values of $\mu$
and $g$ these solutions are explicitly given by
\be
  \psi(x) = A \, \sn \bigg( 4 K(p)  \frac{x+x_0}{L} \bigg| p \bigg),
\ee
where $p$ ($0\le p\le 1$) 
denotes the elliptic parameter and $L$ is the period of the wave function.
The amplitude $A$ and the period $L$ are fixed by
\be
  A^2 = \frac{2\mu p}{g(p+1)} \quad \mbox{and} \quad
  L^2 = \frac{8(p+1) \, K(p)^2}{\mu} \, ,
  \label{eqn-dcomb-repulsive-amp-mu}
\ee
where $K(p)$ denotes the complete elliptic integral of the first kind.
However, it should be kept in mind that other types of solutions are appropriate
for different initial values as illustrated in Fig.~\ref{fig-rep_soltypes1}.

\subsection{Non-diverging solutions for a delta-comb}
\label{sec-dcomb-real-states}

\begin{figure}[htb]
\centering
\includegraphics[width=5.5cm,  angle=0]{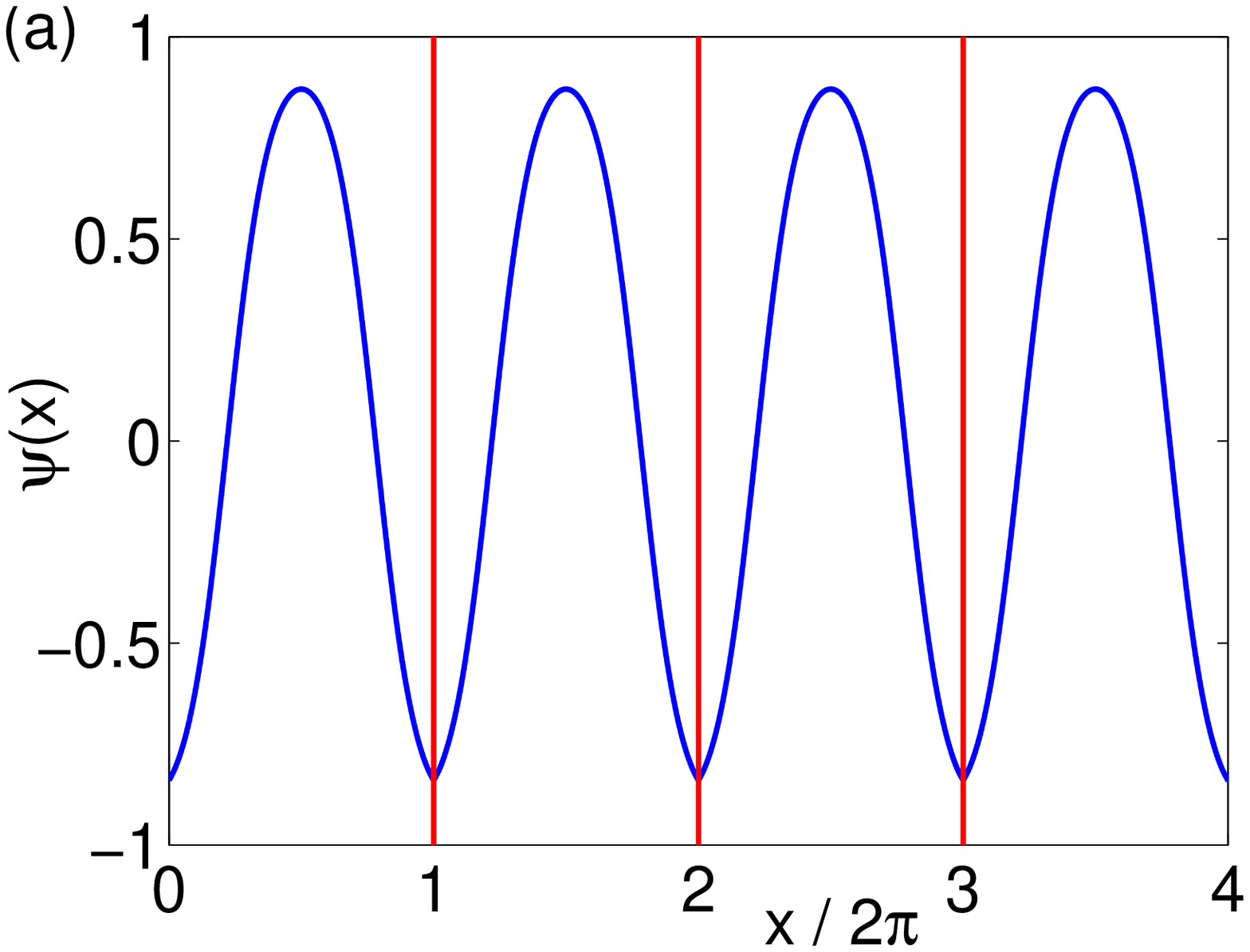}
\hspace{5mm}
\includegraphics[width=5.5cm,  angle=0]{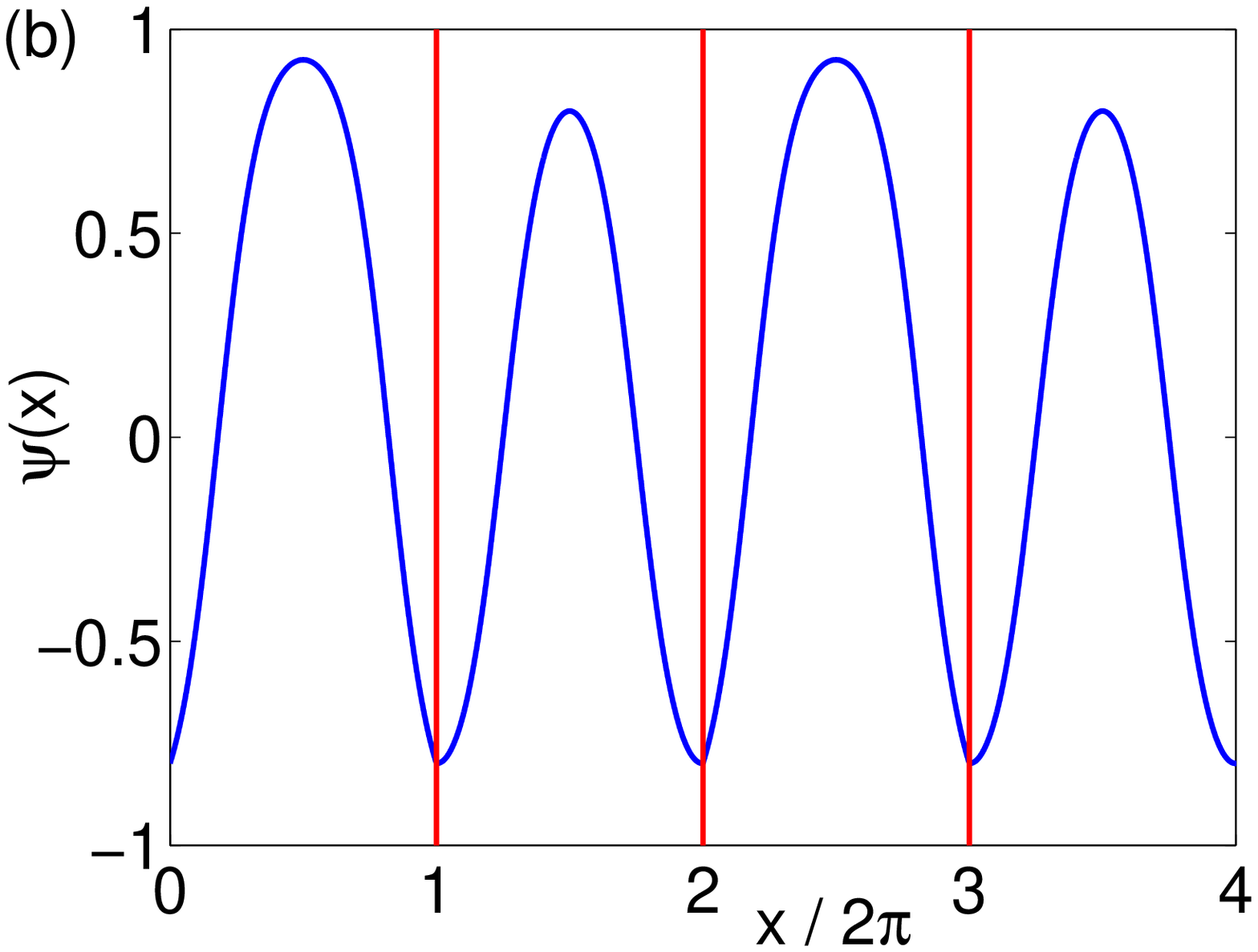}
\vspace{5mm}
\includegraphics[width=5.5cm,  angle=0]{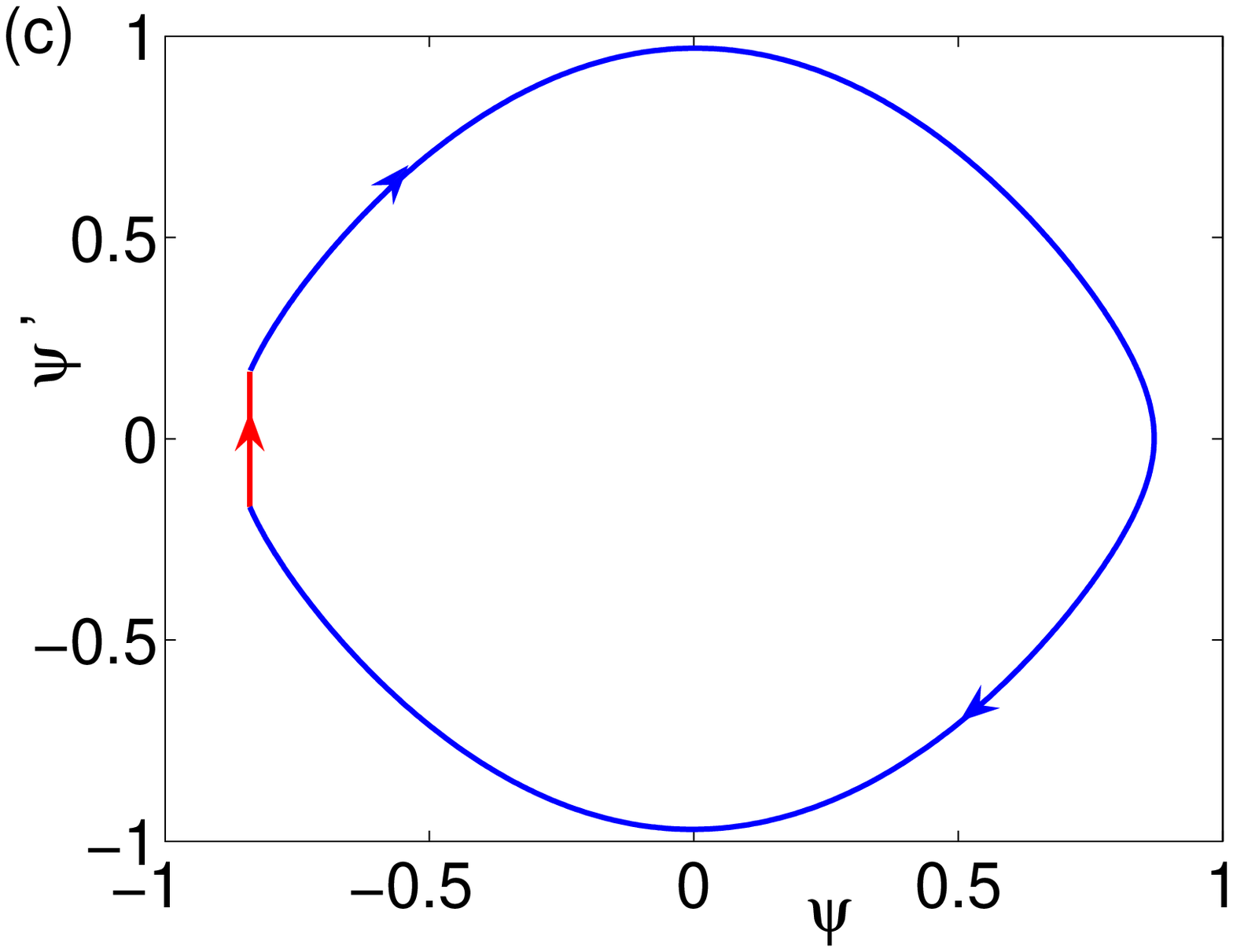}
\hspace{5mm}
\includegraphics[width=5.5cm,  angle=0]{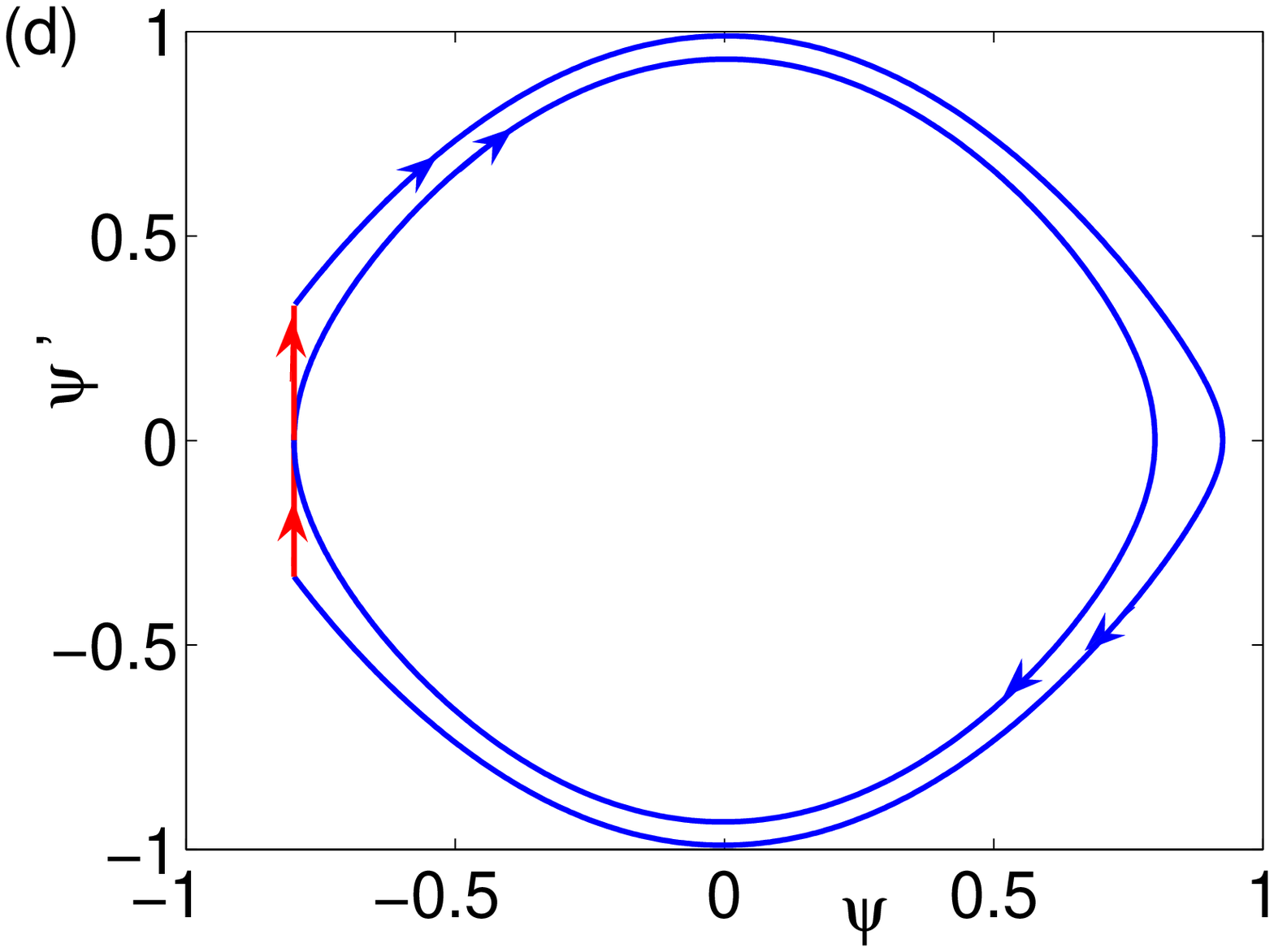}
\caption{\label{fig-dcomb-illu} Periodic trajectories are given by 
solutions of the free nonlinear Schr\"odinger equation connected 
by vertical jumps caused by the delta potentials. The figure 
shows the 'configuration space' representation $\psi(x)$ (a,b)
as well as the phase space representation $(\psi,\psi')$ (c,d).
The period-one orbit for $\lambda=-0.2$ (a,c) bifurcates to a 
period-two orbit (b,d) if the strength of the delta-potentials 
is changed to $\lambda=-0.205$. The remaining parameters 
are $g=\mu=1$. 
}
\end{figure}

The real solutions of the nonlinear Schr\"odinger equation for a delta comb
are essentially the ones of the free equation discussed above.
Hence we make the ansatz
\be
  \psi(x) = A_n \, \sn \bigg( 4 K(p_n)  \frac{x+x_n}{L_n}
  \bigg| p_n \bigg)
  \label{eqn-dcomb-repulsive-ansatz}
\ee
for $x \in (2\pi n,2\pi(n+1))$ with $A_n$ and $L_n$ defined in Eqn.
(\ref{eqn-dcomb-repulsive-amp-mu}).
However, two conditions have to be fulfilled at $x = 2\pi n$: The wave function
is continuous, whereas its derivative is discontinuous according
to Eqn.~(\ref{eqn-delta-condition}). This is illustrated in
Fig.\ref{fig-dcomb-illu} (a,b), where the discontinuity of the the first 
derivative at $x=2\pi n$ is clearly visible.
For the sn-type solutions (\ref{eqn-dcomb-repulsive-ansatz})
the (dis-) continuity conditions are explicitly given by:
\begin{align}
 & {\rm I.}  \;  A_n \, \sn ( u_n | p_n ) = A_{n+1} \, \sn (u_{n+1}  | p_{n+1} ) \\
 & {\rm II.} \; 2 \lambda A_n \, \sn(u_n  | p_n )
   + \frac{4 A_n K_n}{L_n} \cn (u_n  | p_n ) \dn (u_n  | p_n ) \nn \\
 &  \qquad =  \frac{4 A_{n+1} K_{n+1}}{L_{n+1}} \cn (u_{n+1}  | p_{n+1} ) \dn (u_{n+1}  | p_{n+1} )
  \label{eqn-dshell-repres-conditions}
\end{align}
where the abbreviations $u_n = 4 K(p_n) (2 \pi (n+1) + x_n)/L_n$  and
$u_{n+1} = 4 K(p_{n+1}) (2 \pi (n+1) + x_{n+1})/L_{n+1}$ have been used.
If $\abs{A_n \sn(u_n  | p_n )} \le \abs{A_{n+1}} $ the first condition
can always be fulfilled by an appropriate choice of the
''phase shift'' $x_{n+1}$.
Inserting the first condition into the second one and using
the addition theorems of the Jacobi elliptic functions one
arrives at
\begin{align}
  & \frac{p_{n+1}}{(p_{n+1}+1)^2} = \frac{p_n}{(p_n+1)^2}
  + \frac{2\lambda^2}{\mu} \frac{p_n}{p_n+1} \sn^2(u_n  | p_n )
  \label{eqn-dshell-repres-condition2}  \\
  & \qquad + \frac{p_n}{(p_n+1)^{3/2}} \frac{4 \lambda}{\sqrt{2\mu}} \cn(u_n  | p_n )
  \dn (u_n  | p_n ) \sn (u_n  | p_n ) . \nn
\end{align}
These equations define a discrete mapping $f: (p_n,x_n) \rightarrow (p_{n+1},x_{n+1})$
or $\tilde f: (\psi_n,\psi'_n) \rightarrow (\psi_{n+1},\psi'_{n+1})$, respectively,
where we introduced the abbreviations $\psi_n = \psi(2 \pi n + 0)$ and
$\psi'_n = \psi'(2 \pi n + 0)$ for convenience.
In the following we will mainly consider the mapping $\tilde f$ in terms of the
physical variables $\psi$ and $\psi'$. This is the Poincar\'e section of the
area-preserving phase space flow generated by the NLSE. 
However, the mapping $f$ in terms of the abstract parameters $p_n$ and $x_n$ is more
suitable for actual calculations.

Note that the discontinuity of the derivative may lead to a divergence of
the wave function. As stated above, the sn-type solutions are periodic, while
all other solutions of the free NLSE diverge at a finite value of $x$.
Due to the discontinuity of the derivative, the mapping $\tilde f$ can
map a point $(\psi_n,\psi'_n)$ inside the sn-region of phase space to a
point $(\psi_{n+1},\psi'_{n+1})$ outside the sn-region.
This will generally lead to a divergence of $\psi(x)$ at a finite value of
$x$. We will come back to the question of divergences later in this section.

\begin{figure}[htb]
\centering
\includegraphics[width=4.9cm,  angle=0]{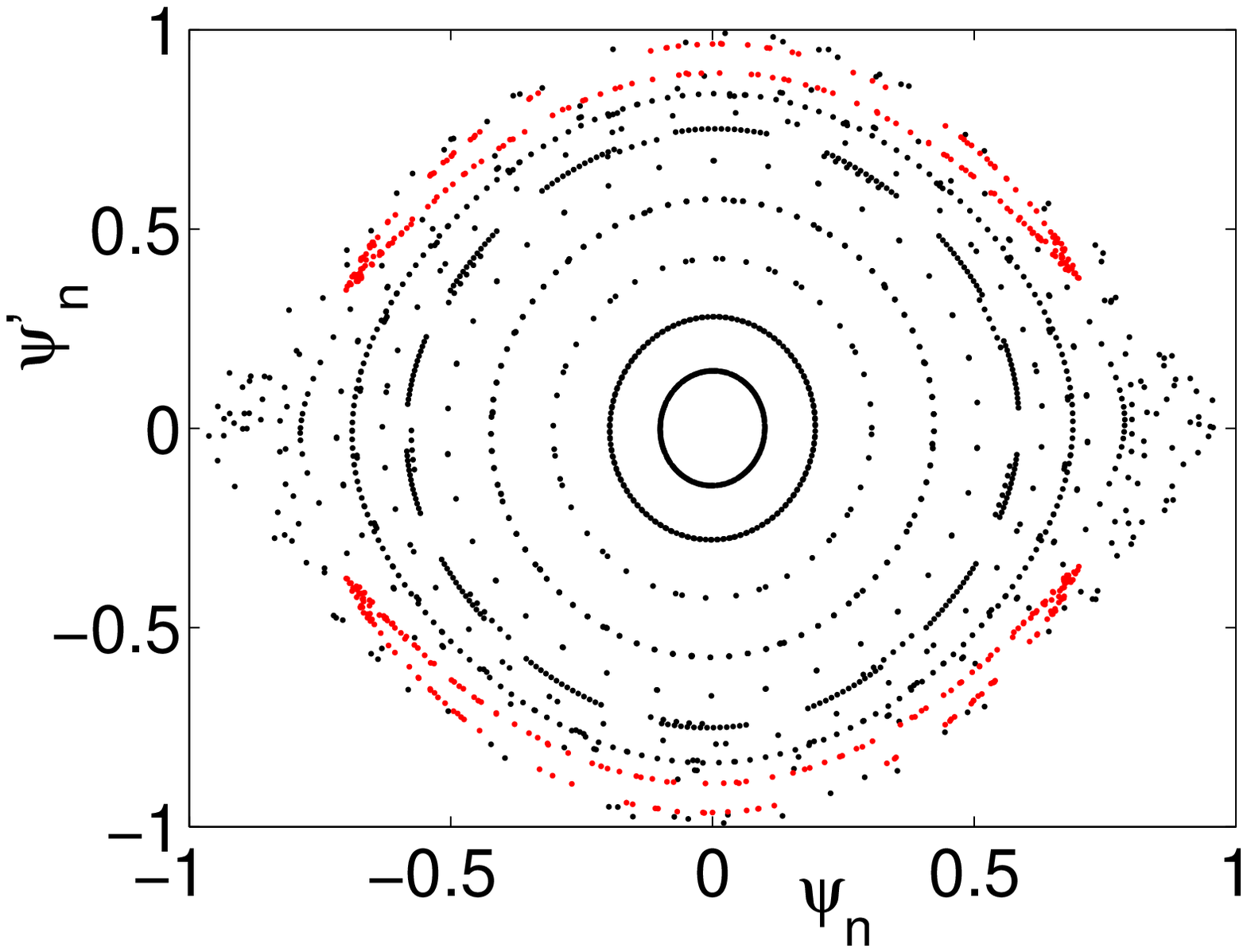}
\includegraphics[width=4.9cm,  angle=0]{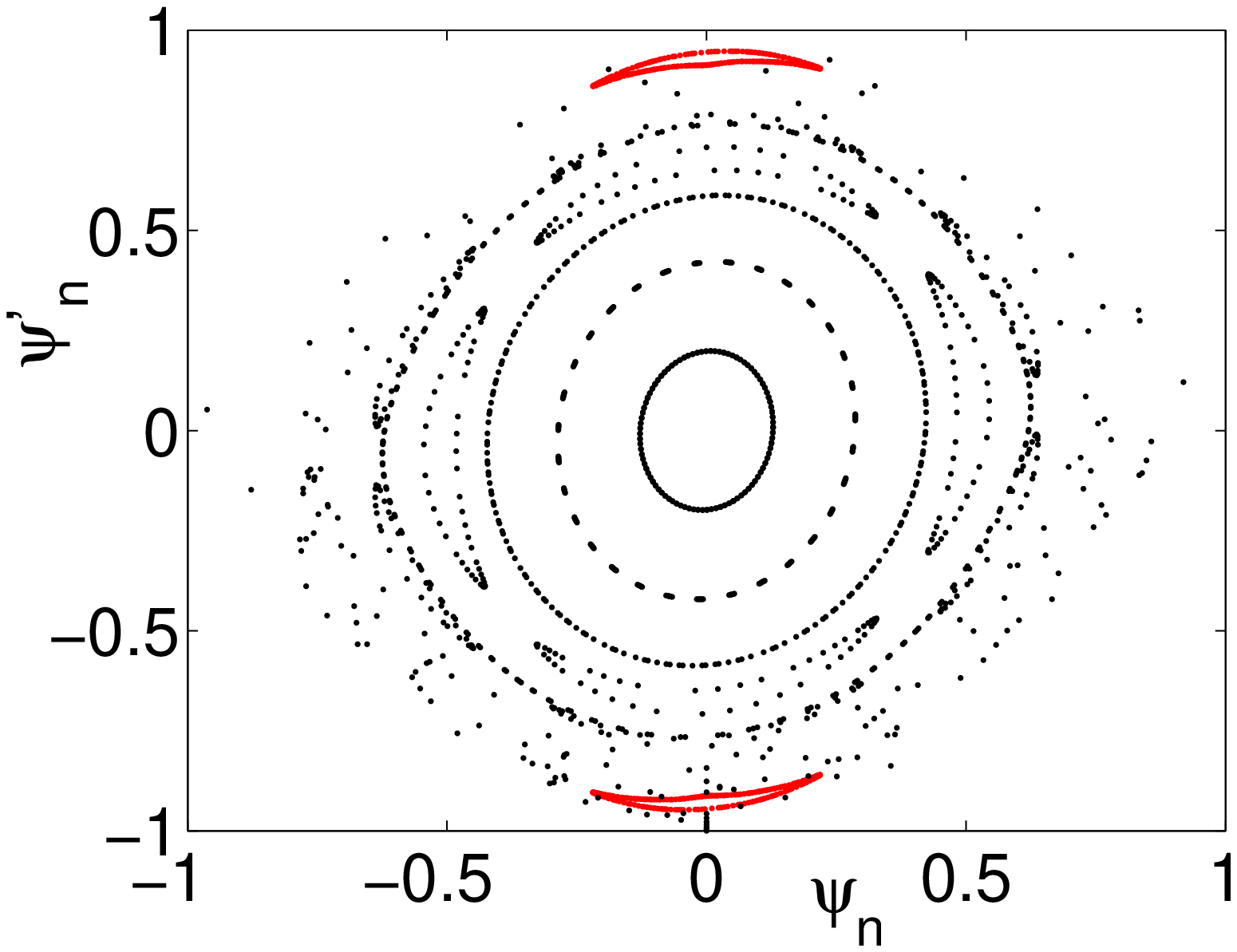}
\includegraphics[width=4.9cm,  angle=0]{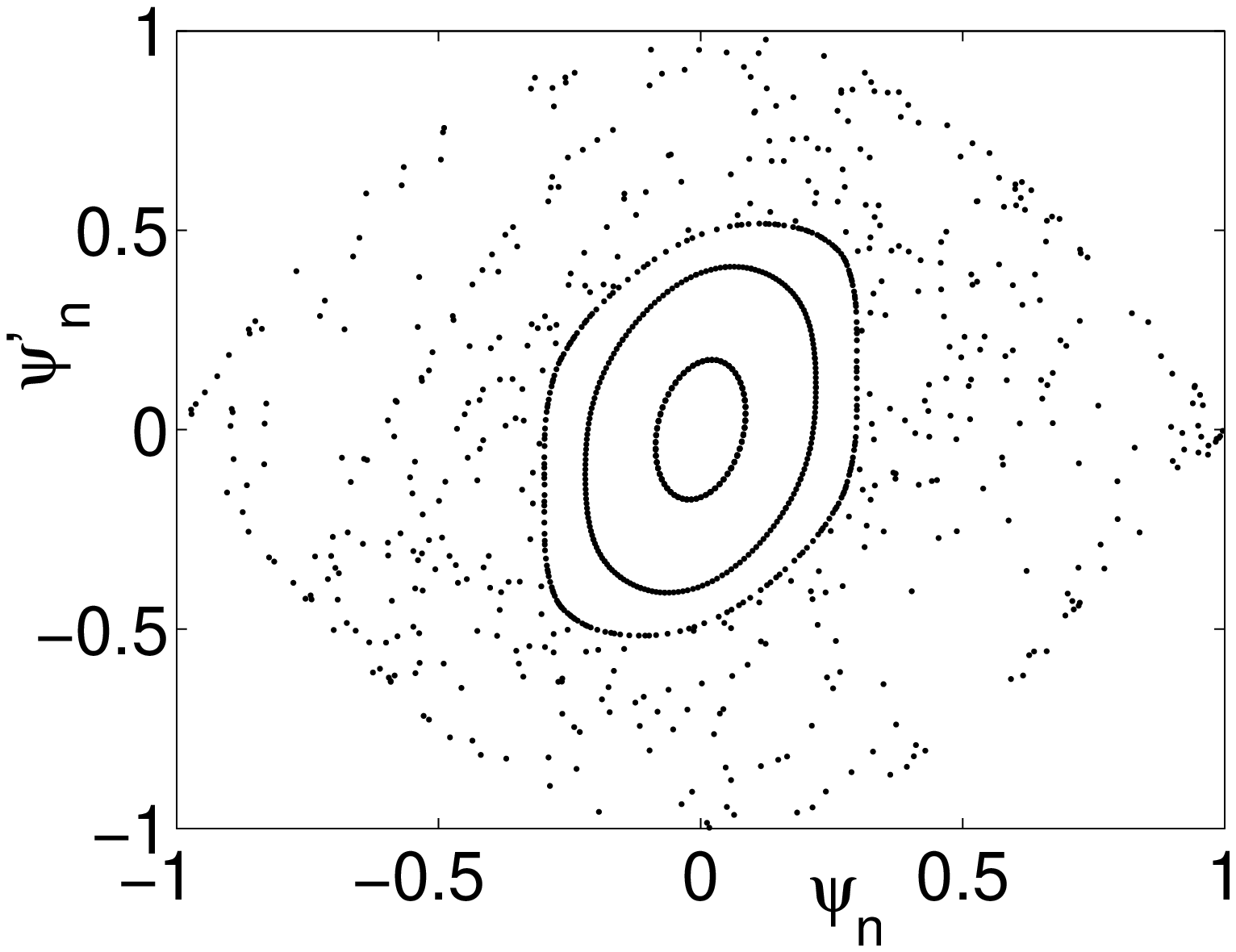}
\caption{\label{fig-dcomb-rep-phase1}
Stroboscopic phase space plots of the mapping
$\tilde f: (\psi_n,\psi'_n) \rightarrow (\psi_{n+1},\psi'_{n+1})$
for $g=1$, $\mu = 1$ and $\lambda = 0.02, \, 0.1, \, 0.5$ (from left to right).}
\end{figure}

\subsection{Dynamics in phase space}
 
It is instructive to first have a look at the flow in phase space
as illustrated in Fig.~\ref{fig-dcomb-illu} (c,d) for periodic solutions
($g=1$, $\mu=1$) in comparison to the respective wave function 
representation (a,b). Panel (c) show a period-one trajectory following
an orbit of the free nonlinear system as discussed in Sec.~\ref{ss-free}
up to a point $\psi\approx -0.84$ where the delta-potential causes a vertical
jump of $\psi'$ according to (\ref{eqn-delta-condition}).  If the potential
strength $\lambda$ is made more negative, the orbit slightly adjusts until it
bifurcates into a double-periodic orbit as shown in  Fig.~\ref{fig-dcomb-illu} (d)
for $\lambda=-0.205$. Here two kicks connect two
different orbits of the free nonlinear system. We will come back to the
period-doubling bifurcation scenario in Sect.~\ref{sec-periodic-stability}
(compare in particular Fig.~\ref{fig-dcomb-4per1}). Similar phase space
plots can be found in Ref.~\cite{Jack04,Komi06,Alfi07}.

\begin{figure}[htb]
\centering
\includegraphics[width=4.9cm,  angle=0]{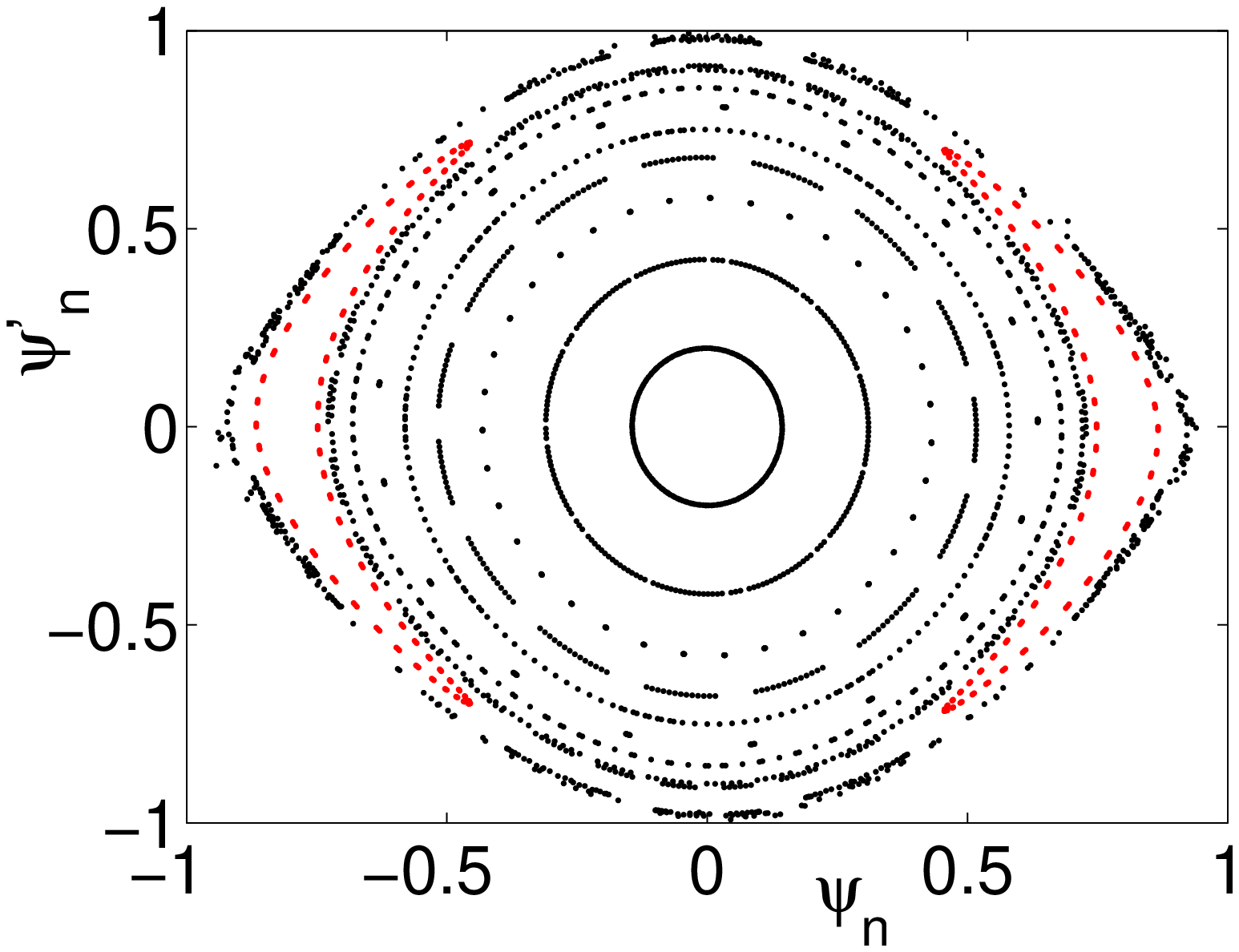}
\includegraphics[width=4.9cm,  angle=0]{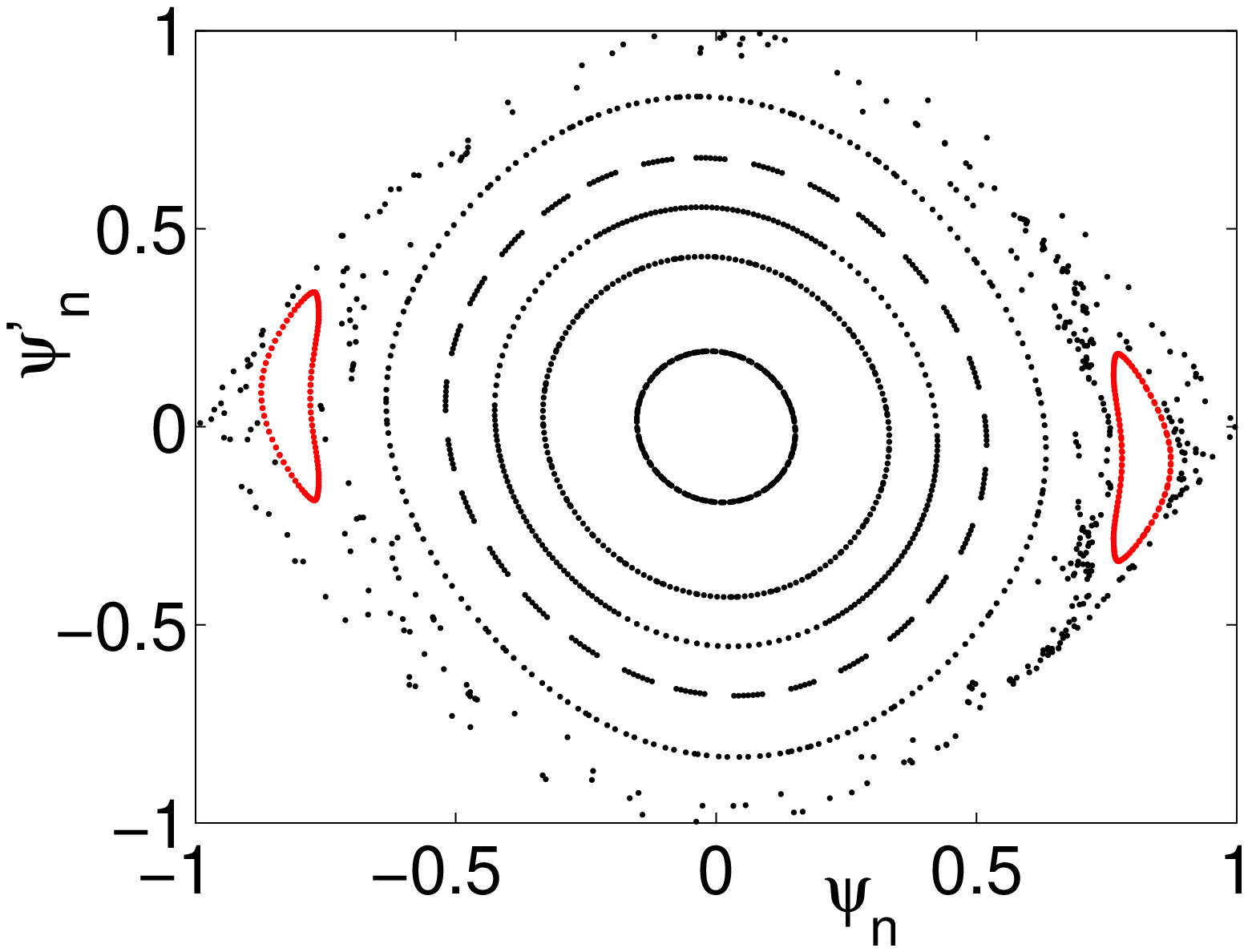}
\includegraphics[width=4.9cm,  angle=0]{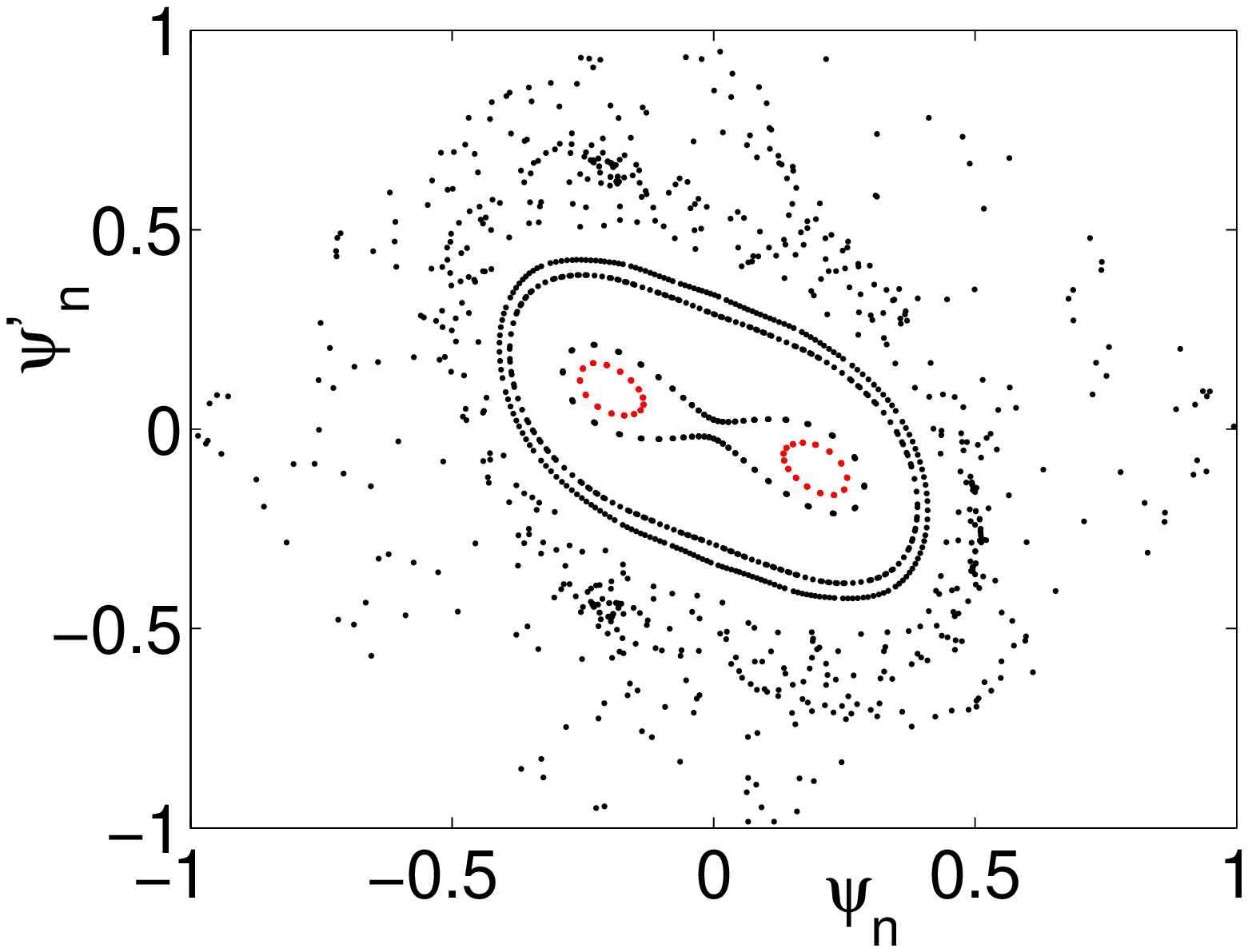}
\caption{\label{fig-dcomb-rep-phase2}
As figure \ref{fig-dcomb-rep-phase1}, however for
$\lambda = -0.02, \, -0.1, \, -0.5$ (from left to right).}
\end{figure}

We now turn to the Poincar\'e section of this this flow given by the discrete 
mapping $f$ or $\tilde f$ respectively, because it offers a global view of the dynamics.
For the actual computation we use the mapping $f$ in terms of $p_n$
and $x_n$, in fact the Eqn.~(\ref{eqn-dshell-repres-condition2}).
We propagate an ensemble of trajectories with randomly chosen initial
conditions within the sn-region of phase space (cf.~Fig.~\ref{fig-rep_soltypes1}).
The dynamics within this region is visualized in
Figs.~\ref{fig-dcomb-rep-phase1} and \ref{fig-dcomb-rep-phase2}
for different values of the potential strength $\lambda$, both for repulsive
and attractive delta-potentials. The remaining parameters are fixed as 
$\mu = 1$ and $g=1$.
First we notice that the dynamics is invariant with respect to a global sign
change $(\psi_n,\psi'_n) \rightarrow (-\psi_n,-\psi'_n)$, hence the phase space
is inversion symmetric. For $\lambda = 0$ the phase space is additionally
symmetric to the coordinate axes. This symmetry is destroyed with increasing
potential strength $\lambda$.

\begin{figure}[htb]
\centering
\includegraphics[width=7cm,  angle=0]{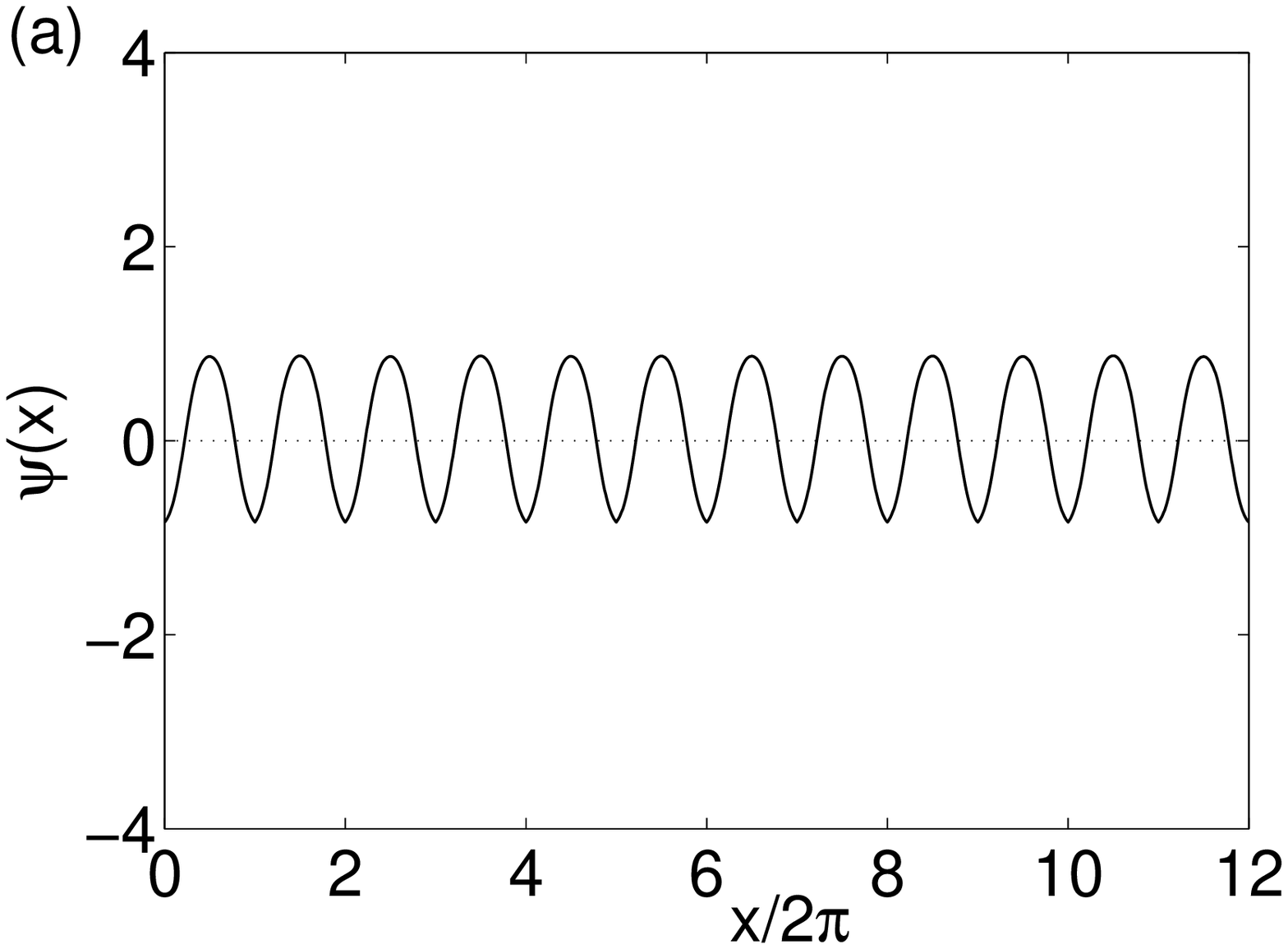}
\includegraphics[width=7cm,  angle=0]{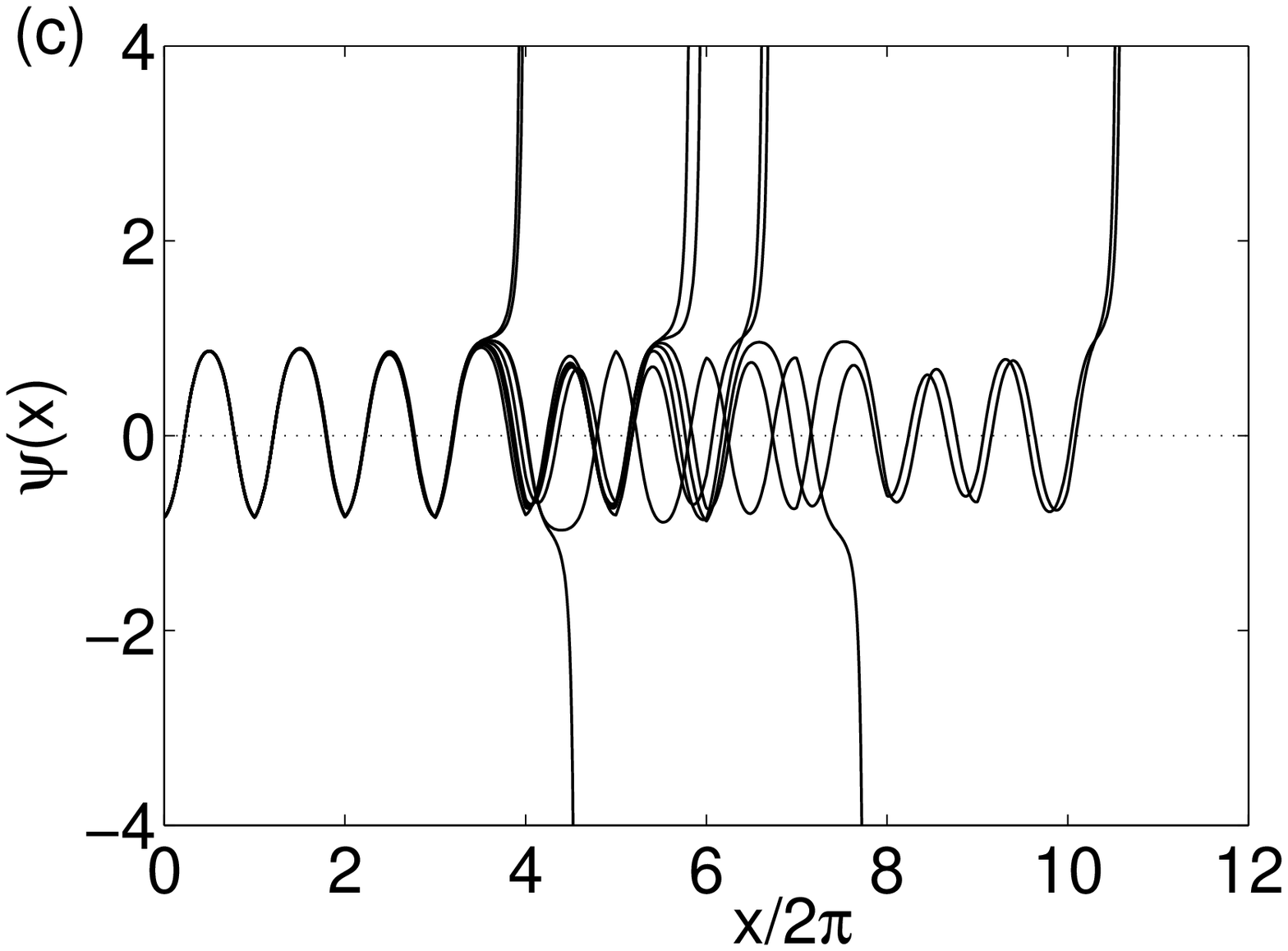}
\includegraphics[width=7cm,  angle=0]{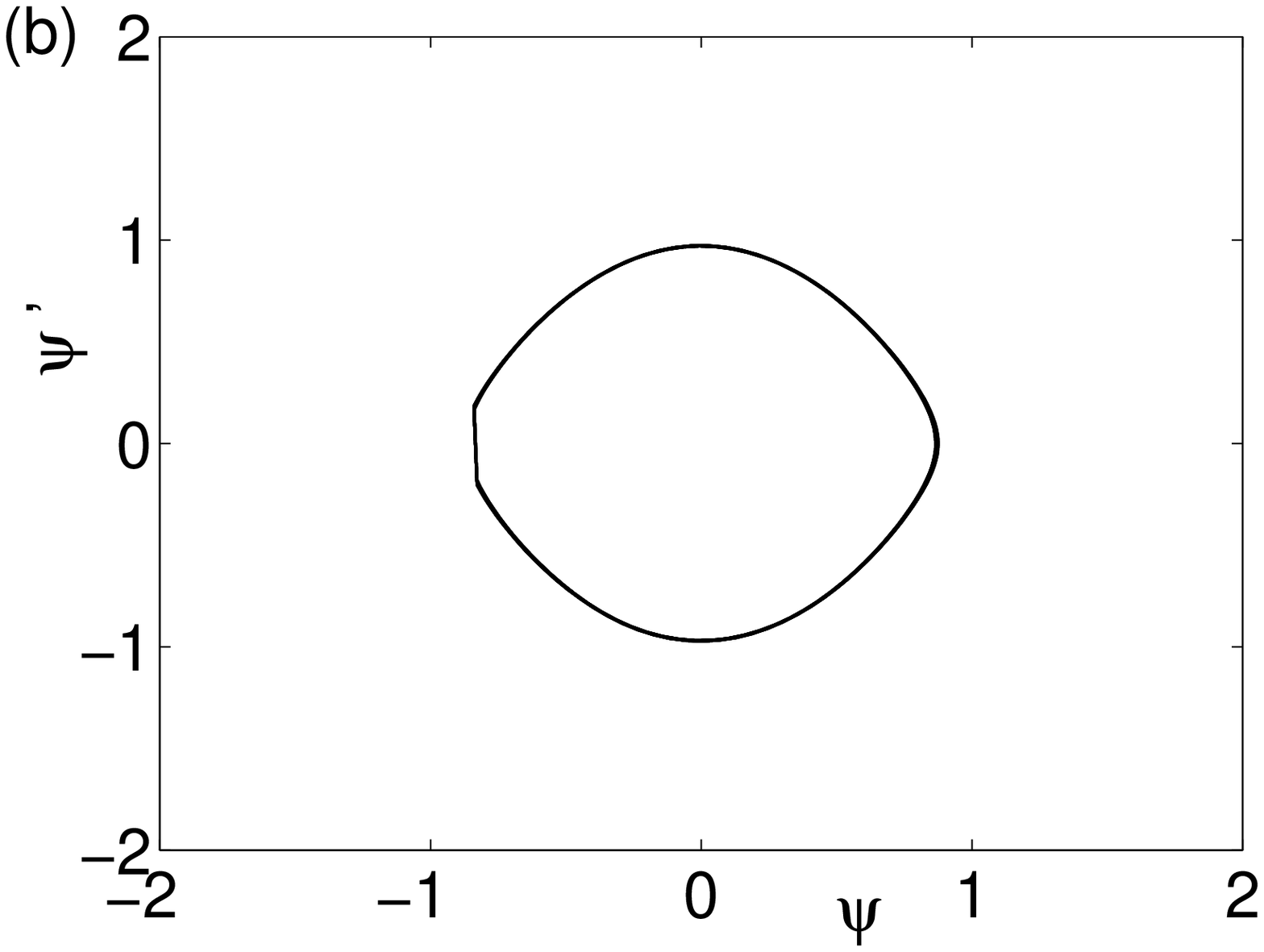}
\includegraphics[width=7cm,  angle=0]{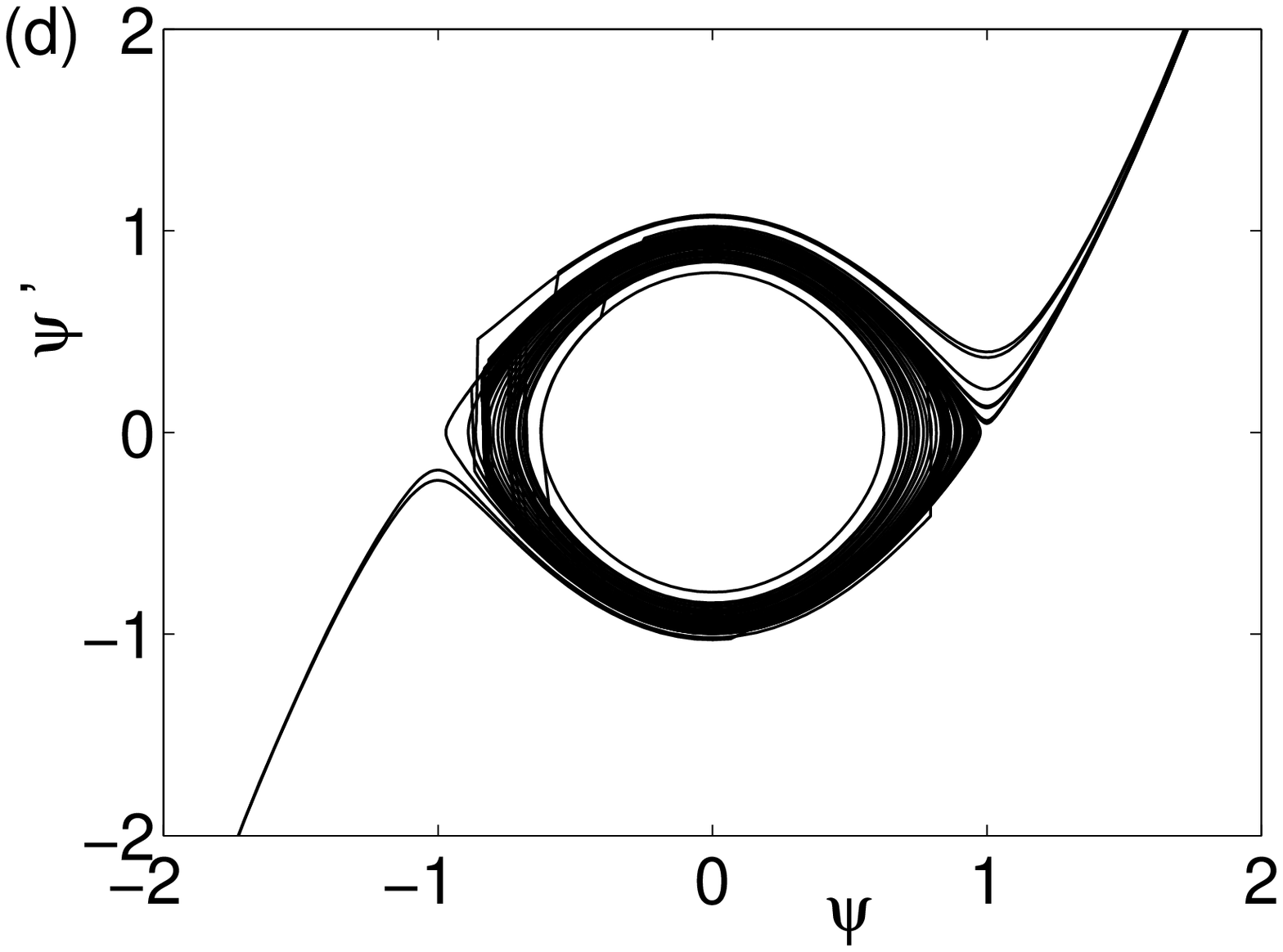}
\caption{\label{fig-instability}
(In)stability of the wave function $\psi(x)$ for $\mu=g=1$ and $\lambda =-0.2$
(a,b) and $\lambda=-0.22$ (c,d). The plots shows the 'dynamics' of the wave 
function $\psi(x)$ in configuration space (a,c) and in phase space (b,d). In all 
cases, the initial state $\psi(0),\psi'(0)$ was chosen as the symmetric periodic
solution plus a small random perturbation (standard deviation $\sigma = 10^{-3}$).
The periodic solution is elliptically for stable $\lambda =-0.2$ (a,b) so that the error
does {\it not} grow and the wave function $\psi(x)$ remains close to it for all $x$. 
The periodic solution becomes hyperbolically unstable for $\lambda = -0.22$
so that the error grows exponentially and the the wave function eventually
diverges. Ten trajectories with different random perturbations are plotted in the
figure, showing the rapid spreading of these trajectories.
}
\end{figure}

The dynamics is quasiperiodic for small amplitudes, i.e. in the vicinity
of the trivial solution $(\psi(x),\psi'(x)) = (0,0)$.
It is chaotic for large amplitudes at the edge of the sn-region,
corresponding to a stronger nonlinear interaction. The chaotic
part of phase space becomes larger with increasing potential strength
$|\lambda|$. The qualitative difference between stability and chaos
is illustrated in Fig.~\ref{fig-instability}, the 'dynamics' of the wave 
function $\psi(x)$ in configuration space and in phase space for 
$\mu=g=1$ and $\lambda =-0.2$  and $\lambda=-0.22$, respectively.
In both cases, the initial state $\psi(0),\psi'(0)$ was chosen as a periodic
solution plus a small random perturbation (standard deviation 
$\sigma = 10^{-3}$). This periodic solution is elliptically stable
for $\lambda = -0.2$ so that the perturbation does not grow with
$x$. It becomes hyperbolically unstable for $\lambda = -0.22$ so 
that the perturbation increases exponentially. Trajectories with
different perturbations spread rapidly, as shown in Fig.~\ref{fig-instability}
(c) and (d). The different forms of fixed points and their stability in 
dependence on the parameters will be discussed in detail in the 
next section.

\begin{figure}[htb]
\centering
\includegraphics[width=8cm,  angle=0]{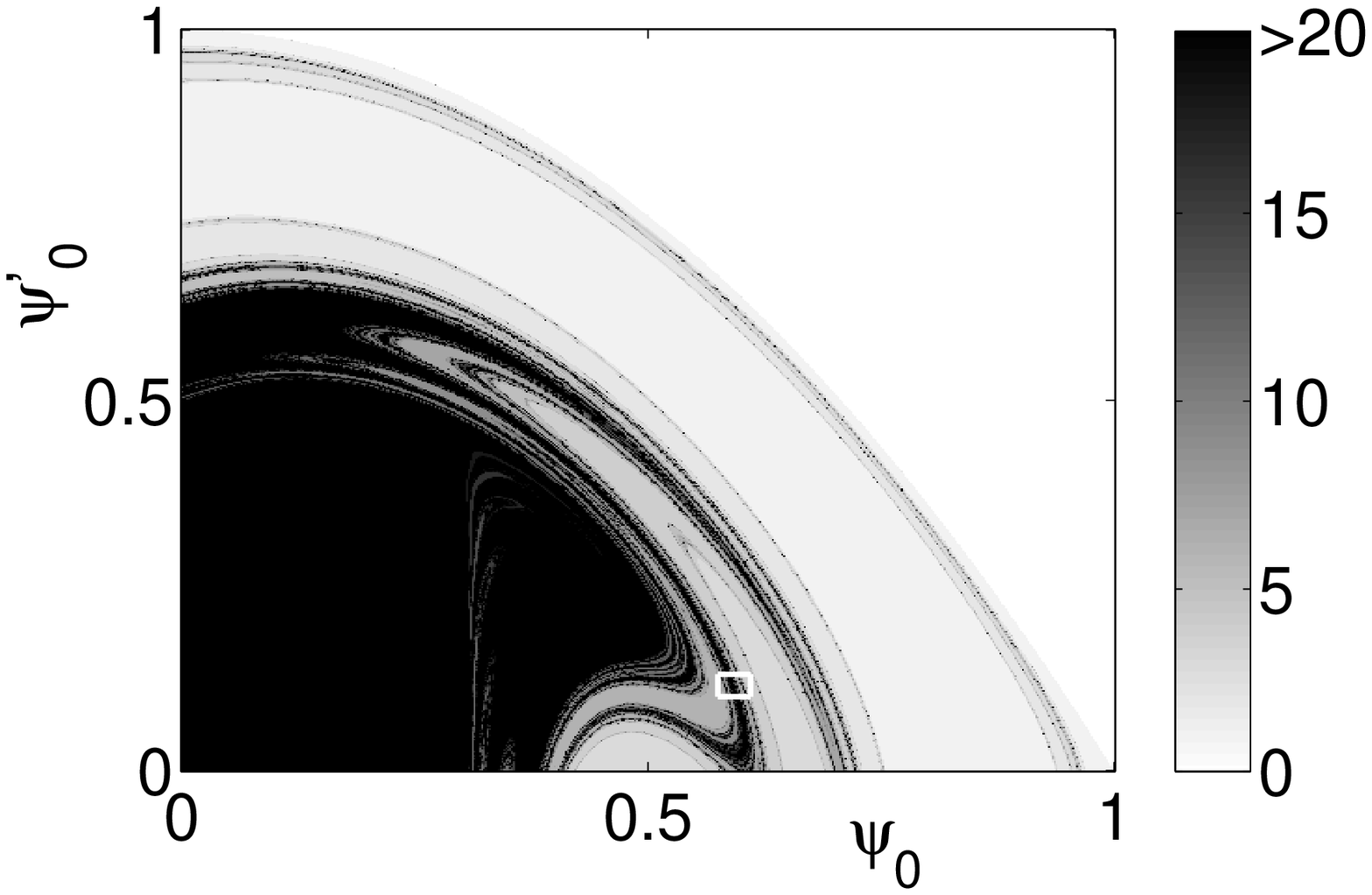}
\vspace{5mm}
\includegraphics[width=8cm,  angle=0]{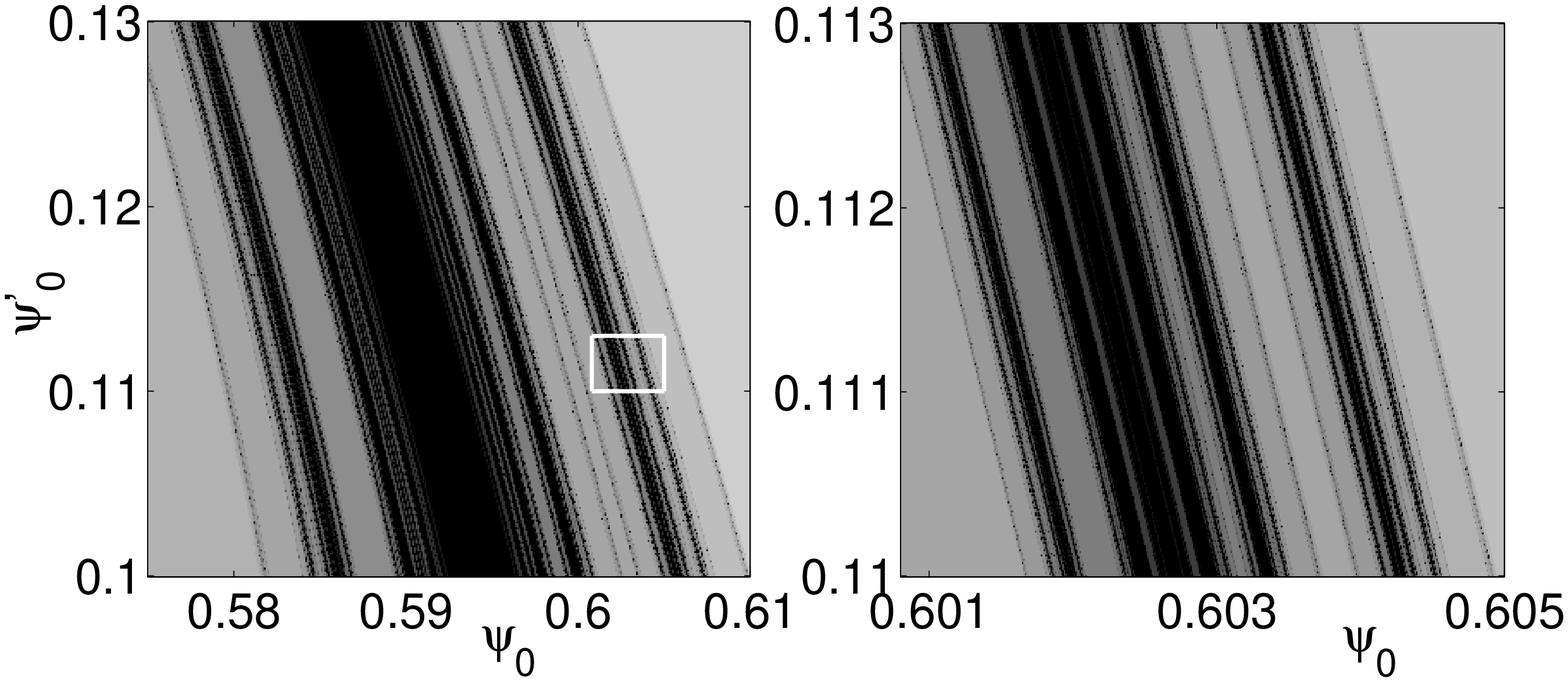}
\caption{\label{fig-rep_lenphase1}
Number of spatial periods, at which a trajectory diverges in dependence
of the initial condition $(\psi_0,\psi'_0)$ for $\lambda =0.5$. 
The lower panels show subsequent magnifications.} 
\end{figure}

As illustrated in Fig.~\ref{fig-instability} (c), a trajectory in the region of spatial chaos 
will generally diverge at some finite value of $x$.
Such a trajectory is usually mapped close to the edge of the sn-region in phase
space, where non-diverging solutions exist (cf.~Fig.~\ref{fig-rep_soltypes1}).
Due to the discontinuity of the derivative $\psi'(x)$ the wavefunction
leaves the sn-region. As mentioned above, all non-sn-type solutions of the NLSE
diverge at a finite value of $x$. Thus also trajectories started in the
chaotic part of the sn-region usually diverge at a finite value of $x$.
An example of such a diverging wavefunction is shown in Fig.
\ref{fig-instability} (c,d) for $\mu = g = 1$ and $\lambda = - 0.22$
(cf.~Fig.~\ref{fig-dcomb-rep-phase2}).
The initial value $(\psi(0),\psi'(0))$ was chosen as an unstable fixed point of
$\tilde f$ inside the chaotic region plus a small random perturbation.

The stability or divergence of a wavefunction depends sensitively on the
initial values $(\psi(0),\psi'(0))$. This is illustrated in Fig.~\ref{fig-rep_lenphase1}.
A grey-scale plot shows the point of divergence (the number of spatial periods until
a trajectory diverges) in dependence of the initial condition
$(\psi_0,\psi'_0)$ for $\lambda = 0.5$.
The left figure shows the position of divergence for one
quadrant of phase space, $\psi_0,\psi'_0 \ge 0$. This figure should be compared
with the Poincar\'e section in Fig.~\ref{fig-dcomb-rep-phase1}.
One clearly recognizes that trajectories with small amplitudes are quasiperiodic
and thus stable.
But one can also find initial values that do not lead to divergences for larger
amplitudes. These values form an approximately self-similar set in phase space.
This is illustrated in Fig.~\ref{fig-rep_lenphase1} in the lower plots, where
magnifications of the upper plot are shown.

\subsection{Attractive nonlinearity}
\label{sec-attractive-nonlin}

For the sake of completeness we also briefly discuss the case of an attractive
nonlinearity $g < 0$ without going into details.

Real solutions of the nonlinear Schr\"odinger equation for a delta comb
can be constructed in terms of the Jacobi elliptic function cn and dn.
The cn-type solutions are given by
\be
  \psi(x) = A_n \, \cn \bigg( 4 K(p_n) \frac{x+x_n}{L_n} \bigg| p_n \bigg)
\ee
for $x \in (2\pi n,2\pi(n+1))$ with
\be
  A^2_n = \frac{2 \mu \, p_n}{g (2 p_n-1)} \; \, \mbox{and} \;
  L_n^2 = \frac{8  (1 - 2p_n)  K(p_n)^2}{\mu} \, .
  \label{eqn-dcomb-cntype-amp-mu}
\ee
The dn-type solutions read
\be
  \psi(x) = A_n \, \dn \bigg( 4 K(p_n) \frac{x+x_n}{L_n} \bigg| p_n \bigg)
\ee
for $x \in (2\pi n,2\pi(n+1))$ with
\be
  A^2_n = \frac{\mu}{g (2-p)} \quad \mbox{and} \quad
  L^2_n = \frac{8 (p-2) K(p)^2}{\mu} .
\ee
Note that cn-type solutions exist for $\mu < 0$, leading to $p \in (0.5, 1]$,
as well as for $\mu > 0$, leading to $p \in [0, 0.5)$, whereas the dn-type solutions
exist only for $\mu < 0$.

\begin{figure}[htb]
\centering
\includegraphics[width=7cm,  angle=0]{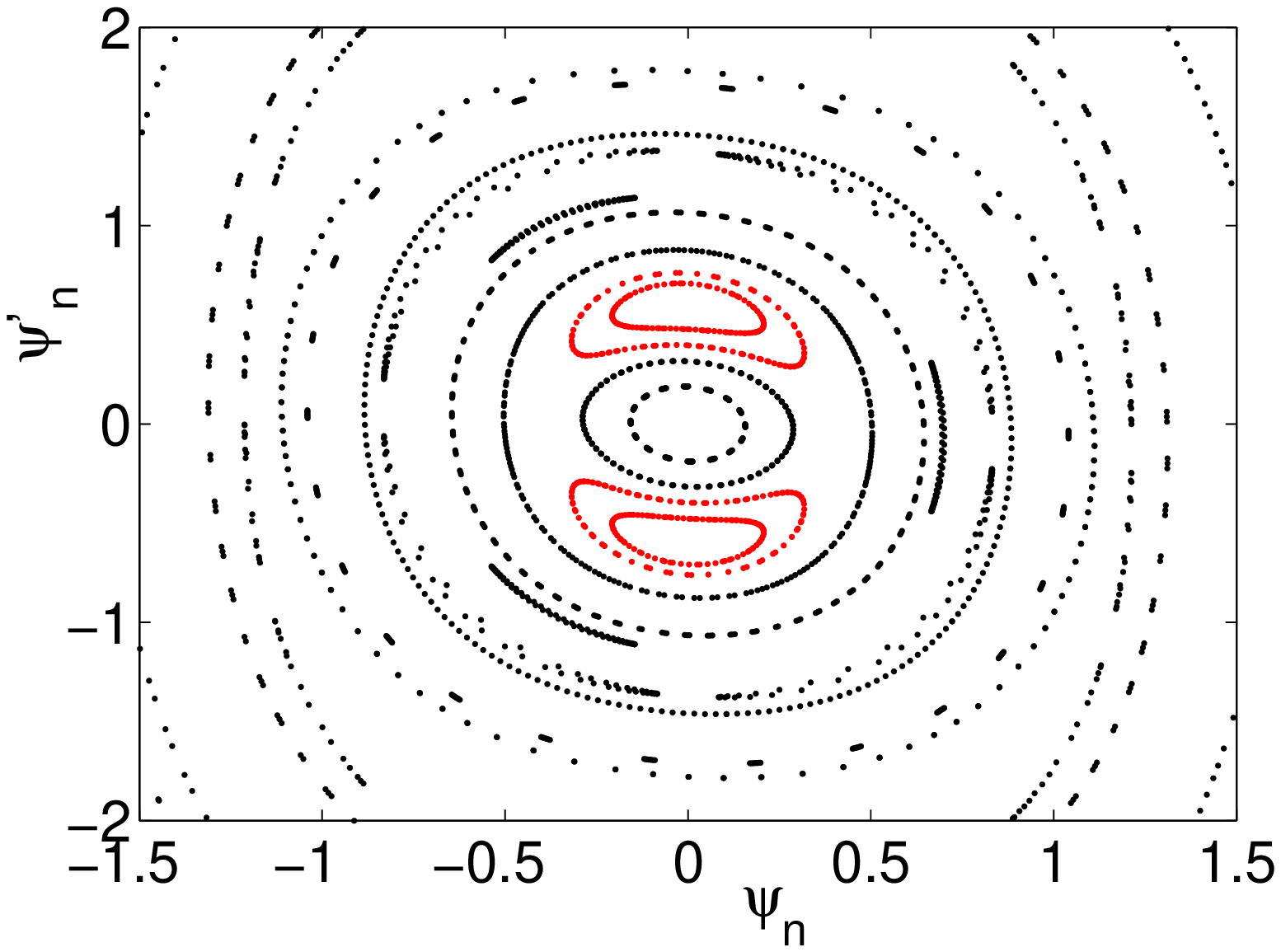}
\includegraphics[width=7cm,  angle=0]{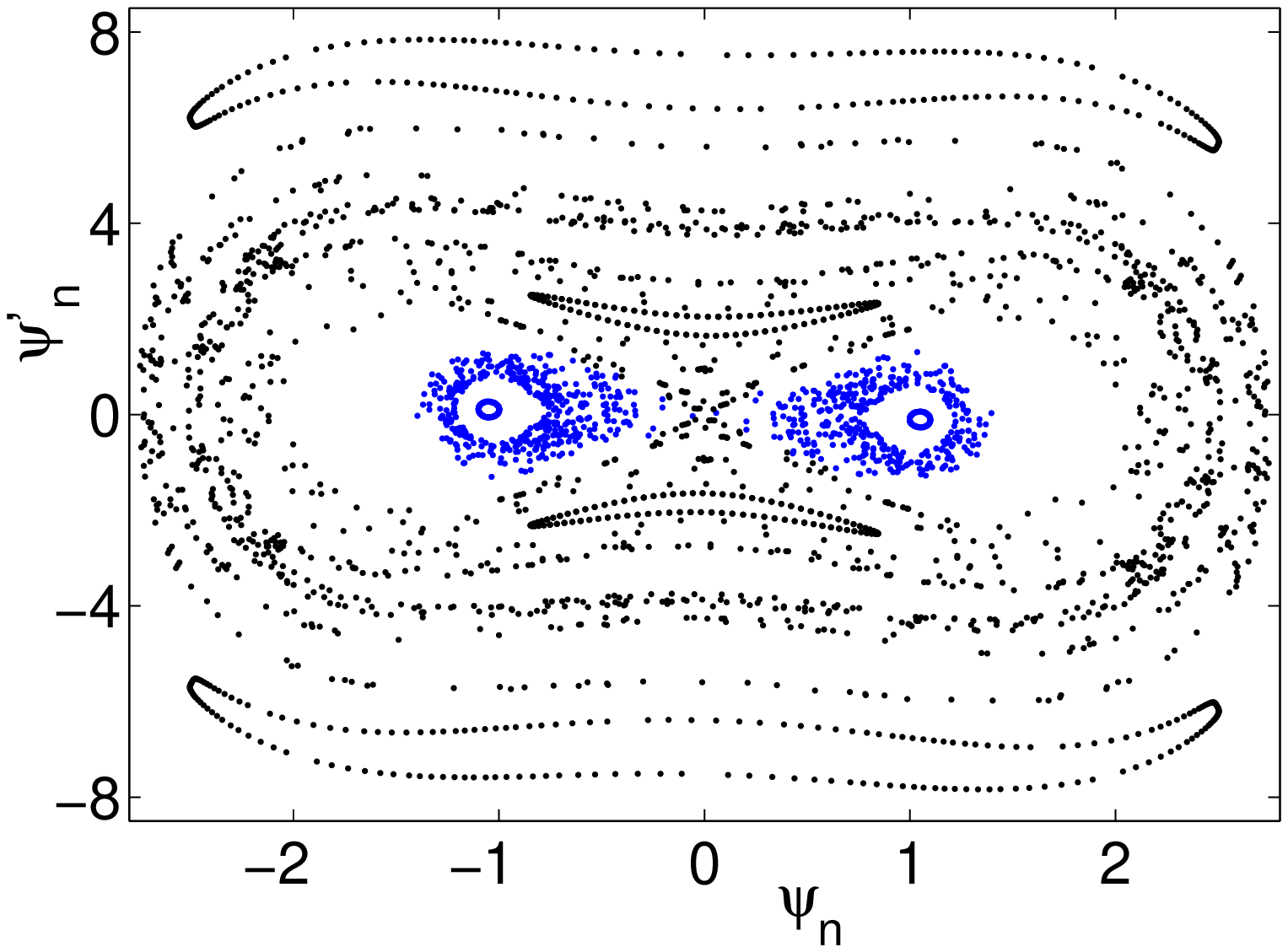}
\caption{\label{fig-dcomb-att-phase}
Stroboscopic phase space plots of the mapping
$\tilde f: (\psi_n,\psi'_n) \rightarrow (\psi_{n+1},\psi'_{n+1})$
for an attractive nonlinearity $g=-1$
and $\mu = +1$ (left panel) and $\mu = -2$ (right panel).}
\end{figure}

Again, two conditions have to be fulfilled at $x = 2\pi n$: The wavefunction
is continuous, whereas its derivative is discontinuous according
to Eqn.~(\ref{eqn-delta-condition}). As above we can consider the
dynamics stroboscopically at and thus arrive at the mapping
$\tilde f: (\psi_n,\psi'_n) \rightarrow (\psi_{n+1},\psi'_{n+1})$.
Examples of the dynamics in phase space in a delta comb of strength
$\lambda = -0.1$ are plotted in Fig.~\ref{fig-dcomb-att-phase}.

An important difference to the case of a repulsive interaction is that one
can find a solution of the free NLSE in terms of the non-diverging Jacobi
elliptic functions cn and dn for {\it all} initial values of a wave function
$\psi(x_0), \psi'(x_0)$. This should be compared to the case of a repulsive
interaction, in particular to Fig.~\ref{fig-rep_soltypes1}.
Hence the solutions for a delta-comb do not diverge at a finite value of $x$,
regardless of the starting point $(\psi_0,\psi'_0)$. The dynamics in phase space
around $(\psi_n,\psi'_n) =(0,0)$ is illustrated in Fig.~\ref{fig-dcomb-att-phase}
for $\lambda = -0.1$ and two different values of $\mu$.

The solutions are always given by the Jacobi elliptic function cn for $\mu > 0$.
For $\mu=1$ and $\lambda = -0.1$, the dynamics is still completely regular and one we can
identify fixed points again (cf.~the left panel of Fig.
\ref{fig-dcomb-att-phase}).
The dynamics becomes partly chaotic for higher values of $\lambda$, but
still a wave function cannot diverge at a finite value of $x$.
However, a totally different mechanism of divergence exists in this case.
The wave function $\psi(x)$ diverges for $x \rightarrow \infty$, if its phase
is such that the changes of the amplitude due to the delta kicks accumulate.
Revisiting Eqn.~(\ref{eqn-dcomb-cntype-amp-mu}) one recognizes that this
divergence of the amplitude $A_n$ implies $p_n \rightarrow 0.5$.
Indeed one finds numerically that most trajectories in the chaotic region of
phase space show such a behavior.

For $\mu < 0 $ the solution is described either by the function cn or dn, depending
on the initial value $(\psi_0,\psi'_0)$. The dynamics in phase space is illustrated
in the right panel of Fig.~\ref{fig-dcomb-att-phase} for $\lambda = -0.1$.
Solutions of dn-type are found around the elliptic fixed points at
$(\psi_n,\psi'_n) = (\pm 1,0)$.
One can identify regular islands in a chaotic sea, surrounded by quasi-periodic orbits
for large amplitudes. Again the cn-type solutions can diverge for $x \rightarrow \infty$
if $p_n \rightarrow 0.5$.

\subsection{Complex solutions}

Complex solutions of the free NLSE can be found by decomposing the wave function
into amplitude and phase
\be
  \psi(x) = \sqrt{S(x)} \, \re^{\ri \phi(x)}
\ee
with real-valued functions $\phi(x)$ and $S(x) \ge 0$.
The phase is then given by
\be
   \phi'(x) = \frac{\alpha}{S(x)}
   \label{eqn-complex-phase}
\ee
with a real integration constant $\alpha$.
In the limit $\alpha \rightarrow 0$ one recovers the real-valued sn- and cn-type
solutions described above, which change sign at the zeros of the density. This 
limit is discussed in detail in \cite{06nl_transport} where interaction-induced transitions 
from scattering states to bound states are analyzed for a finite square well potential.

We focus on the special solution \cite{Carr00,06nl_transport}
\be
  S(x) = B_n - \frac{\varrho_n^2}{g} \,  \dn^2 \bigg(\varrho_n (x+x_n) \bigg| p_n \bigg)
  \label{eqn-ansatz-complex}
\ee
of the free NLSE with $\varrho_n  = 4K(p_n)/L_n$.
Inserting this ansatz into the free NLSE yields the following conditions for the
remaining constants
\begin{align}
   & 2\mu = 3 g B_n - (2-p_n) \varrho_n^2 \nn \\
   & \varrho_n^4 (p_n-1) B_n + g \alpha_n^2 + 2 g_n^2 B_n - 2g \mu B_n^2 = 0. \nn
\end{align}

For the delta-comb potential, one can make the ansatz (\ref{eqn-ansatz-complex})
separately for each interval $x \in (2\pi n,2\pi(n+1))$.
The derivative $\psi'$ of the the wave function is again discontinuous at
$x = 2 \pi n$ according to Eqn.~(\ref{eqn-delta-condition}).
In terms of the density $S(x)$ this yields
\be
   \lim_{\epsilon \to 0+}\big( S'(2\pi n+\epsilon) - S'(2 \pi n-\epsilon)\big)
   = 4 \lambda S(2 \pi n).
   \label{eqn-delta-condition-density}
\ee
Complex periodic solutions of this type will be discussed in Sec.
\ref{sec-per-complex}. These solutions are of great importance, since they
occur for nonlinear Bloch bands (cf.~Sec. \ref{sec-bloch-bands}).


\section{Periodic solutions}
\label{sec-periodic-states}

Real periodic solutions can be viewed as fixed points of the mapping $f$ resp.
$\tilde f$ defined in Sec.~\ref{sec-dcomb-real-states}.
In the following we calculate these solutions explicitly and discuss their
spatial stability.

\subsection{Symmetric and antisymmetric real periodic solutions}
\label{sec-trivial-symmetric-states}

One kind of periodic solutions is found for $\psi(2 \pi n) = 0$.
They will be named antisymmetric periodic solutions in the
following since they fulfill $\psi(2\pi n - x)  = -\psi(2\pi n + x)$.
In this case the wave function and its derivative are continuous everywhere.
The period length of the Jacobi elliptic functions $L$ must fulfill
\be
  L = 4 \pi / m
  \label{eqn-period-antismym}
\ee
for $m \in \mathbb{N}$.
The antisymmetric solutions are $2\pi$-periodic only for even $m$, since
their fundamental period is $4\pi/m$. Note that $m$ is the number of nodes
of the wave function in $[0,2\pi)$.
In the following we will mostly consider the ''primary resonance'' $m = 2$
with the fundamental period $2 \pi$. However, we will show in 
Sec.~\ref{sec-periodic-stability} that this solution bifurcates when $\lambda$ is
varied and new period doubled solutions emerge.

In terms of the elliptic parameter $p$ Eqn.~(\ref{eqn-period-antismym}) reads
\be
  (p+1) \, K(p)^2 = \frac{2 \pi^2 \mu}{m^2} .
  \label{eqn-def-trivial-per-states}
\ee
The left side of this equation is bounded from below by
$(p+1) \, K(p)^2 \ge K(0)^2 = \pi^2/4$. Thus antisymmetric periodic solutions
only exist if $\mu \ge m^2/8$.
Comparing with the case of a harmonic potential, i.e. the nonlinear Mathieu equation
\cite{Nayf79,Mont82}, one recognizes that an antisymmetric periodic solution comes
into being just at the $2:m$ resonance $2 \mu = m^2/4$.

\begin{figure}[tb]
\centering
\includegraphics[width=7cm,  angle=0]{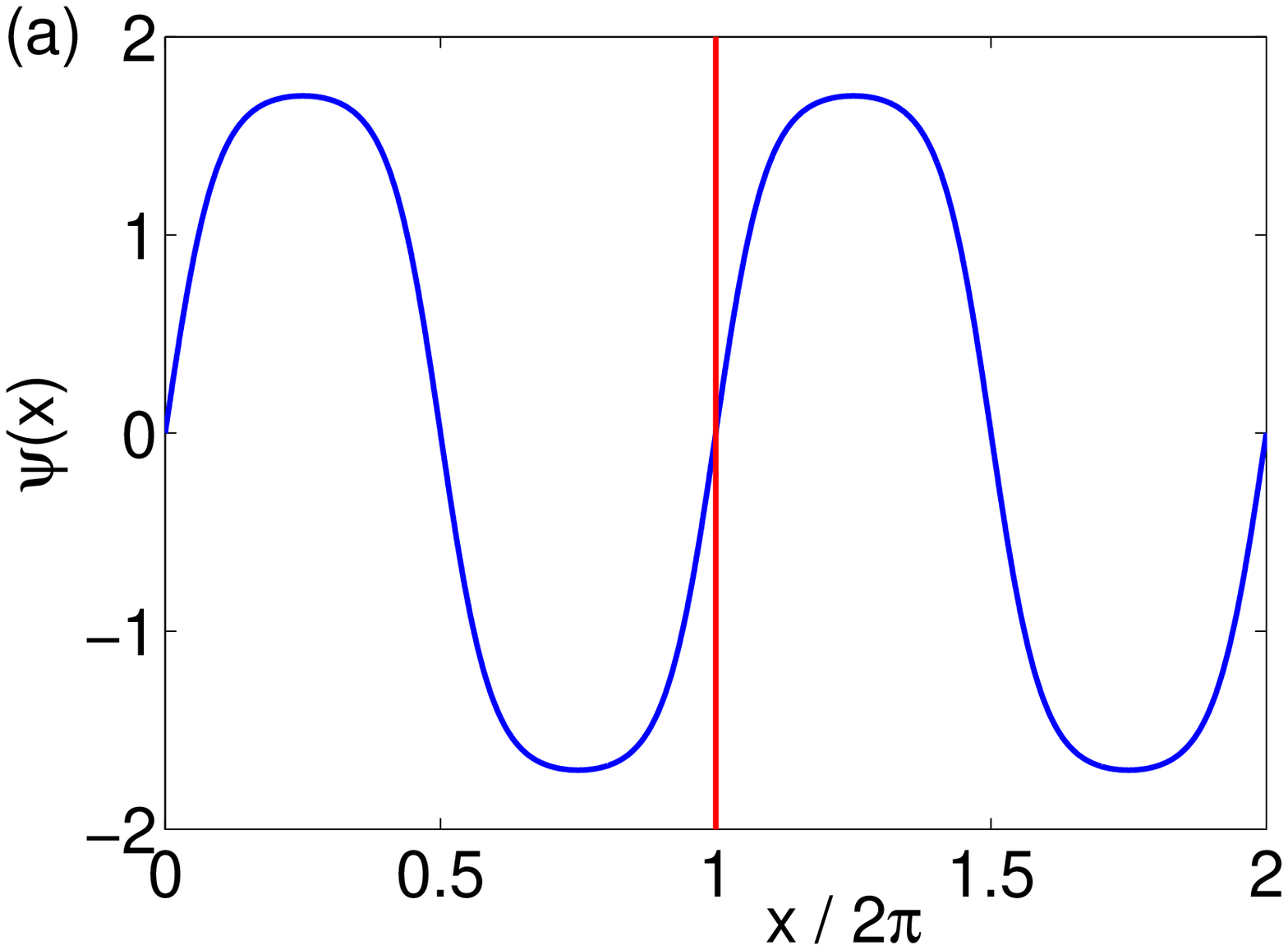}
\hspace{5mm}
\includegraphics[width=7cm,  angle=0]{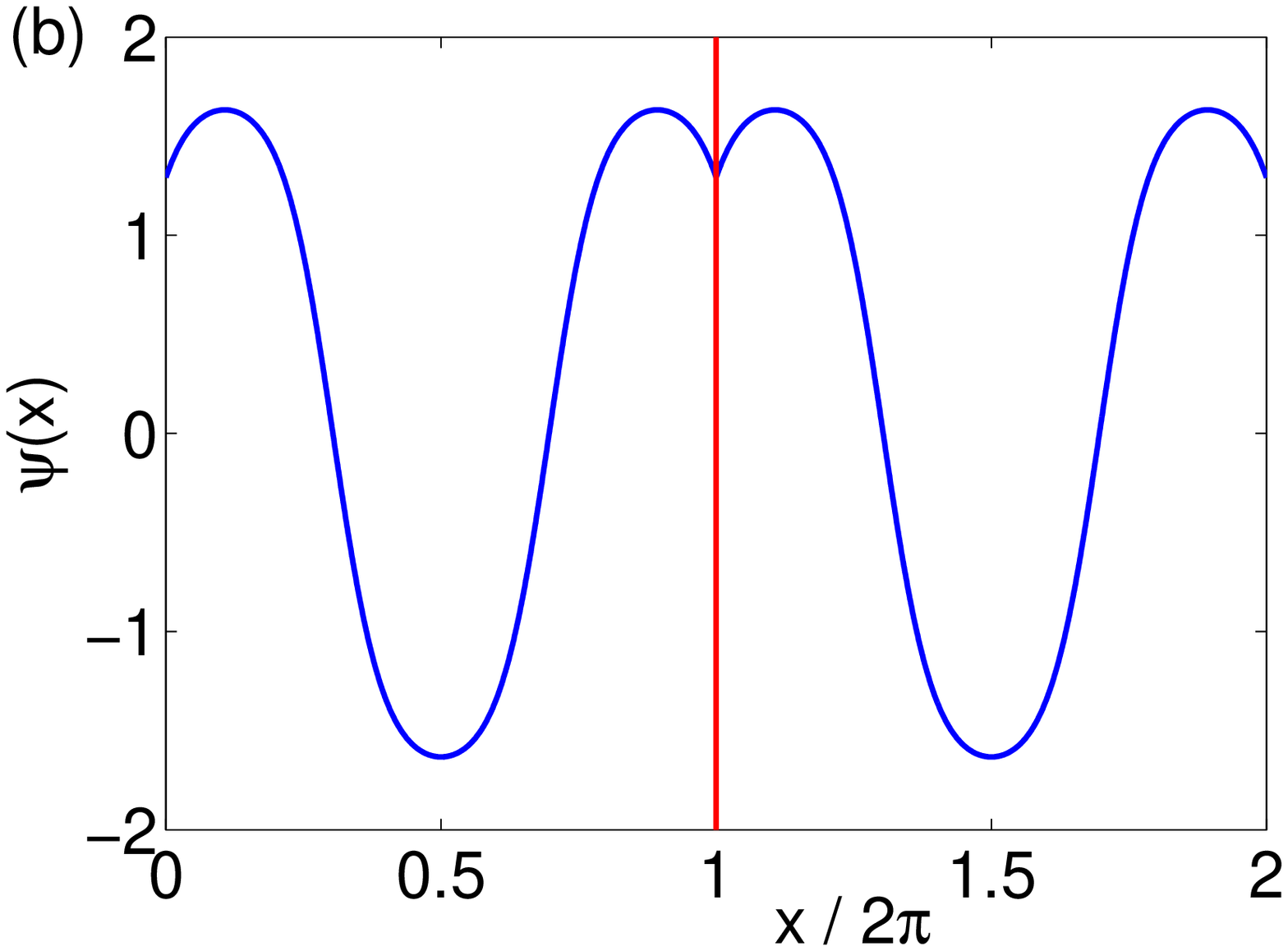}
\hspace{5mm}
\includegraphics[width=7cm,  angle=0]{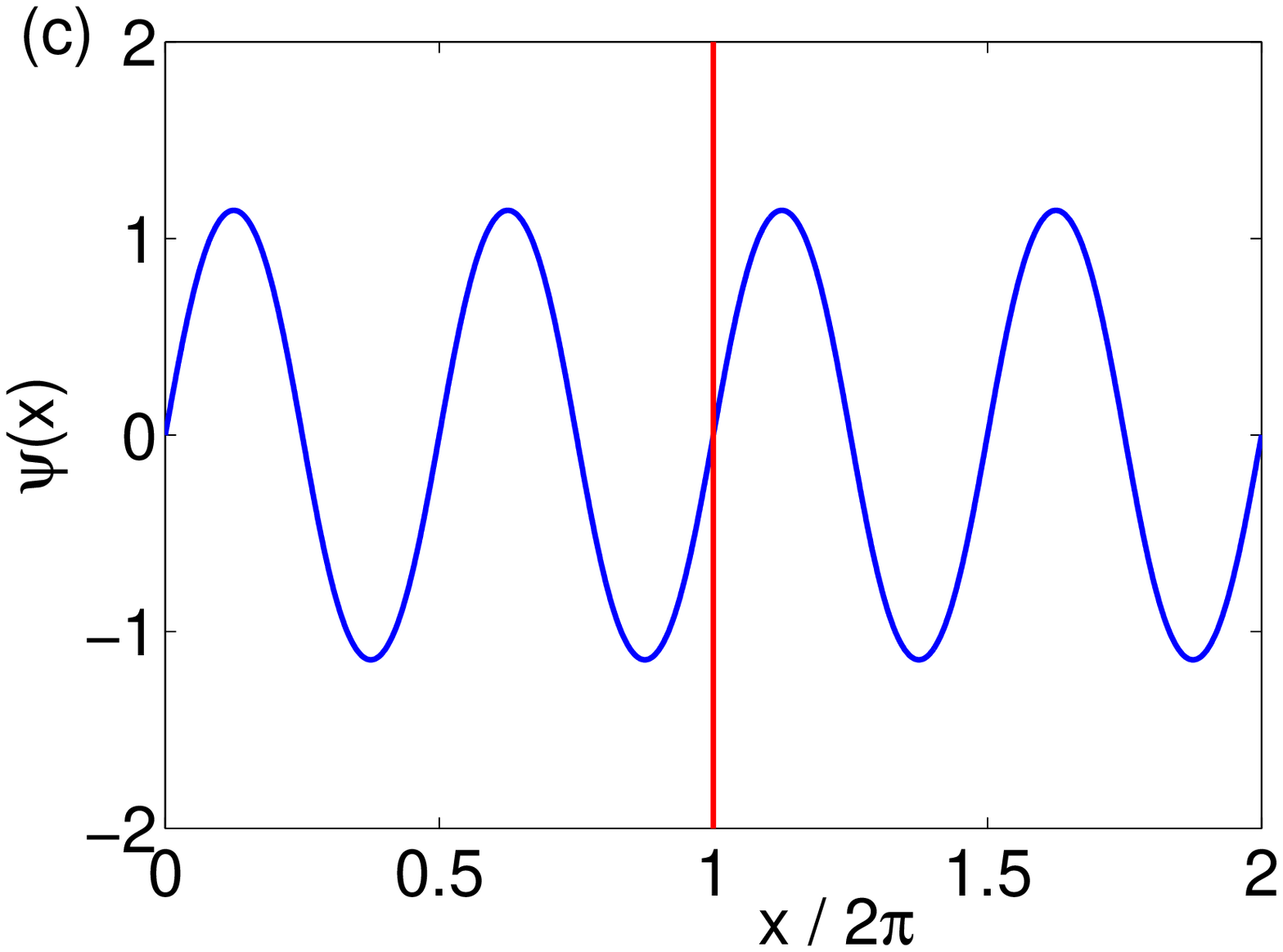}
\hspace{5mm}
\includegraphics[width=7cm,  angle=0]{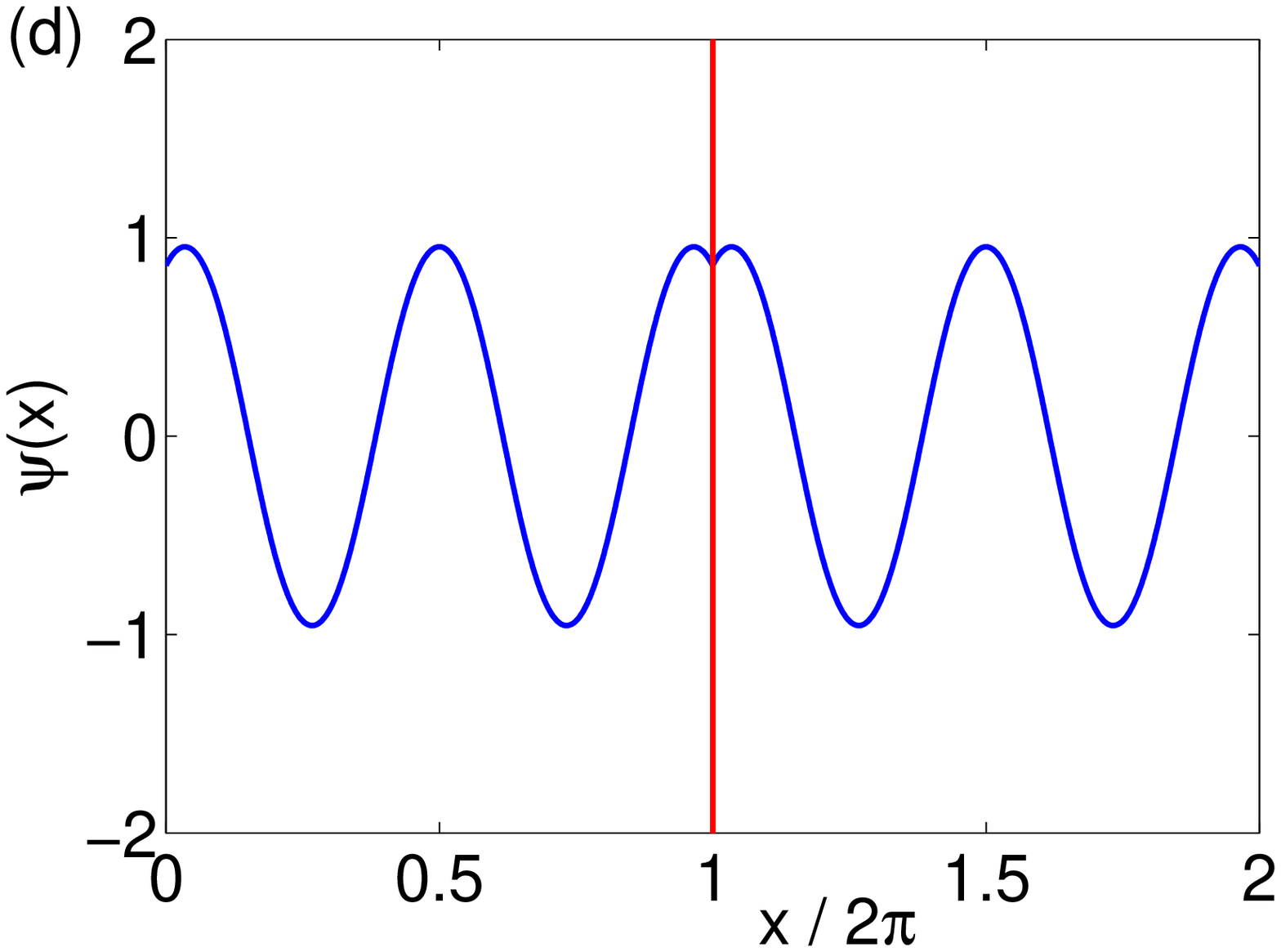}
\caption{\label{fig-fixed_points_examples}
Periodic solutions $\psi(x)$ of the NLSE for a delta-comb potential of strength 
$\lambda = 1$ for $\mu = 3$ and $g = 1$. The red vertical lines indicate the 
positions of the delta potentials.
Antisymmetric solutions (a,c) have a node at the positions of the delta potential,
$\psi(2\pi n) =  0$, so that they experience no discontinuity of the first derivative 
and are thus independent of  $\lambda$. Symmetric solutions (b,d) are non-zero
at $x = 2\pi n$ and symmetric with respect to a reflexion at this point.  
Furthermore the solutions are characterized by the number of nodes in the
interval $[0,2\pi)$ - the examples have $m=2$ (a,b) and $m=4$ (c,d) nodes,
respectively.}
\end{figure}

The shift $x_n$ is given by $x_n = 0$ resp. $x_n = \pi$.
However, these solutions differ only by an overall sign.
Note that the solution of Eqn.~(\ref{eqn-def-trivial-per-states})
does not depend on $\lambda$.
As an example the antisymmetric periodic solutions with
$m = 2$ and $m=4$ re shown in Fig.~\ref{fig-fixed_points_examples}
(a) and (c), respectively,  for $\mu = 3$ and $g = 1$.

Furthermore, $2\pi$-periodic solutions can be found which are symmetric
around the positions of the delta potentials at $x = 2 \pi n, \, n \in \mathbb{Z}$.
This symmetry and the periodicity of the wave function imply
that the solution is also symmetric around $x = (2n+1)\pi$,
as one can easily see:
\be
  \psi((2n+1)\pi-x) =  \psi(-(2n+1)\pi+x) = \psi((2n+1)\pi+x).
\ee
Hence the the wave function assumes a maximum or minimum at
$x = (2n+1)\pi$, in the middle between two delta potentials.
Without loss of generality we conclude that
\be
  x_n = -(2n+1)\pi + L/4 \quad \mbox{or} \quad x_n = (2n-1)\pi + 3L/4.
\ee
These solutions differ only by an overall sign, hence only one of them
must be considered.
Again condition (\ref{eqn-delta-condition}) must be fulfilled at the
positions of the delta-comb.
Because of the periodicity it is, however, sufficient to evaluate this
equation at $x = 0$ only.
A further simplification arises from the symmetry of the wave function,
yielding the condition
\be
  \lambda \sn(u|p) = \frac{4 K(p)}{L} \, \cn(u|p) \, \dn(u|p)
  \label{eqn-def-symmetric-sols}
\ee
with $u = 4 K(p) x_0 /L = K(p) (1 - 4 \pi/L)$.
Eqn.~(\ref{eqn-def-symmetric-sols}) has a solution for every $\mu >0$.
For higher values of $\mu$ it may have several solutions, differing in the
period $L$ and correspondingly in the number of nodes of the wave function.
Again, these solutions will be labeled by the number $m$ of nodes of the
wave function in the interval $[0,2\pi)$. In the following we consider
the solution with $m=2$ nodes. Exemplarily the solutions with $m=2$ and
$m=4$ are shown in  Fig.~\ref{fig-fixed_points_examples} (b) and 
(d), respectively, for $\mu = 3$, $g=1$ and $\lambda = 1$.

One can easily see that these two solution classes are the only possible
$2\pi$-periodic sn-type wave functions.
Due to the periodicity the elliptic parameter $p_n$ is the same for all $n$,
i.e. $p_n = p$.  As the elliptic parameter $p$ is fixed, the derivative $\psi'(x_0)$
at a point $x_0$ is determined by the wave function $\psi(x_0)$ up to a sign.
Considering $x_0 = 2 \pi n$ this leaves the possibilities
$\psi'(2\pi n+\epsilon) = \psi'(2 \pi n-\epsilon)$, for which one recovers
the antisymmetric periodic solutions, whereas the other possibility
$\psi'(2\pi n+\epsilon) = - \psi'(2 \pi n-\epsilon)$ leads to the
symmetric periodic solutions.

\subsection{Spatial Stability of the real periodic solutions}
\label{sec-periodic-stability}

As already stated, real periodic solutions of the NLSE in a delta-comb
are the fixed points of the mapping $f$ resp. $\tilde f$, defined in 
Sec.~\ref{sec-dcomb-real-states}. Similarly, we can find solutions with a
periodicity of $2\pi r$ as fixed points of the mapping $f^r$.
In the following we discuss the characteristics of these fixed points,
especially their spatial stability, in dependence of the control-parameter
$\lambda$.

\begin{figure}[tb]
\centering
\includegraphics[width=7cm,  angle=0]{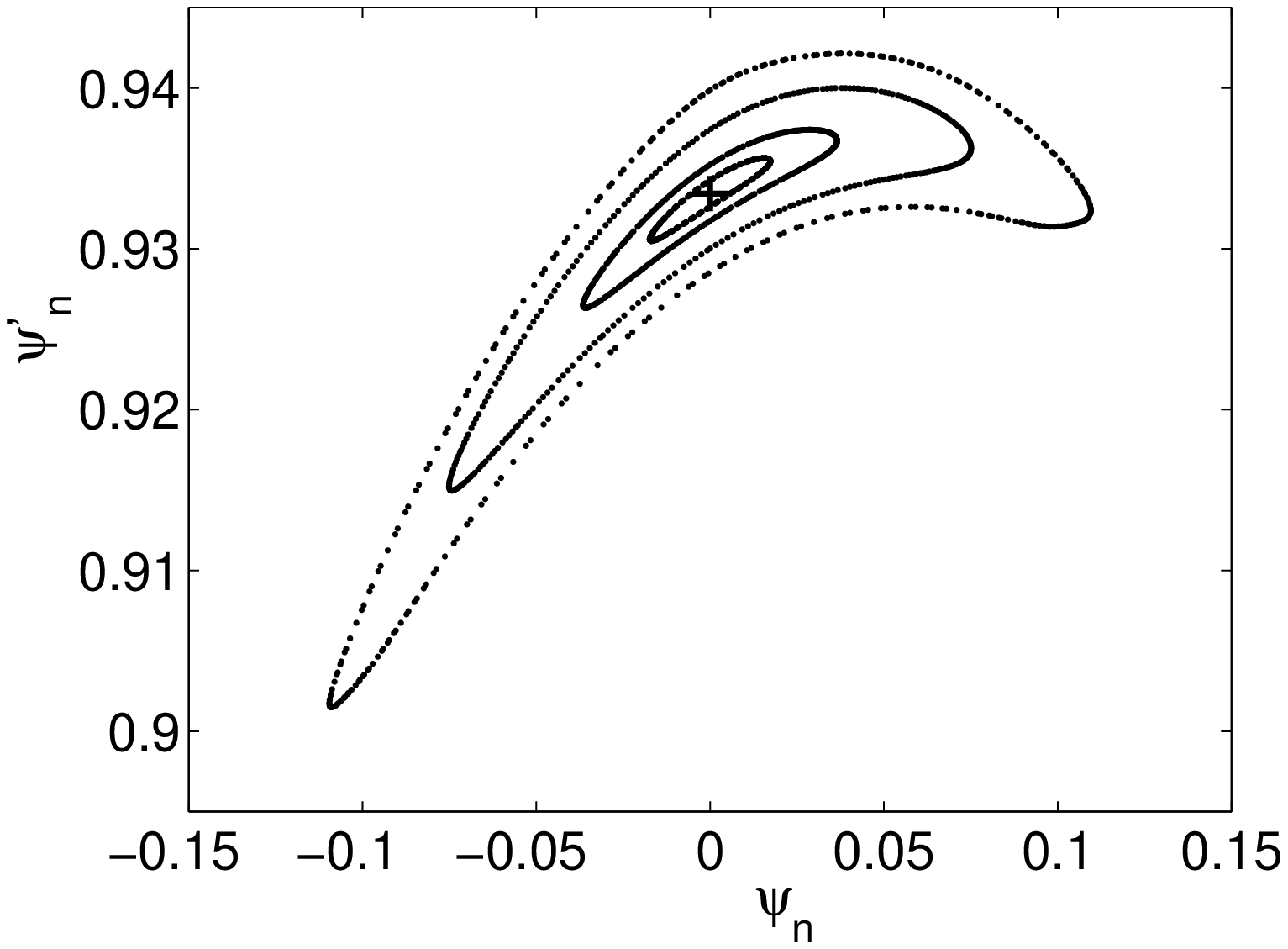}
\hspace{5mm}
\includegraphics[width=7cm,  angle=0]{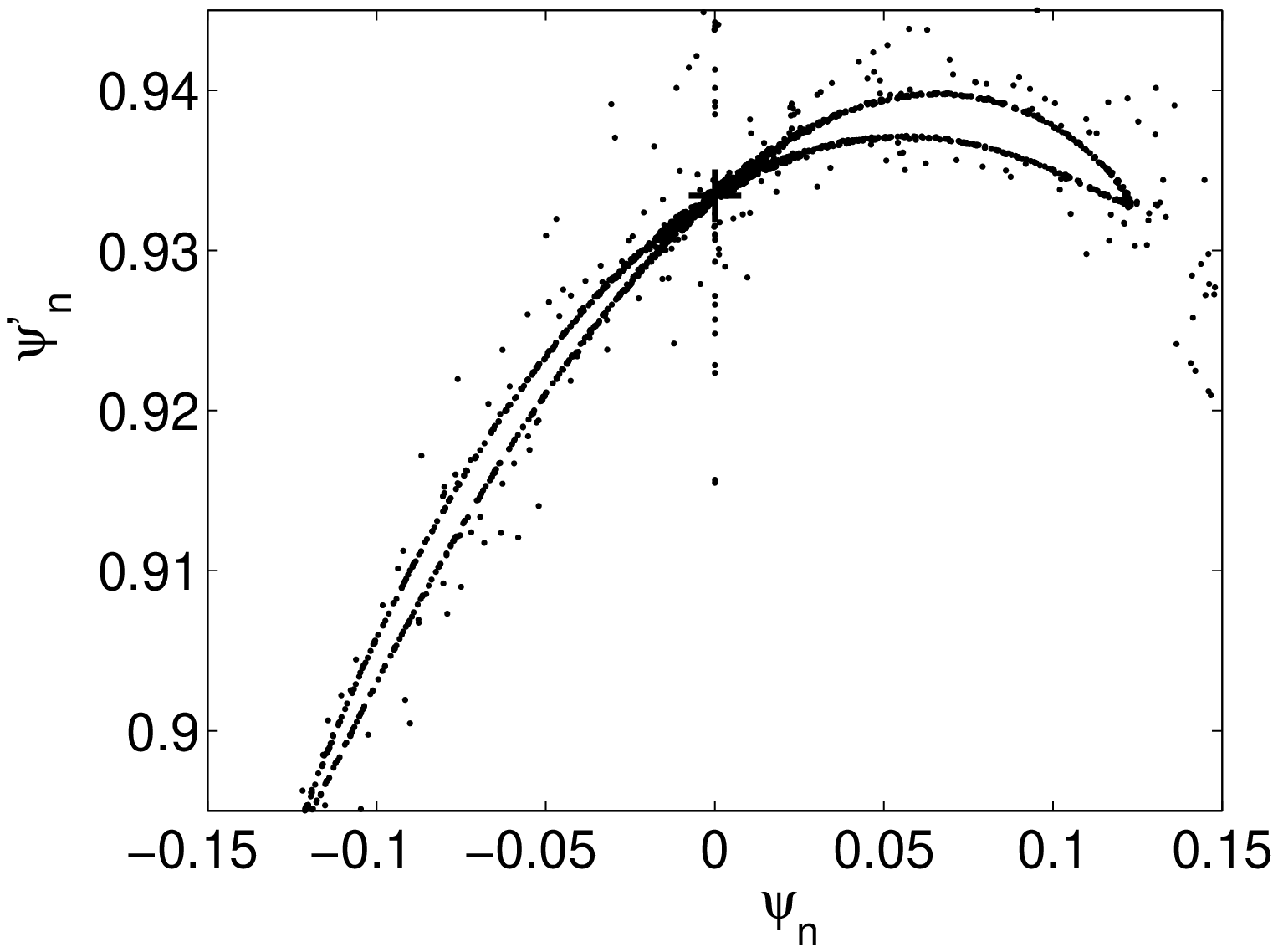}
\caption{\label{fig-rep-phase-stab1}
Phase space around the antisymmetric fixed point of $f$
(marked by a $+$) before the bifurcation ($\lambda = 0.14$, left)
and after the bifurcation ($\lambda = 0.16$, right).}
\end{figure}

First we discuss the antisymmetric periodic solutions with $L = 2\pi/m$ for a
repulsive nonlinearity $g=1$ and $\mu=1$.
As an example, we consider the solutions with $m=2$, which have nodes at
$x =  \pi n, \, n \in \mathbb{Z}$, i.e. at the positions of
the delta-potentials $x = 2\pi n$ and additionally in the middle between
them. These periodic solutions are found for $p \approx 0.4719$
and $x_n = 0$, resp. $x_n=\pi$, differing only by an overall sign.

To determine the stability of a fixed point numerically we iterate
the mapping $f$, where the initial value $(p_0,x_0)$ is chosen as
the fixed point plus a small random perturbation.
Furthermore we calculate the stability (or monodromy) matrix of the mapping
$\tilde f$
\be
  M(\psi_*,\psi'_*) = \left[ \frac{\partial(\psi_{n+1},\psi'_{n+1})}{\partial(\psi_n,\psi'_n)}
   \right]_{\psi_*,\psi'_*}
\ee
at the fixed point $(\psi_*,\psi'_*)$.
As the mapping $\tilde f$ is area-preserving, the product of the
eigenvalues $\exp(\pm \gamma)$ of $M$ is unity.
The mapping is unstable if the stability exponent $\gamma$ is real.
Otherwise it is purely imaginary and the mapping is elliptically stable.
Then $|\gamma|$ is called the stability angle.

Exemplarily we consider the antisymmetric fixed point with $m = 2$, which is
found to be elliptically stable if $\lambda > 0$ and $\lambda$ is not too large.
The stability exponents $\gamma_{\pm}$ are
purely imaginary and one finds quasiperiodic orbits around the fixed point.
This is shown in the left panel of Fig.~\ref{fig-rep-phase-stab1} for
$\lambda = 0.14$. The fixed point becomes unstable when the control-parameter
$\lambda$ is increased above a critical value $\lambda_c \approx 0.156$.
This is illustrated in the right panel of Fig.~\ref{fig-rep-phase-stab1}
for $\lambda = 0.16$.

\begin{figure}[tb]
\centering
\includegraphics[width=7cm,  angle=0]{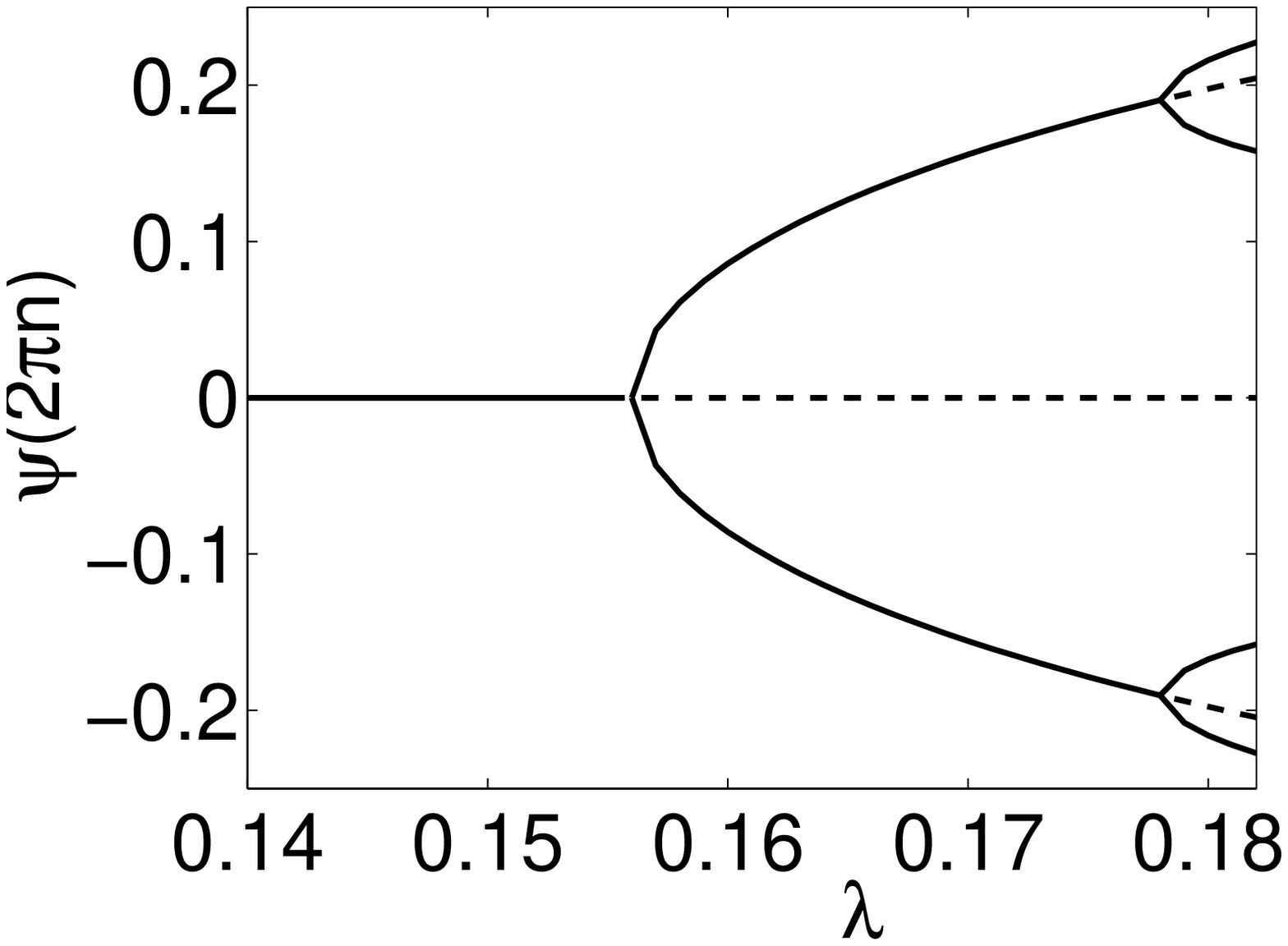}
\includegraphics[width=7cm,  angle=0]{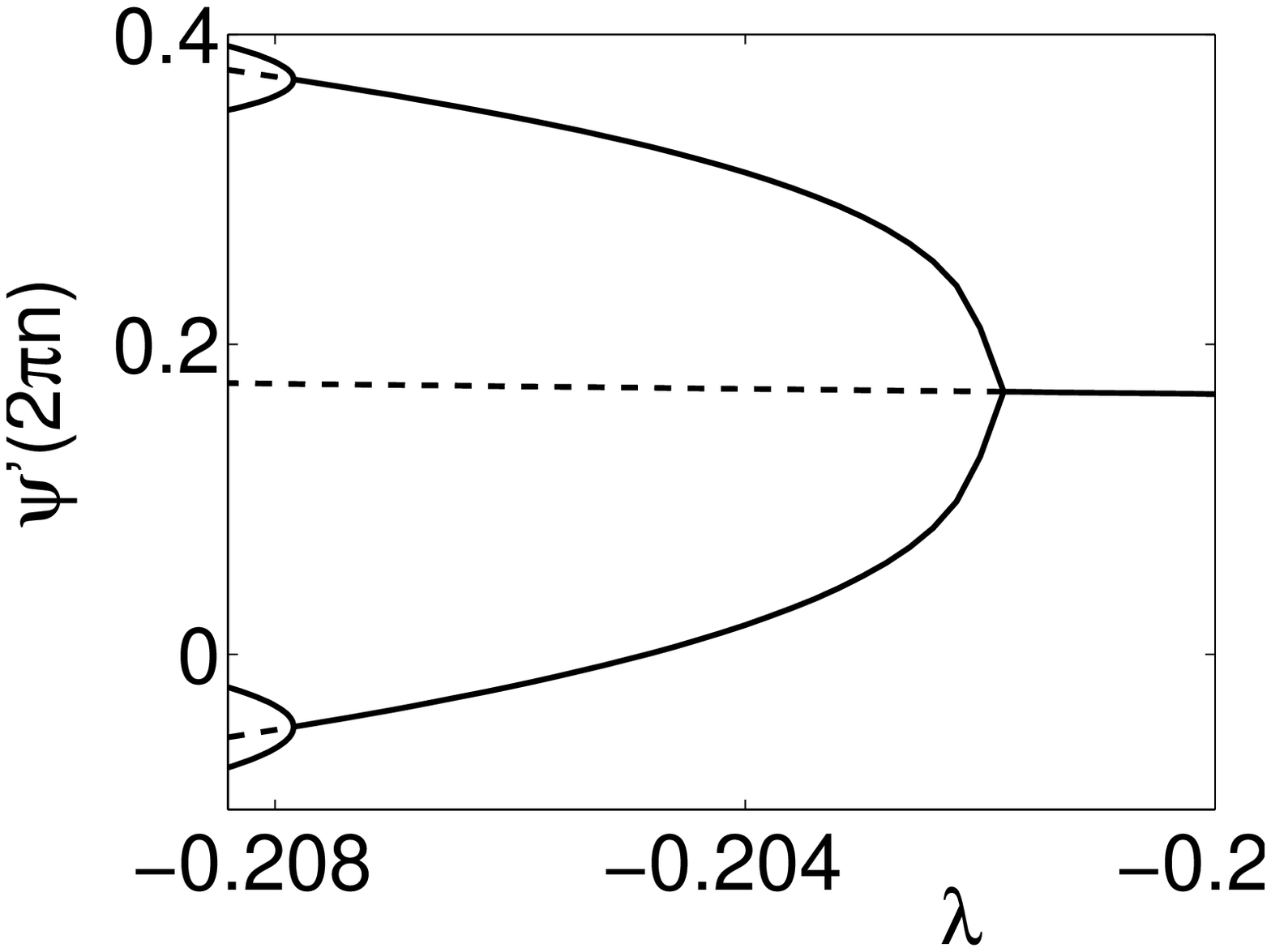}
\caption{
\label{fig-dcomb-4per1} \label{fig-dcomb-sym-4per1}
Period doubling of the antisymmetric fixed point with $m=2$ (left panel)
and of the symmetric fixed point (right panel) for $\mu=g=1$.
The Fixed points of $f$, $f^2$ and $f^4$ are plotted as a function of
the control-parameter $\lambda$. Stable fixed points are indicated by a
solid line, unstable fixed points by a dashed line.}
\end{figure}

In fact, a bifurcation occurs at the critical value $\lambda_c$
and two new fixed points of $f^2$ emerge. This is illustrated in
Fig.~\ref{fig-dcomb-4per1}, where $\psi_n = \psi(2\pi n)$
is plotted for the fixed points of $f^r, \, r = 1,2,4$  as a function
of $\lambda$.
One clearly observes a pitchfork bifurcation scenario.
The new fixed point of $f^2$ no longer fulfills $\psi(2\pi n) = 0$,
but instead $\psi(2\pi n) = - \psi(2\pi (n+1))$.
A second bifurcation occurs at $\lambda'_c \approx 0.178$, where
the fixed points of $f^2$ become unstable and four new elliptically
stable fixed points of $f^4$ emerge. These fixed points again become
unstable at $\lambda''_c \approx 0.182$.

One should keep in mind that the NLSE represents a Hamiltonian system
and that the period doubling scenario of Hamiltonian systems
(cf.~\cite{Meye70,Agui87}) is different from the familiar Feigenbaum
scenario for dissipative systems. For instance, the stable fixed points
are always elliptically (resp. neutrally) stable and not asymptotically
stable.

\begin{figure}[tb]
\centering
\includegraphics[width=8cm,  angle=0]{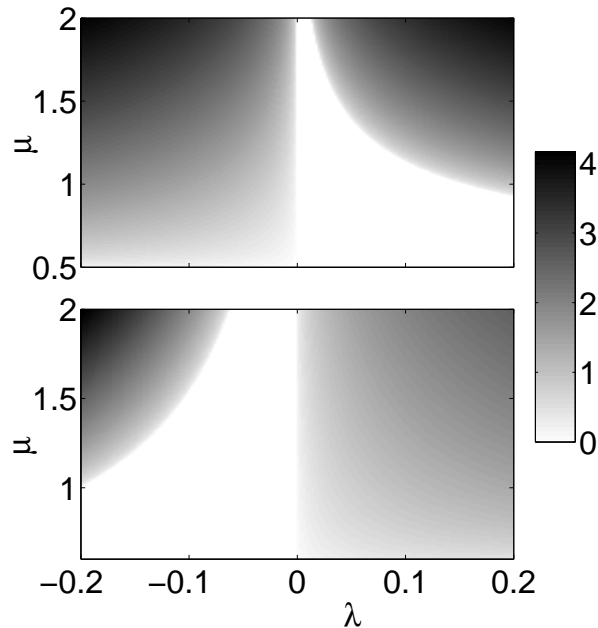}
\caption{\label{fig-rep_fixp_gamma2}
Stability exponent Re$(\gamma)$ for the antisymmetric fixed point
with $m=2$ (two nodes in $[0,2\pi)$, upper panel) and the symmetric
fixed point (lower panel) as a function of $\mu$ and $\lambda$ in a
greyscale plot. Wave functions corresponding to these fixed points
are illustrated in Fig.~\ref{fig-fixed_points_examples}.
Stable regions are colored in white (Re$(\gamma) = 0$).}
\end{figure}

For $\lambda <0$ the antisymmetric fixed points are unstable. However, large
regular regions exist in phase space around the trivial solution $\psi(x) \equiv 0$
and the symmetric fixed points as long as $|\lambda|$ is not too large
(cf.~Fig.~\ref{fig-dcomb-rep-phase2}). A ''trajectory'' started in the vicinity
of an antisymmetric fixed point is unstable but it may get trapped in the
chaotic sea between the large regular regions and consequently show
quasi-regular dynamics over many periods.
For stronger potentials, i.e. larger values of $|\lambda|$, the regular regions
shrink and cannot trap such a trajectory any longer so that it will in general
diverge.

A similar stability behavior is found for the symmetric fixed point.
However, this solution is always unstable for a repulsive delta-comb, $\lambda > 0$,
and shows a bifurcation scenario for an attractive delta-comb, $\lambda <0$.
For small repulsive potentials, a similar trapping scenario occurs
as described above for the antisymmetric fixed points in an attractive delta-comb.
Again this trapping is lost for stronger potentials, since the regular regions
of phase space shrink and the trajectories finally diverge in general.

The symmetric fixed points are elliptically stable for an attractive
delta-comb as long as $\lambda  > \lambda_c$ . A bifurcation occurs
if $\lambda$ is decreased below the critical value $\lambda_c \approx -0.203$.
The $4\pi$-periodic states are again elliptically stable up to the next
bifurcation. This bifurcation route is illustrated in Fig.
\ref{fig-dcomb-sym-4per1} for the symmetric periodic solution
with $m=2$ nodes in the interval $|0,2\pi)$.

Note that also the trivial solution $\psi(x) \equiv 0$ of the NLSE shows a
bifurcation when $\lambda$ is decreased. The trivial solution is stable
for $\lambda \ge \lambda^{(0)}_c \approx - 0.391$,
where two new elliptically stable period doubled fixed points emerge.
Quasiperiodic orbits around the emerging fixed points are illustrated
in the lower panel of Fig.~\ref{fig-dcomb-rep-phase2}
for $\lambda = -0.5$.

\begin{figure}[tb]
\centering
\includegraphics[width=8cm,  angle=0]{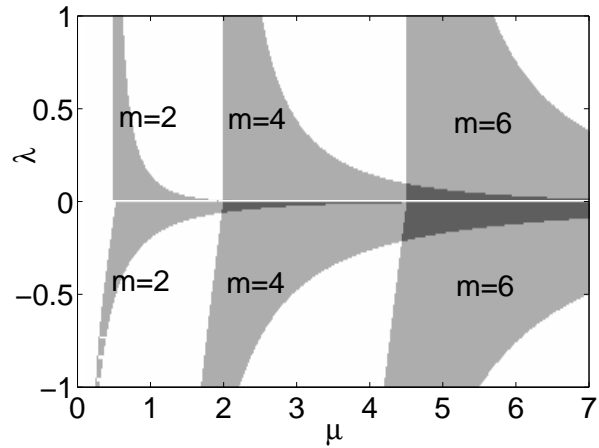}
\caption{\label{fig-rep_stab_map}
Stability map of the periodic states for a delta-comb with $g=1$.
The periodic solutions are stable in the shaded areas of
the parameter plane (antisymmetric states for $\lambda > 0$ and
symmetric states for $\lambda < 0$). The different solutions are
labeled by the number of nodes $m$ in the interval $[0,2\pi)$.}
\end{figure}

The stability of the antisymmetric and the symmetric fixed point with $m=2$
is summarized in Fig.~\ref{fig-rep_fixp_gamma2}, where the stability exponent
Re$(\gamma)$ is plotted as a function of $\mu$ and $\lambda$.
Fig.~\ref{fig-rep_stab_map} shows a stability map of several fixed points
with different $m$.

One can easily understand the qualitative differences of the stability between
$\lambda < 0$ and $\lambda > 0$  by considering the full phase-space dynamics
of the classical analogon (\ref{eqn-hill-nonlin}).
An antisymmetric periodic solution is not affected by the delta-comb because of
$\psi(2 \pi n) = 0$. However, a trajectory with a slightly larger classical energy
moves ahead of the periodic solution, hence it experiences a kick.
This kick will increase the energy of the classical oscillator if $\lambda > 0$.
The effect of the kicks accumulate and consequently the trajectory becomes unstable.
If $\lambda < 0$ the kick will lower the energy and stabilize the trajectory.
The contrary effect occurs for the symmetric periodic solution. However, the situation
is a bit more involved here since the periodic solution itself experiences the delta-kicks.

\subsection{Complex periodic solution}
\label{sec-per-complex}

Complex periodic solutions can be found using the ansatz
(\ref{eqn-ansatz-complex}).
By similar arguments as above one finds that the density
$S(x) = |\psi(x)|^2$ must be symmetric around $x = 2 \pi n$.
Hence it must assume a maximum or minimum at $x = (2n+1)\pi$,
in the middle between two delta-peaks. Therefore the ''phase shift''
$x_n$ is given by
\be
  x_n = -(2n+1) \pi \quad \mbox{or} \quad x_n = -(2n+1) \pi + L/4.
\ee
Using the symmetry of the density, the continuity of 
Eqn.~(\ref{eqn-delta-condition-density}) yields the condition
\be
  \frac{\varrho^2 p}{g\lambda } \sn(u|p) \cn(u|p) \dn(u|p)
  = B - \frac{\varrho^2}{g} \dn^2(u|p)
   \label{eqn-complex-per-cond}
\ee
with $u = 4 K(p) x_n/L$ for a periodic complex solution.
An example of such a periodic solution is shown in Fig.
\ref{fig-blochbands_wavefun} (dashed line).

For $\alpha \rightarrow 0$ the complex solutions tend to the real-valued sn-type
solutions which give rise to antisymmetric periodic solutions as discussed
above.
In fact, this is exactly what happens at the occurrence of loop structures
in nonlinear Bloch bands at $\kappa = 0$ (cf.~Sec.~\ref{sec-bloch-bands}).


\section{Nonlinear Bloch bands}
\label{sec-bloch-bands}

Recently, nonlinear Bloch states and Bloch bands have attracted considerable 
attention. New features such as looped Bloch bands \cite{Wu00,Wu03,06zener_bec} 
and period doubled Bloch states \cite{Mach04} were found.  Nonlinear Bloch bands 
for the delta-comb were first calculated by Seaman et al. \cite{Seam05}. In this 
section we want to link these results to the properties of periodic solutions 
discussed in Sec.~\ref{sec-periodic-states}.

The appearance of looped bands, or more generally the bifurcation
of eigenvalues in nonlinear systems, is intimately related to branch point
singularities, i.e. exceptional points (see \cite{Cart08} and references 
therein). Those exceptional points are eigenvalue and eigenvector degeneracies, which for
linear systems can only appear in a non-hermitian description.
As a boundary of regions of purely real eigenvalues they appear in so called
PT-symmetric models. In this context,
bifurcation phenomena of Bloch bands are also discussed
in linear, however non-hermitian periodic potentials with
PT-symmetry \cite{Cerv03,Cerv04, Makr08}. A full understanding of the 
interrelation of these phenomena has not yet been achieved and may be
provided by an analysis of the full nonlinear {\it and\/} non-hermitian
periodic potential in an extension of the analysis of the
corresponding two-state case (see, e.g., \cite{06nlnh,07nlres}). This is, however, 
only in the beginning (see
\cite{Muss08} for a very recent study).

Looped Bloch bands arise in nonlinear systems when a space-periodic stationary solution of 
the NLSE (a Bloch state with quasimomentum $\kappa=0$) undergoes a bifurcation associated
with the emergence of a \textit{dynamical} instability. 
Before the bifurcation, this Bloch state is neutrally stable with respect to small
perturbations in the sense that the time-dependent wave function $\psi(x,t)$ will 
remain close to the initial state $\psi(x,t=0)$. In the bifurcation, two novel neutrally
stable Bloch states emerge, while the old state becomes hyperbolically unstable
in the the sense that the time-dependent wave function $\psi(x,t)$ rapidly spreads
from the the initial state $\psi(x,t=0)$ leading to full \textit{spatio-temporal} chaos, 
an example of which is shown in Fig.~\ref{fig-gauss-wavefun}. Such a
dynamical instability leads to a divergence of the Bose-Einstein condensate mode
so that a simple mean-field approximation is no longer valid. Nevertheless, the 
time-dependent NLSE is still meaningful since it accurately predicts the occurence 
of the instability and the rate of the depletion \cite{Cast97,Cast98,Trim08,09phaseappl}.

On the contrary, period-doubled Bloch bands come into existence when
a usual $2\pi$-periodic solution $\psi(x)$ of the NLSE undergoes a period-doubling
bifurcation to form a $4\pi$-periodic solution. Consequently, the $2\pi$-periodic
wave function is \textit{spatially} stable up to the bifurcation where it becomes 
spatially unstable as illustrated in Fig.~\ref{fig-instability}.

\subsection{Linear and nonlinear Bloch states}

Bloch states are nonlinear eigenstates of the form
\be
  \psi_\kappa(x) = \re^{\ri \kappa x} u_\kappa(x),
\ee
where $u_\kappa(x)$ is $2\pi$-periodic and $\kappa$ denotes the quasimomentum.
The periodic functions $u_\kappa(x)$ fulfill the differential equation
\be
    - \frac{1}{2} \left( \frac{\rd}{\rd x} + \rmi \kappa\right)^2 u_\kappa(x)
   + V(x) u_\kappa(x)
      + g |u_\kappa(x)|^2 u_\kappa(x) = \mu(\kappa) u_\kappa(x)
   \label{eqn-nlse-blochu}
\ee
The eigenenergies $\mu(\kappa)$ form the nonlinear Bloch bands.
In the preceding sections we analyzed the solutions of the NLSE in
dependence of the parameter $\mu$, whereas the normalization of the
wave function was arbitrary.
For example, periodic states, which are Bloch states for $\kappa = 0$, exist
for any value of $\mu > 0$ (cf.~Sec.~\ref{sec-trivial-symmetric-states}).
Considering $\mu$ as a parameter is of course not appropriate here, since
this would lead to a continuous and multiply degenerate spectrum of
$\mu(\kappa)$ for every $\kappa$, whereas the corresponding Bloch states
differ by their normalization.
Consequently we consider $2\pi N$-periodic Bloch states here, whose
normalization is fixed as
\be
  \frac{1}{2 \pi N} \int_0^{2 \pi N} |u_{\kappa}(x)|^2 \rd x = 1
  \label{eqn-blochbands-norm}
\ee
throughout this section, leading to a discrete Bloch spectrum.
The Bloch states and the bands are calculated as described in \cite{Wu03}.

The set of Bloch states is characterized by one continuous parameter,
the quasimomentum $\kappa$, and an integer number counting the bands.
General non-diverging stationary states are given by four continuous
parameters, the real and imaginary parts of the initial value
$(\psi_0, \psi'_0)$.
Thus the Bloch states are of measure zero within the the set of all
possible non-diverging stationary states. The nonlinear stationary
states in a periodic potential generally do not fulfill Bloch's
theorem.

\begin{figure}[htb]
\centering 
\includegraphics[width=7cm,  angle=0]{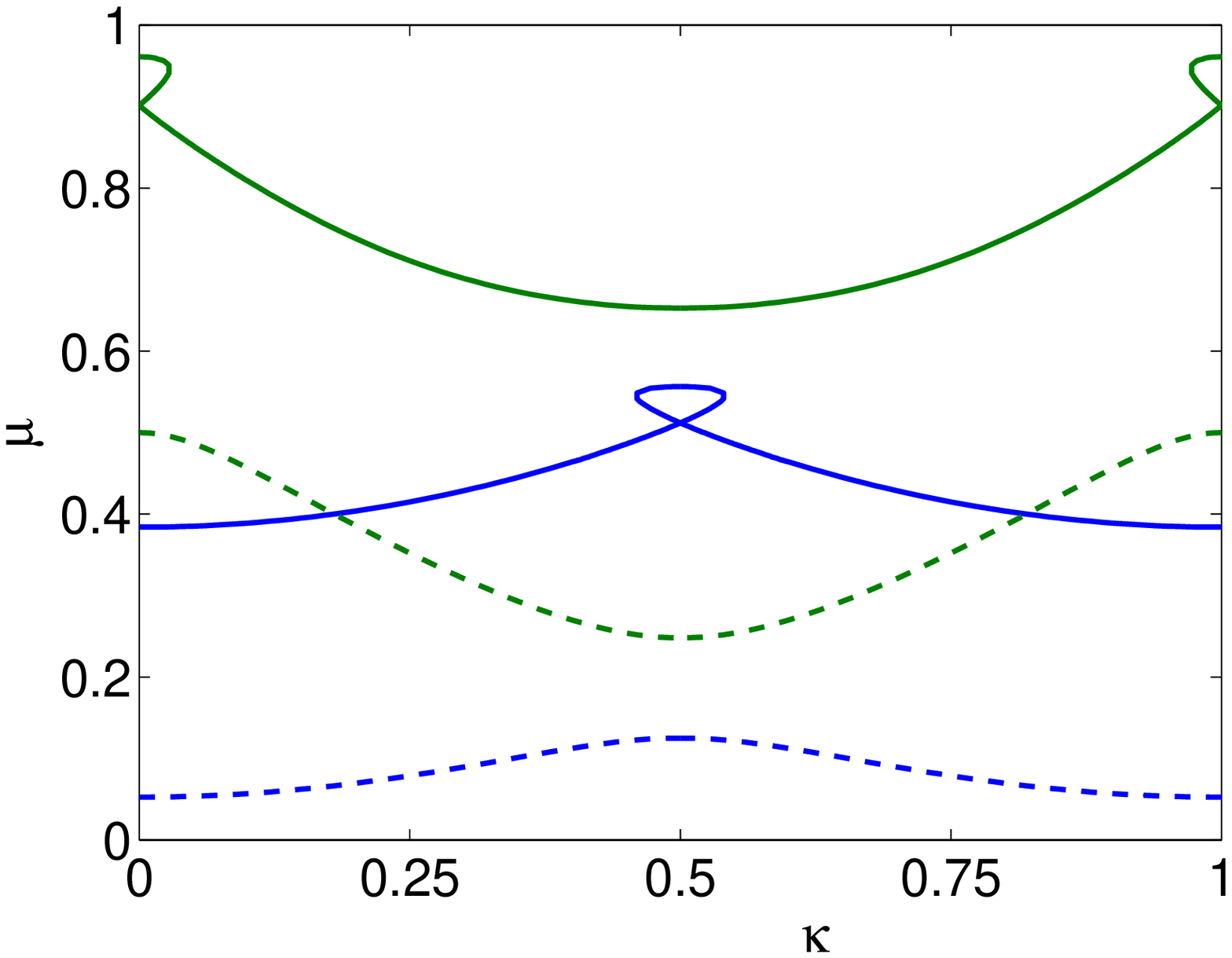}
\hspace{5mm}
\includegraphics[width=7cm,  angle=0]{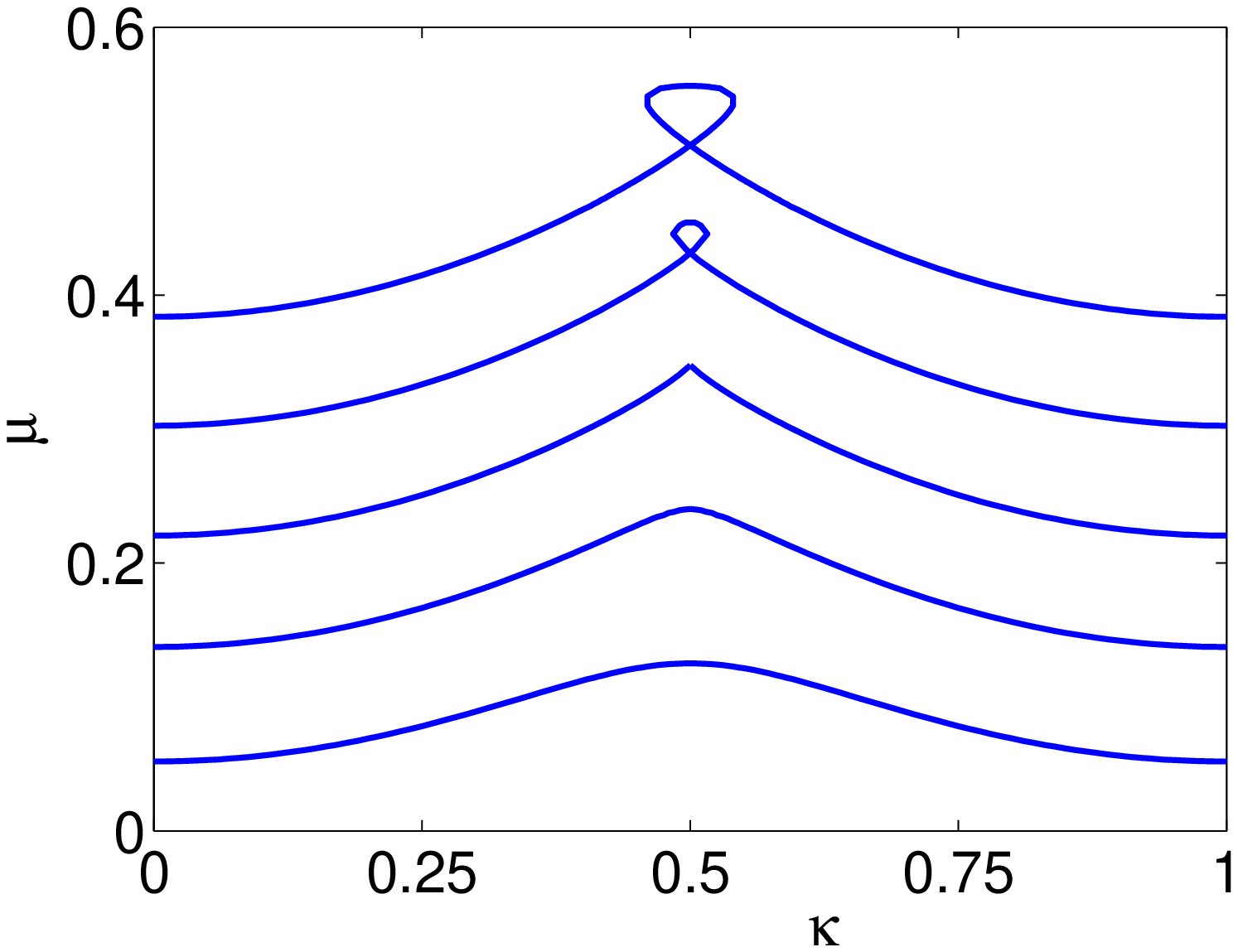}
\caption{\label{fig-blochbands1}
Nonlinear Bloch bands and emergence of loop structures for a
delta-comb potential.
Left: The lowest nonlinear Bloch bands for $g=1/\pi$ (solid lines) in comparison
to the linear case $g=0$ (dashed lines). Right: The lowest Bloch band for
$2 \pi g = 0, 0.5, 1, 1.5, 2$ (from bottom to top) and $\lambda = 0.5$.}
\end{figure}
\begin{figure}[htb]
\centering 
\includegraphics[width=7cm,  angle=0]{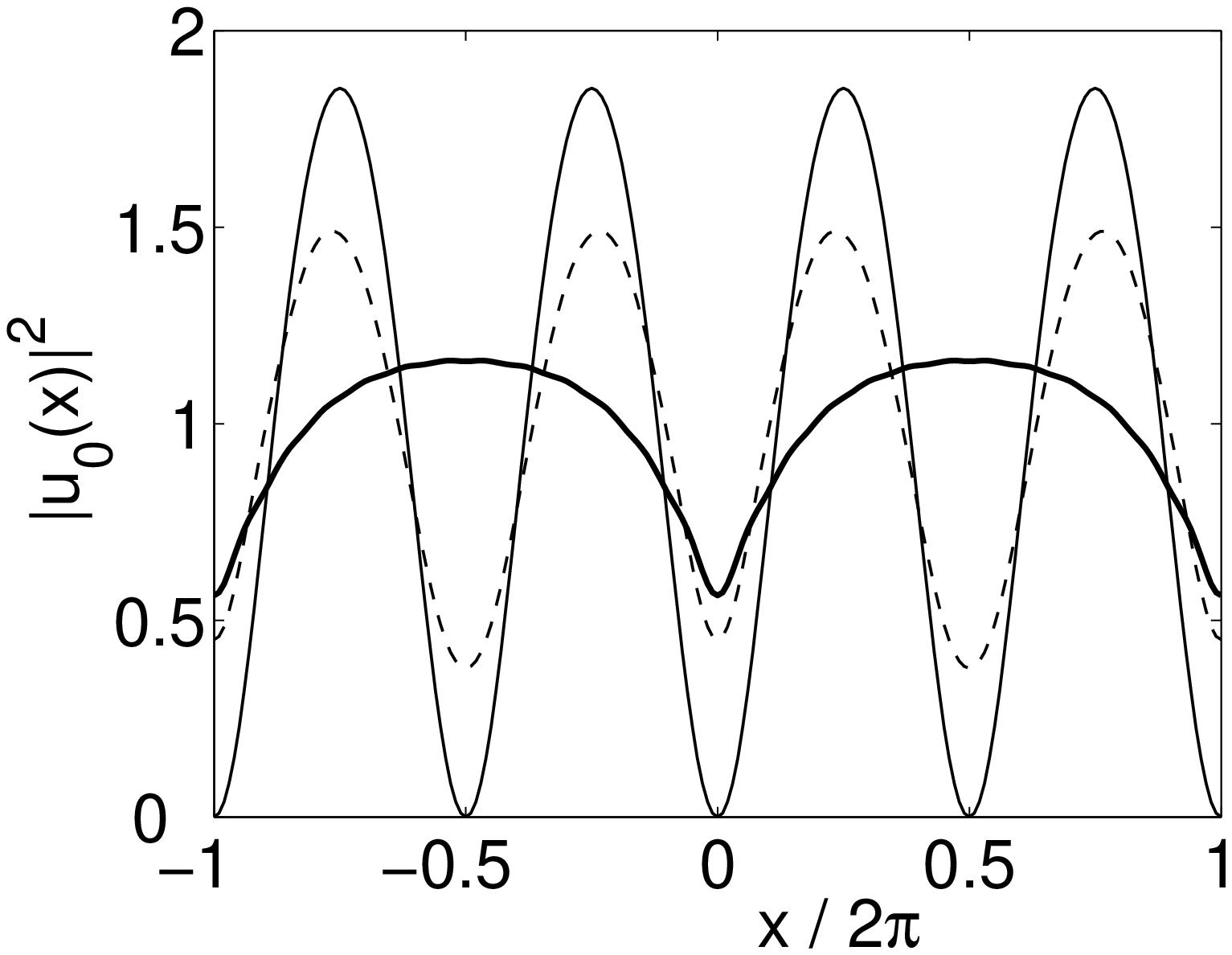}
\hspace{5mm}
\includegraphics[width=7cm,  angle=0]{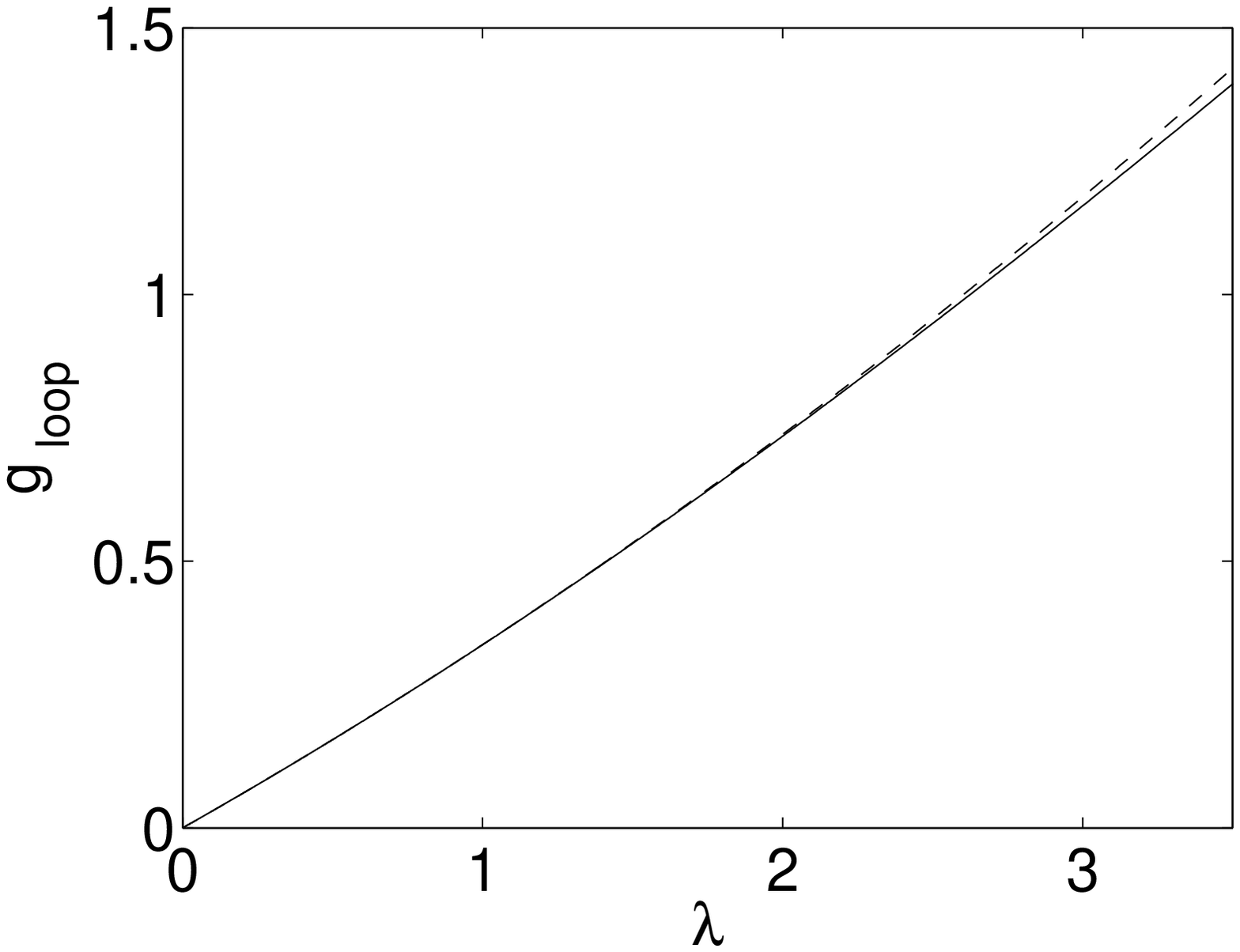}
\caption{\label{fig-blochbands_wavefun}
Left: Squared modulus of the wave function of nonlinear Bloch states for $\kappa = 0$,
$g=1/\pi$ and $\lambda = 0.5$ (bold line: lowest band,
dashed line: state at the self-crossing of the loop in the first excited band,
thin solid line: state at the top of the loop).
Right: Dependence of the critical nonlinearity $g_{\rm loop}$ for the emergence of
loops in dependence of the potential strength $\lambda$ (solid line) and the
approximation (\ref{eqn-loop-gc-approx}) (dashed line).}
\end{figure}

\subsection{Looped Bloch bands}

The nonlinear Bloch bands for a delta-comb potential $V(x) = \lambda \delta_{2\pi}(x)$
are shown in Fig.~\ref{fig-blochbands1} for $g=1/\pi$ and $\lambda = 0.5$.
The left-hand side shows the lowest bands in comparison to the linear case $g=0$,
whereas the right hand side illustrates the emergence of loop structures at the upper
edge of the lowest band at $\kappa = 0.5$. Similarly loops emerge at the edge of the
first excited band at $\kappa = 0$.
The chemical potential is larger than in the linear case
by an amount of approximately
\be
   \frac{g}{2\pi} \int_0^{2\pi N} |u_{\kappa}(x)|^4 \rd x
\ee
because of the repulsive mean-field potential.
Considering the energy per particle instead of the chemical potential
one finds swallow's tail structures instead of the loops \cite{Seam05,06zener_bec}.

The Bloch states with $\kappa = 0$ are strictly periodic and can be chosen real
up to the emergence of loops at a critical nonlinearity $g_{\rm loop}$.
These states can be identified with the symmetric
resp. antisymmetric periodic states described in Sec.~\ref{sec-trivial-symmetric-states}.
It is found that the Bloch state in the lowest band is a symmetric
periodic solution with no node, while the state in the first excited band
is an antisymmetric periodic solution with $m=2$ nodes in the interval $[0,2\pi)$.

Loops emerge for $g > g_{\rm loop}$, hence additional Bloch states with $\kappa = 0$
come into being.
A state corresponding to the self-crossing of a looped band at $\kappa = 0$
cannot be chosen purely real, whereas the state at the top of the loop can.
It is found that the state at the top of the loop in the first excited band
still corresponds to a real antisymmetric periodic solution as for $g < g_{\rm loop}$.
The Bloch state at the self-crossing corresponds to a complex periodic solution
as introduced in Sec.~\ref{sec-per-complex}.
The Bloch states in the two lowest bands at $\kappa = 0$  are shown in Fig.
\ref{fig-blochbands_wavefun} for $g = 1/\pi$ and $\lambda = 0.5$.
The squared modulus of a symmetric periodic solution (ground band)
and the antisymmetric periodic solution with $m = 2$ (first excited band)
are plotted as a bold (symmetric) resp. solid line (antisymmetric).
The complex solutions that degenerate at the self-crossing at $\kappa = 0$
transform into each other by a sign change of $\kappa$ and consequently
by a sign change of $x$ (cf.~Eqn.~(\ref{eqn-nlse-blochu})).
The squared modulus of these two solutions is the same and hence symmetric
around $x=0$ (dashed line in Fig.~\ref{fig-blochbands_wavefun}), whereas
the phases are antisymmetric: $\phi(-x) = - \phi(x)$.

Using the explicit form of the real and complex periodic solutions
one can derive the critical nonlinearity $g_{\rm loop}$ for the emergence of loops
in the first excited Bloch band.
To this end we reconsider condition (\ref{eqn-complex-per-cond}) for
the complex periodic solutions. For $g \rightarrow g_{\rm loop}$ the complex periodic
solution and the real antisymmetric solution merge. In this limit the wave
function tends to zero at the positions of the delta-peaks,
$\psi(2\pi n) \rightarrow 0$, and its derivative becomes continuous.
Hence both sides of Eqn.~(\ref{eqn-complex-per-cond}) tend to zero.

Consider a complex periodic solution slightly above the critical point.
The ''wave number'' $\varrho$ is written as $\varrho = \varrho_{\rm loop} + \delta \varrho$,
where $\varrho_{\rm loop} = m K(p)/\pi$ is the wave number at the critical
point $g = g_{\rm loop}$. Expanding Eqn.~(\ref{eqn-complex-per-cond})
around $\varrho_{\rm loop}$ with fixed $B$ then yields
\be
   - \frac{2 p \varrho_{\rm loop}^3}{g} \pi \, \delta \varrho
  + \mathcal{O}(\delta \varrho^2)  =
    \underbrace{2 \lambda B  - \frac{2 \lambda \varrho_{\rm loop}^2}{g}}_{=0}
  - \frac{4 \lambda \varrho_{\rm loop}}{g} \delta \varrho +
  \mathcal{O}(\delta \varrho^2) \, .
\ee
Comparing the coefficients of the linear term yields an equation for the elliptic
parameter $p_{\rm loop}$ at the critical point
\be
  p_{\rm loop} K(p_{\rm loop})^2 = \frac{2 \pi \lambda}{m^2} \, .
  \label{eqn-loop-pc}
\ee
The normalization of the wave function at the critical point yields
\be
  2 \pi \stackrel{!}{=} \int_0^{2\pi} A^2 \, \sn^2 \bigg( 4 K(p) \frac{x+x_n}{L} \bigg| p \bigg)
  = \frac{2 m^2}{\pi g} K(p) \left[K(p) - E(p)\right],
\ee
where $K(p)$ and $E(p)$ denote the complete elliptic integrals of the first
and second kind, respectively.
Hence the critical nonlinearity is given by
\be
  g_{\rm loop} = \frac{m^2}{\pi^2} K(p_{\rm loop}) \left[K(p_{\rm loop}) - E(p_{\rm loop})\right],
  \label{eqn-loop-gc}
\ee
where $p_{\rm loop}$ is determined by Eqn.~(\ref{eqn-loop-pc}).
One can derive an approximate explicit expression for the critical nonlinearity
parameter $g_{\rm loop}$ by expanding the left-hand side of Eqn.~(\ref{eqn-loop-pc}) 
up to second order in $p_{\rm loop}$ and substituting
the result into Eqn.~(\ref{eqn-loop-gc}).
This yields in second order in $p_{\rm loop}$
\be
  g_{\rm loop} \approx \frac{\lambda}{\pi} + \frac{\lambda^2}{\pi^2 m^2} \, .
  \label{eqn-loop-gc-approx}
\ee
The dependence of $g_{\rm loop}$ on $\lambda$ is shown in the lower panel
of Fig.~\ref{fig-blochbands_wavefun} for $m=2$, which corresponds
to the first excited Bloch band.

\begin{figure}[tbh]
\centering
\psfrag{cr}{\tiny pd}
\includegraphics[width=7cm, angle=0]{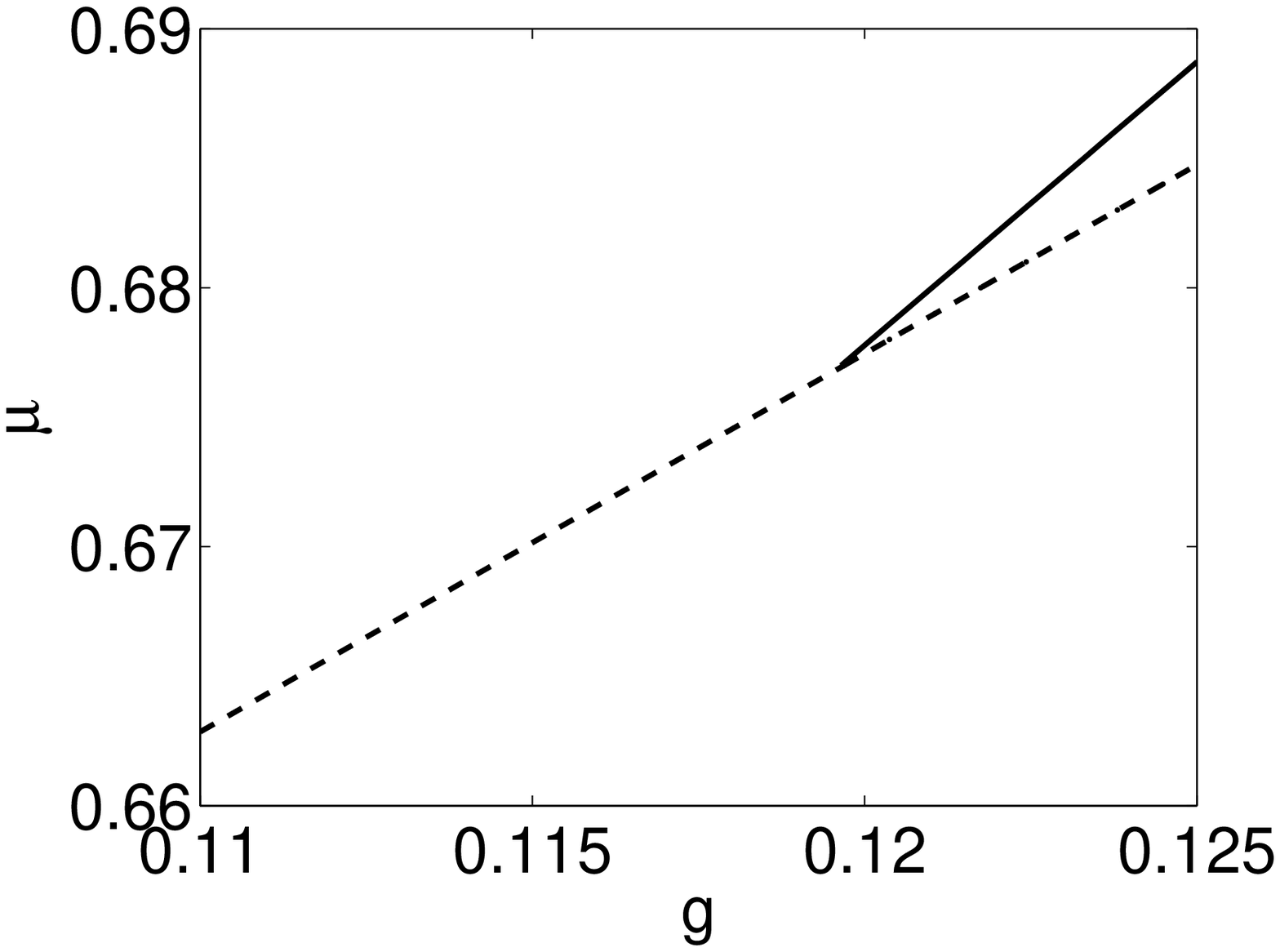}
\includegraphics[width=7cm, angle=0]{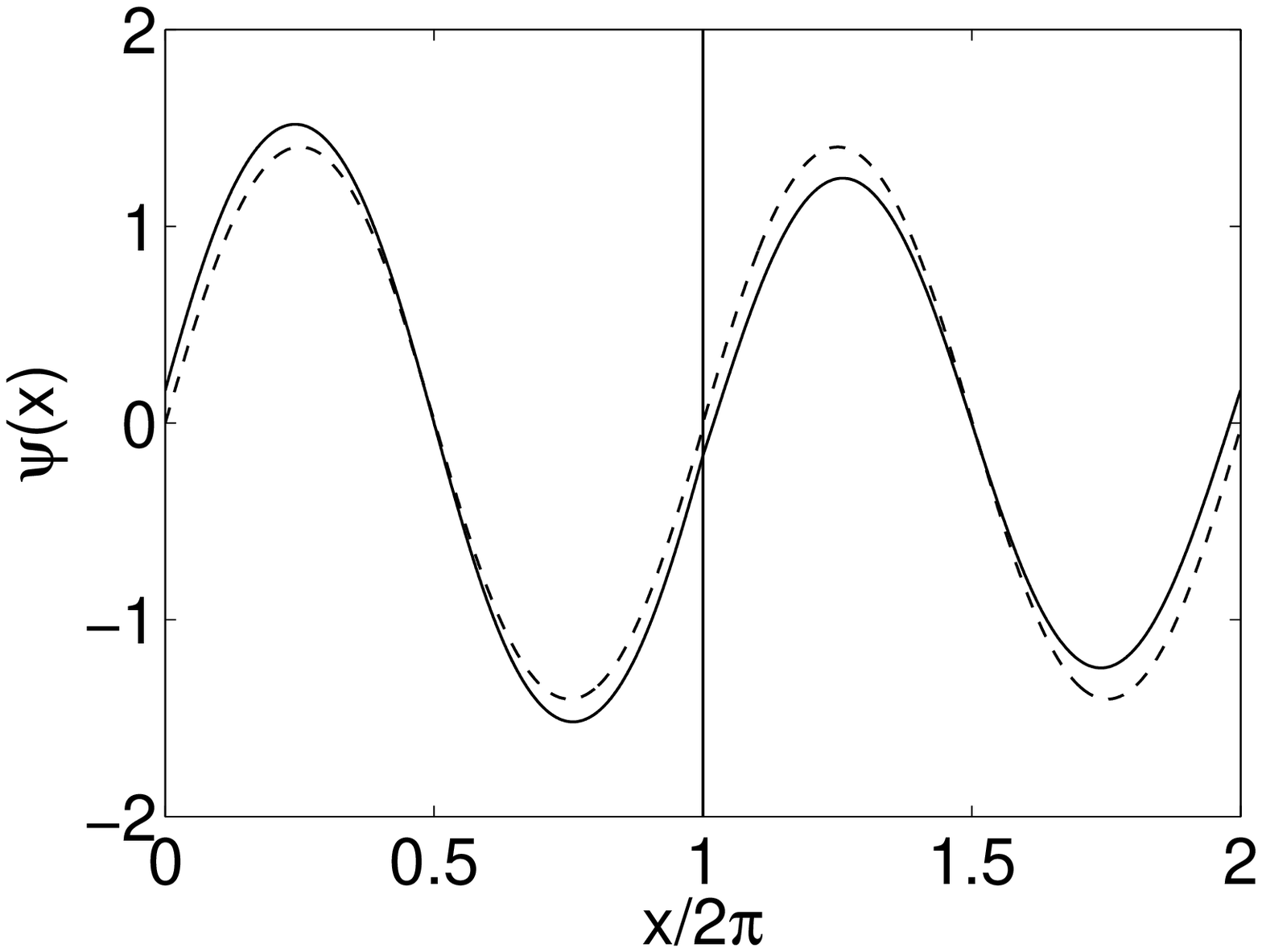}
\caption{\label{fig-bbands_bifur}
Left panel: Bifurcation of the $2\pi$-periodic  state corresponding to the first 
excited Bloch band by a variation of $g$. Shown is the chemical potential $\mu$ 
of the $2\pi$-periodic state ($---$) and the emerging $4\pi$-periodic 
state (---).
Right: Wave function of the $2\pi$-periodic ($---$) and the 
$4\pi$-periodic (---) state for $g = 1/8$.}
\end{figure}

\subsection{Period doubled Bloch bands}

Let us finally discuss another new feature of nonlinear Bloch bands -- the existence of
period doubled Bloch bands \cite{Mach04}. These Bloch bands come into being if a
periodic state -- a Bloch state with $\kappa = 0$ -- becomes spatially unstable in 
a period doubling bifurcation, giving birth to a $4\pi$-periodic state. This state is
now again embedded into a (period-doubled) Bloch band.

The period doubling bifurcation has already been introduced in 
Sec.~\ref{sec-periodic-stability} for the case of a variation of the potential 
strength $\lambda$. As shown on the left-hand side of 
Fig.~\ref{fig-dcomb-4per1}, the antisymmetric $2\pi$-periodic
state bifurcates when $\lambda$ is increased above a critical value 
$\lambda_{\rm pd}$. (We now use the index 'pd' for period doubling to 
distinguish it from the critical value for the emergence
of looped Bloch bands).

The period-doubling bifurcation can also take place if the nonlinearity $g$ is 
increased while $\lambda$ and the normalization (\ref{eqn-blochbands-norm}) 
remain fixed. This is illustrated on the left-hand side of Fig.~\ref{fig-bbands_bifur}, 
where we have plotted the chemical potential $\mu$ as a function of $g$ for the
$2\pi$-periodic state (green) and the $4\pi$-periodic state (red). The wave functions
of these two states are compared on the right-hand side of Fig.~\ref{fig-bbands_bifur}.
for $g = 1/8$. 

\begin{figure}[tbh]
\centering
\includegraphics[width=7cm,  angle=0]{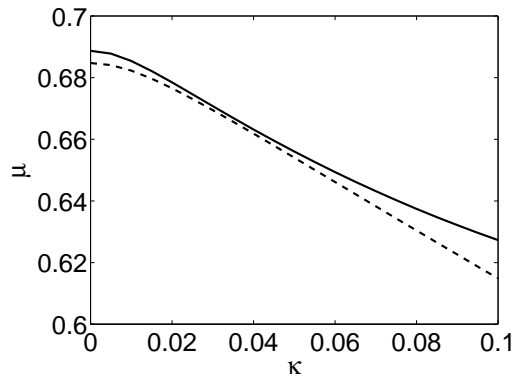}
\caption{\label{fig-blochbands_4pi}
A $4\pi$-periodic Bloch band (---) around $\kappa = 0$ and the corresponding
corresponding first excited $2\pi$-periodic Bloch band ($---$) for 
$\lambda= 0.7$ and $g = 1/8$.}
\end{figure}

The asymmetric $2\pi$-periodic state is just the Bloch state with $\kappa=0$ in the
first excited band which is shown in Fig.~\ref{fig-blochbands_4pi} (green line).
Similarly, the $4\pi$-periodic state is embedded in a 'period-doubled' Bloch band 
which is also shown in Fig.~\ref{fig-blochbands_4pi} as a red line.

The connection to the period-doubling bifurcation of an antisymmetric periodic
state has several implications for the existence and stability of nonlinear
Bloch states:
Firstly, period-doubled Bloch bands obviously do not exist for $g=0$, they
come into being when $g$ is increased over the critical value $g_{\rm pd}$
for a period doubling bifurcation.
Secondly, it was already argued in Sec.~\ref{sec-periodic-stability} that the 
corresponding $2\pi$-periodic state becomes spatially unstable in this bifurcation.
The lowest symmetric periodic state corresponding to the ground band is always 
unstable for $\lambda > 0$, while the anti-symmetric state corresponding to
the first excited band is is always unstable for $\lambda < 0$ (cf. the stability map
in Fig.~\ref{fig-rep_stab_map}).

This has several implications for the existence and stability of nonlinear
Bloch states.
The $4\pi$-periodic states exist only after a bifurcation, i.e. for
$g > g_{\rm pd}$. 
They are stable until the next bifurcation.
As argued in Sec.~\ref{sec-periodic-stability}, the corresponding
$2\pi$ periodic state becomes spatially unstable at this bifurcation.


\section{The Gaussian comb}
\label{sec-gauss}

Finally we present some numerical results for a more realistic potential,
consisting of a periodic structure of narrow Gaussian peaks,
\be
  V(x) = \lambda \sum_n \frac{1}{\sqrt{2\pi}\Delta x} \exp\left[ -
  \frac{(x - 2 \pi n)^2}{2 \Delta x^2} \right].
\ee
with $\Delta x = 0.1$. The area of each Gaussian peak is normalized to unity.
The paramters are such that $\Delta x \ll \xi \ll d$, where $\xi = 1/\sqrt{2 g \rho}$
  is the healing length and $\rho$ is the averaged condensate density. This means that
  changes in the condensate density are 'healed' on a length scale which is much longer
  than the width of the Gaussian potential barriers, but smaller than the separation of
  the barriers. This gives a rough argument why the Gaussian comb is indeed well approximated
  by a delta-comb discussed above. Such smooth potentials
can be implemented, at least in principle, in current atom-chip
experiments (see \cite{Fort07} for a recent review, see also the 
discussion in \cite{Paul05}).

\subsection{Solutions of the time-independent NLSE}

\begin{figure}[htb]
\centering
\includegraphics[width=4.9cm,  angle=0]{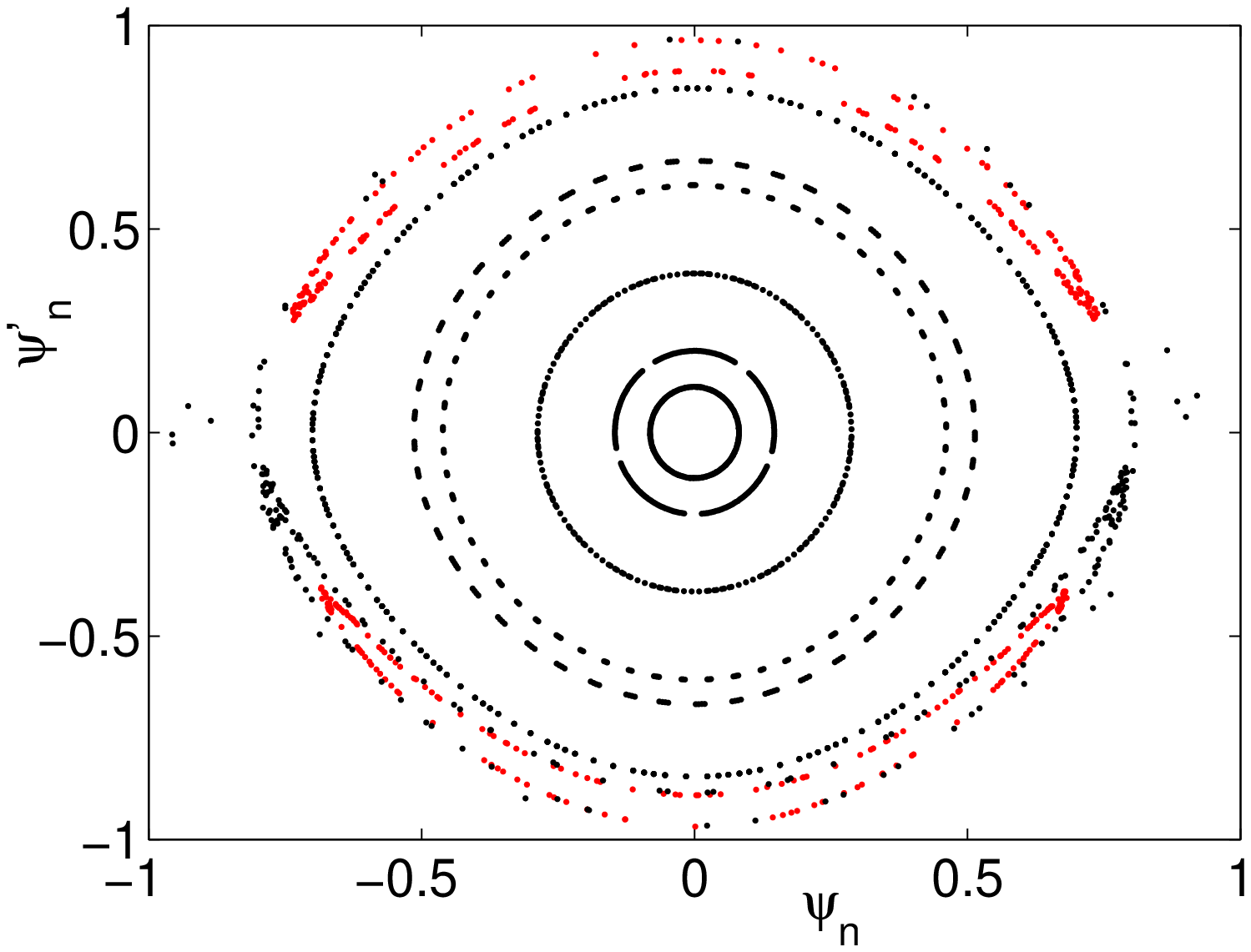}
\includegraphics[width=4.9cm,  angle=0]{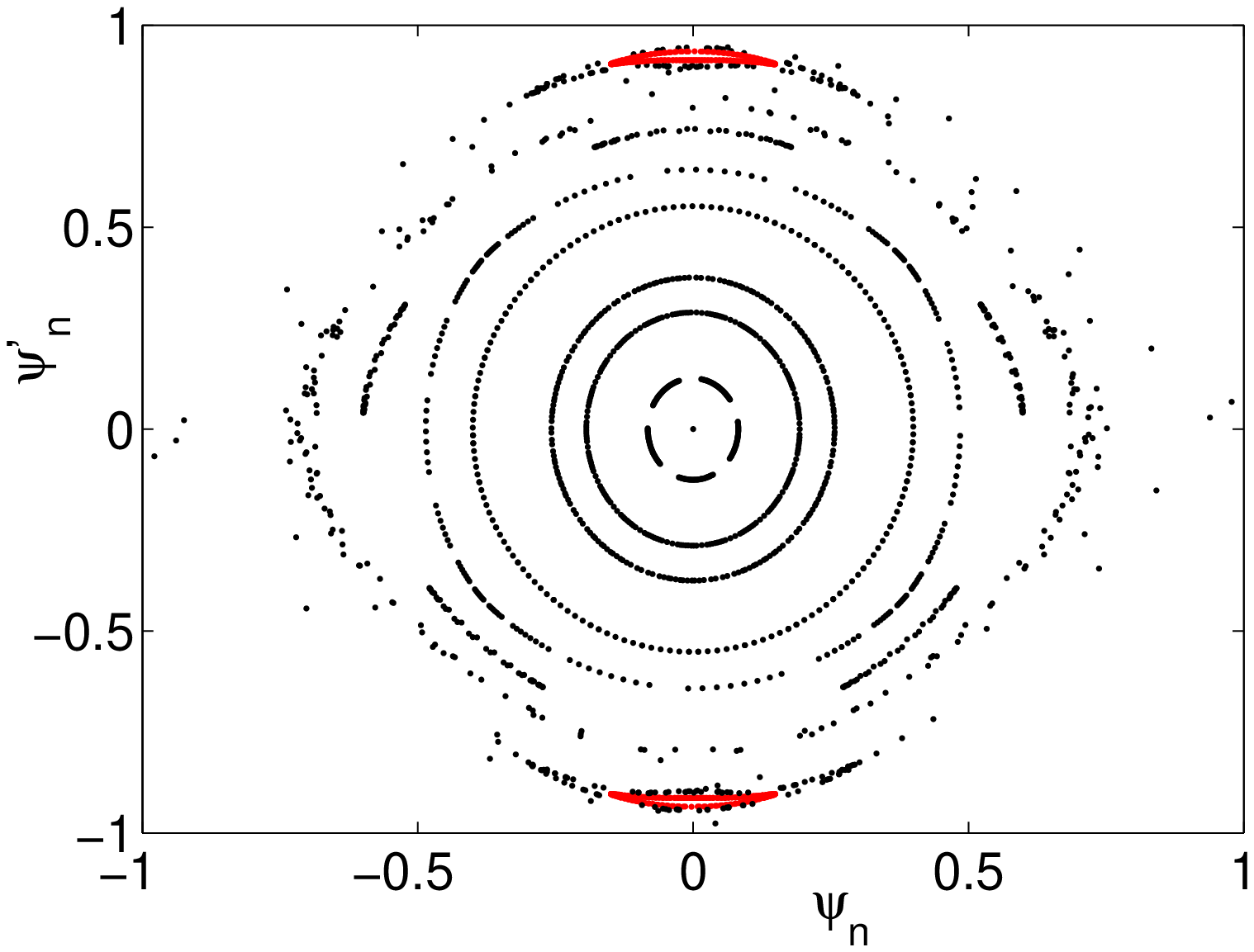}
\includegraphics[width=4.9cm,  angle=0]{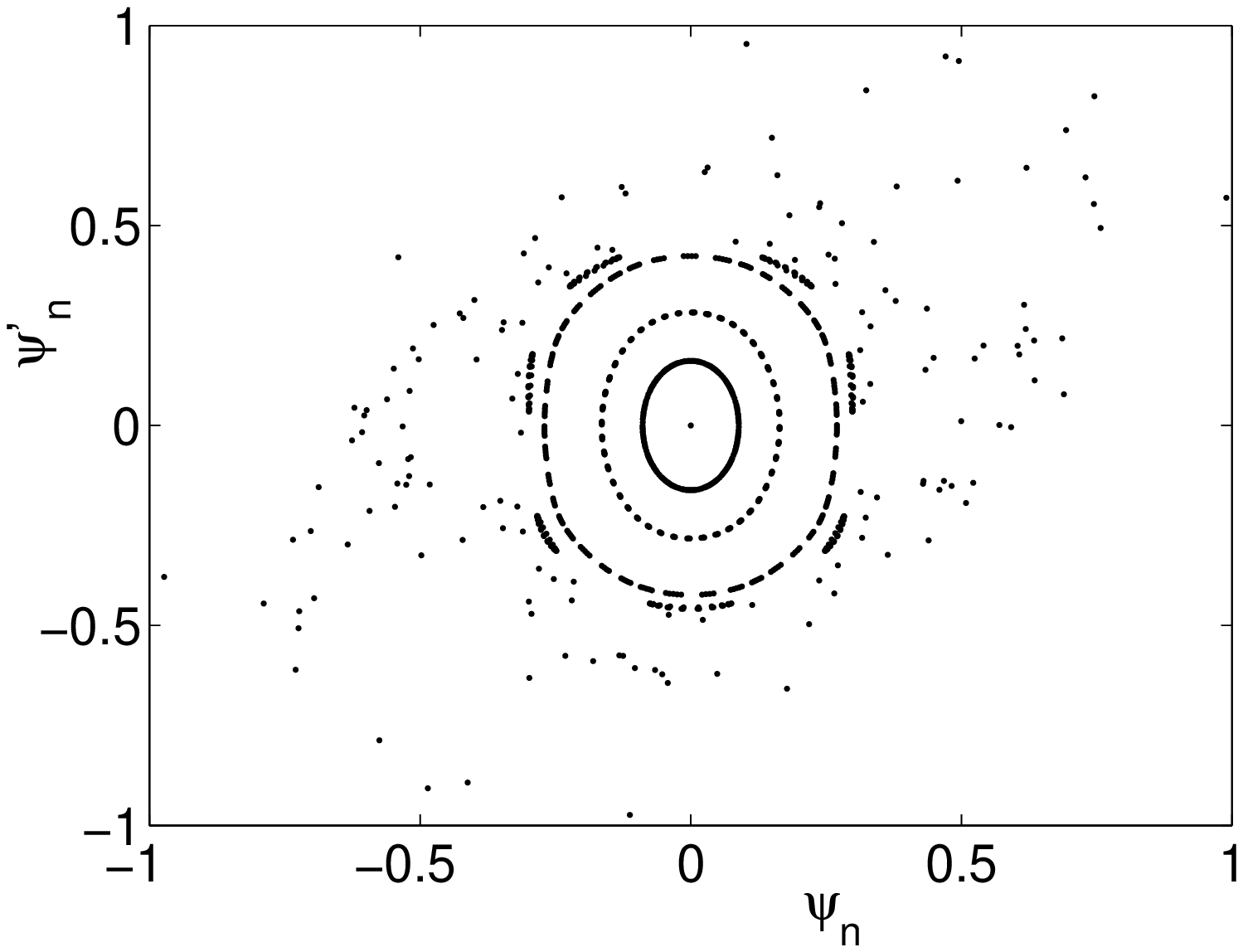}
\caption{\label{fig-gauss-rep-phase1}
Poincar\'e section in phase space $(\psi_n,\psi'_n)$ for the NLSE in a
Gaussian comb with $g=1$, $\mu = 1$ and $\lambda = 0.02, \, 0.1, \, 0.5$
(from left to right).}
\end{figure}

\begin{figure}[htb]
\centering
\includegraphics[width=4.9cm,  angle=0]{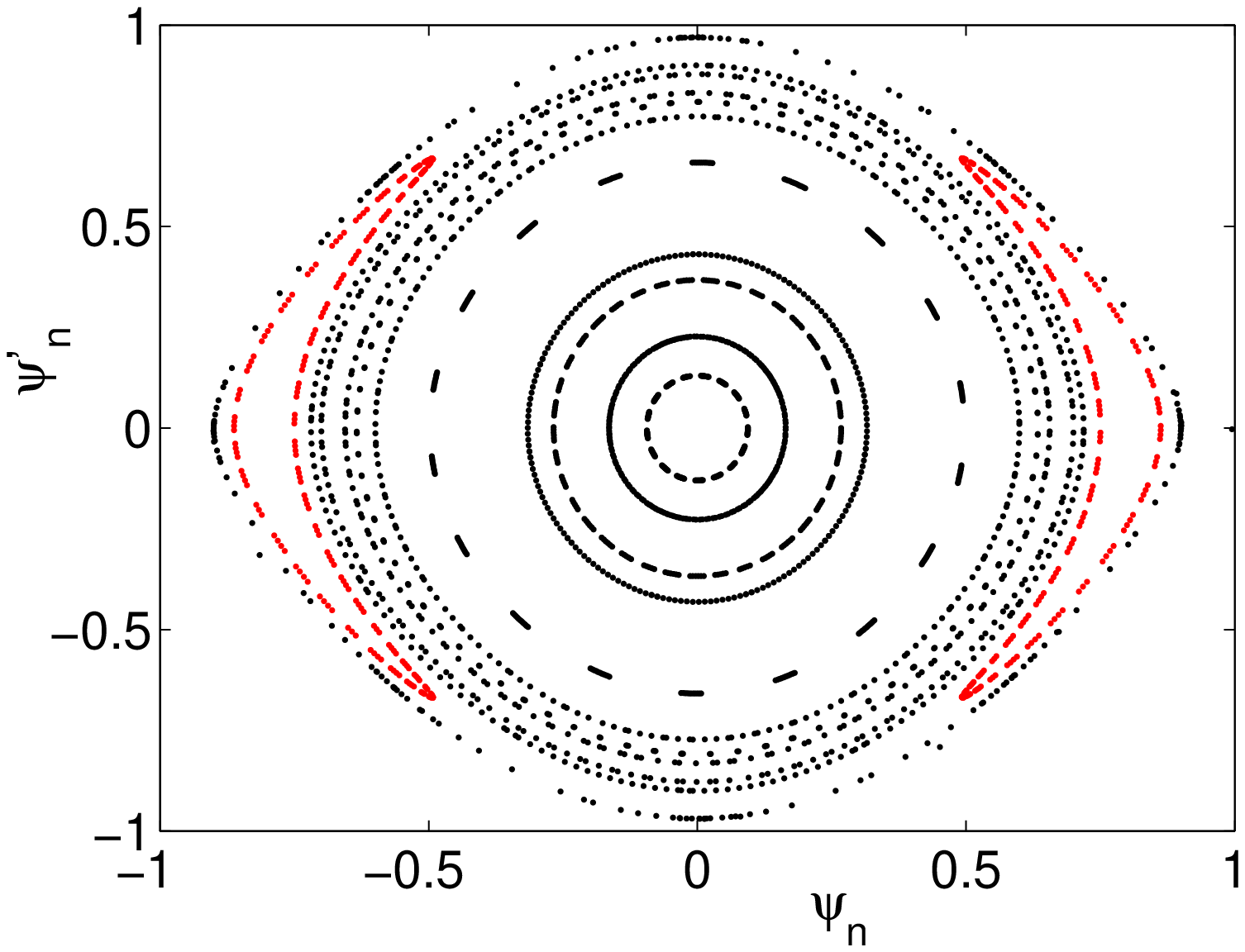}
\includegraphics[width=4.9cm,  angle=0]{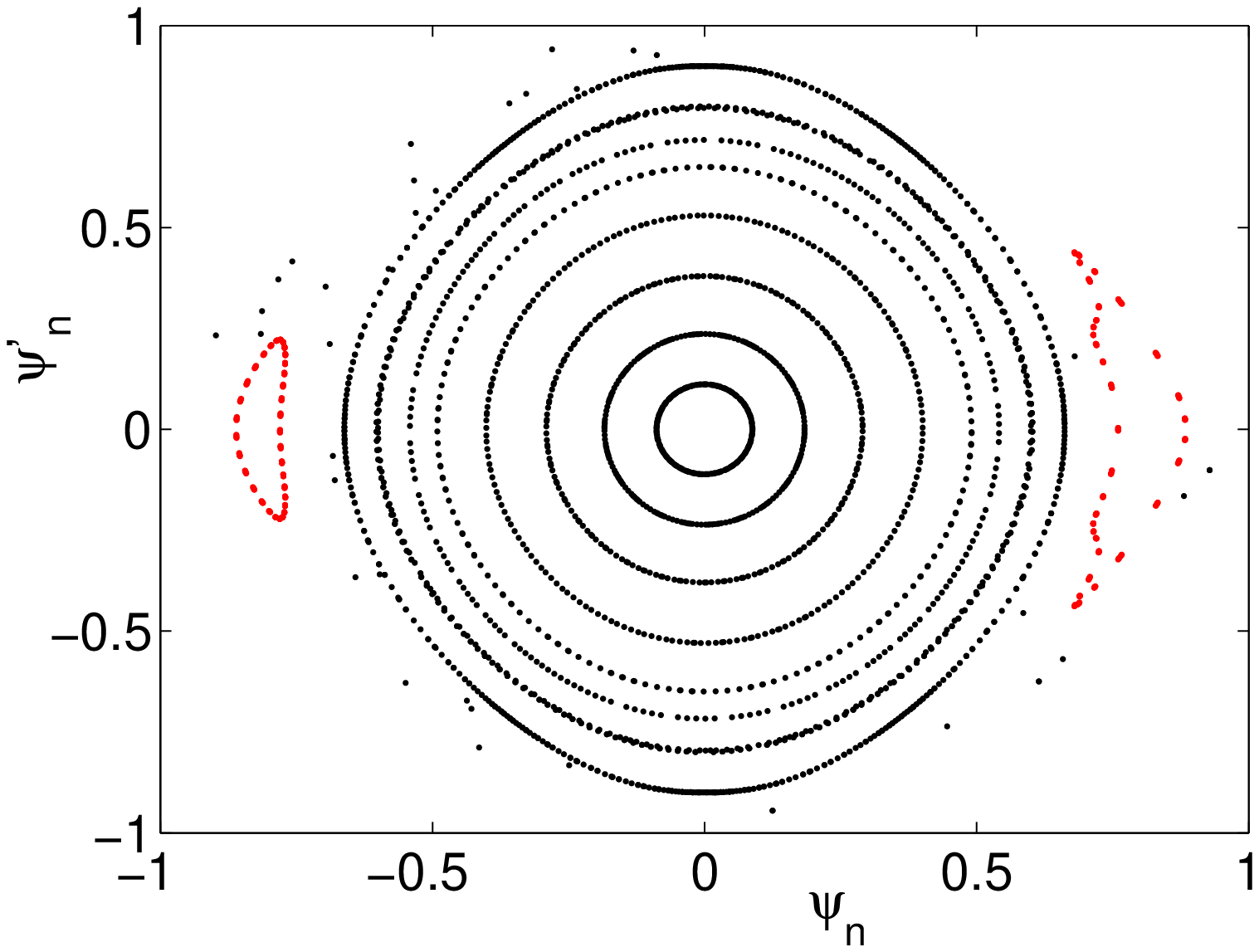}
\includegraphics[width=4.9cm,  angle=0]{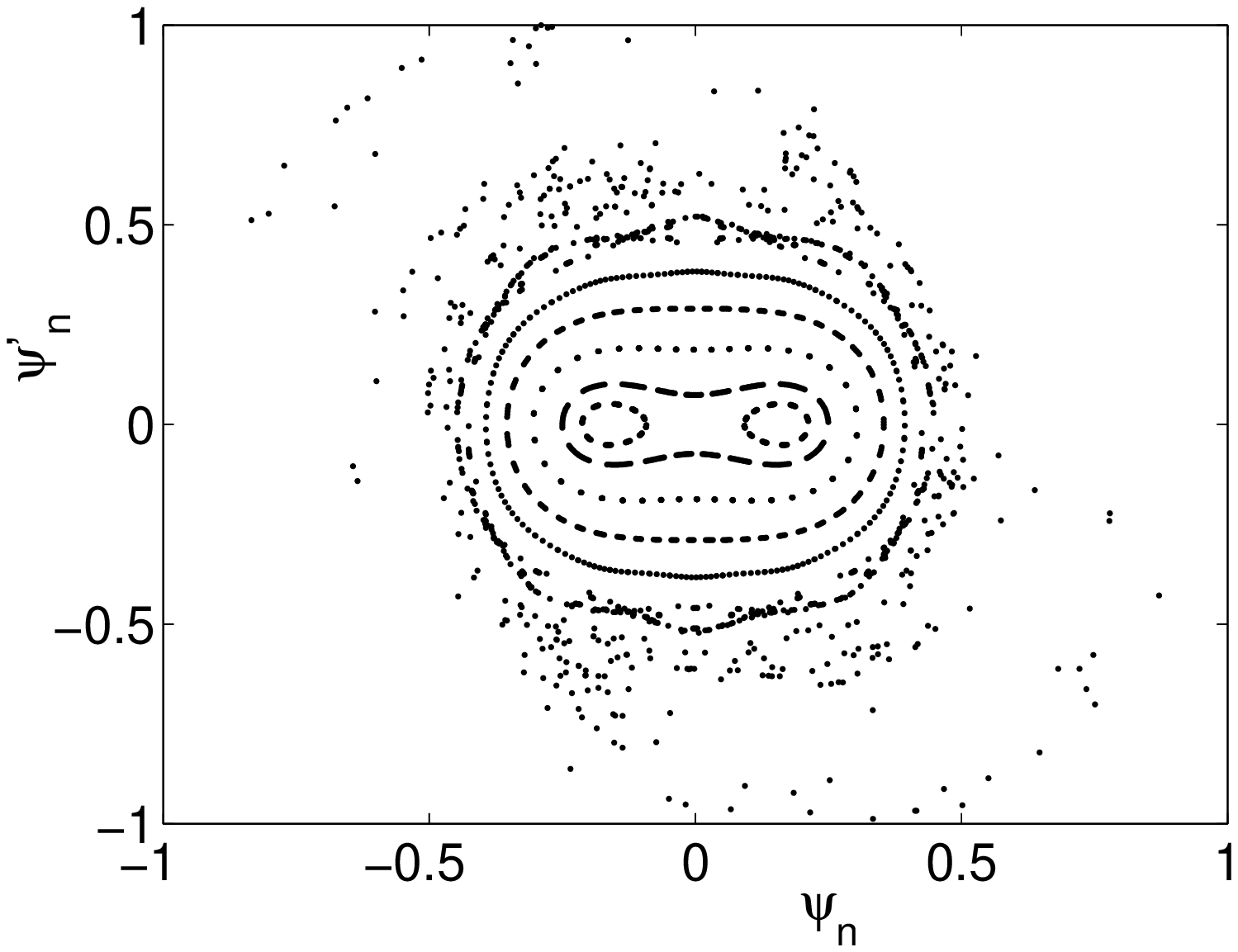}
\caption{\label{fig-gauss-rep-phase2}
As figure \ref{fig-gauss-rep-phase1}, however for
$\lambda = -0.02, \, -0.1, \, -0.5$ (from left to right).}
\end{figure}

We consider the solutions of the time-independent nonlinear Schr\"odinger
equation in a spatial Poincar\'e section in phase space as in
Sec.~\ref{eqn-sol-general}.
The Figs. \ref{fig-gauss-rep-phase1} and \ref{fig-gauss-rep-phase2} show
$(\psi_n,\psi'_n) = (\psi(2\pi n),\psi'(2\pi n))$ obtained by a numerical
integration for different values of the potential strength $\lambda$.
The dynamics in phase space strongly resembles the one for the delta-comb 
illustrated in the Figs.~\ref{fig-dcomb-rep-phase1} and
\ref{fig-dcomb-rep-phase2}.

\begin{figure}[htb]
\centering
\includegraphics[width=7cm,  angle=0]{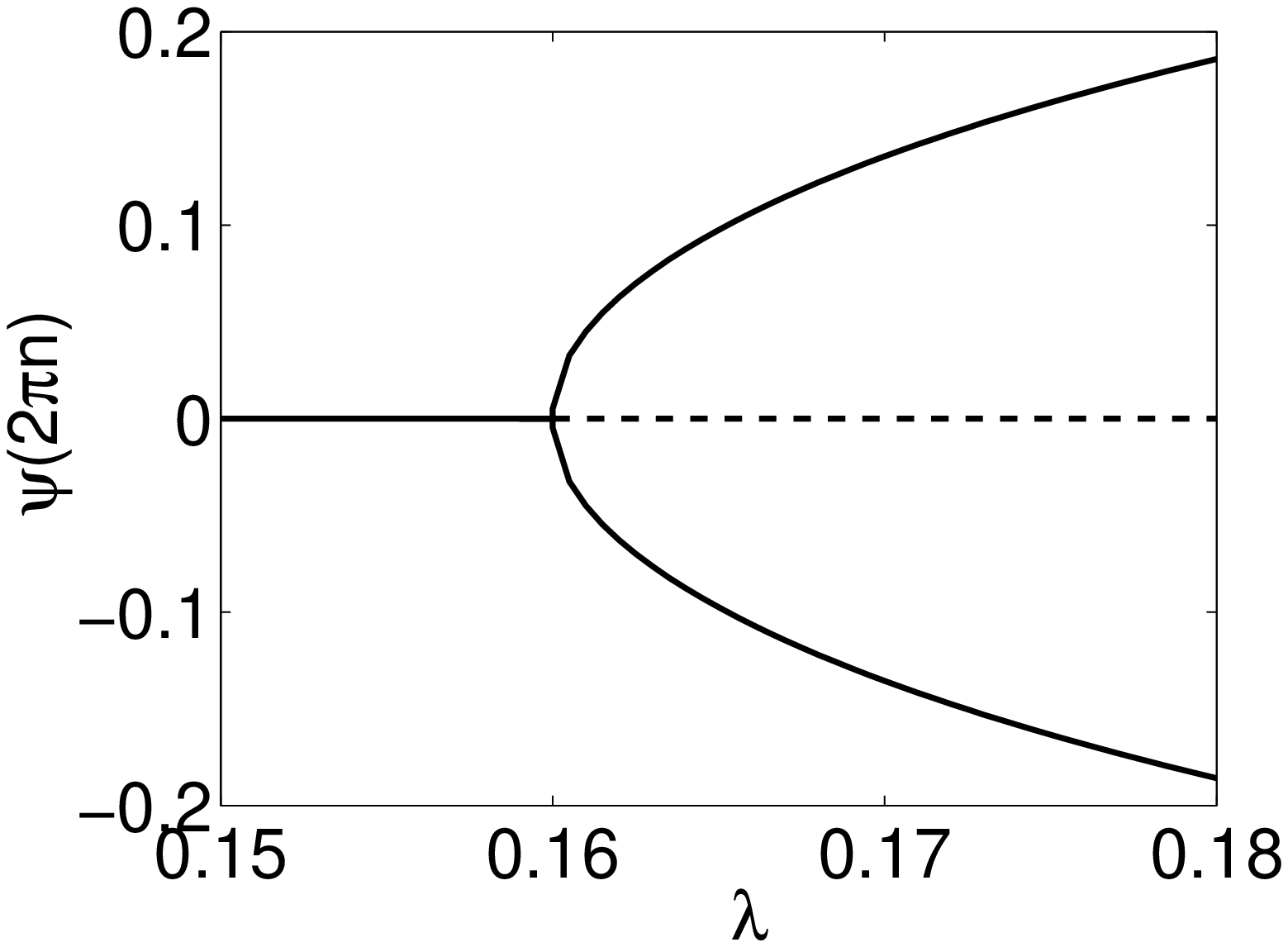}
\hspace{5mm}
\includegraphics[width=7cm,  angle=0]{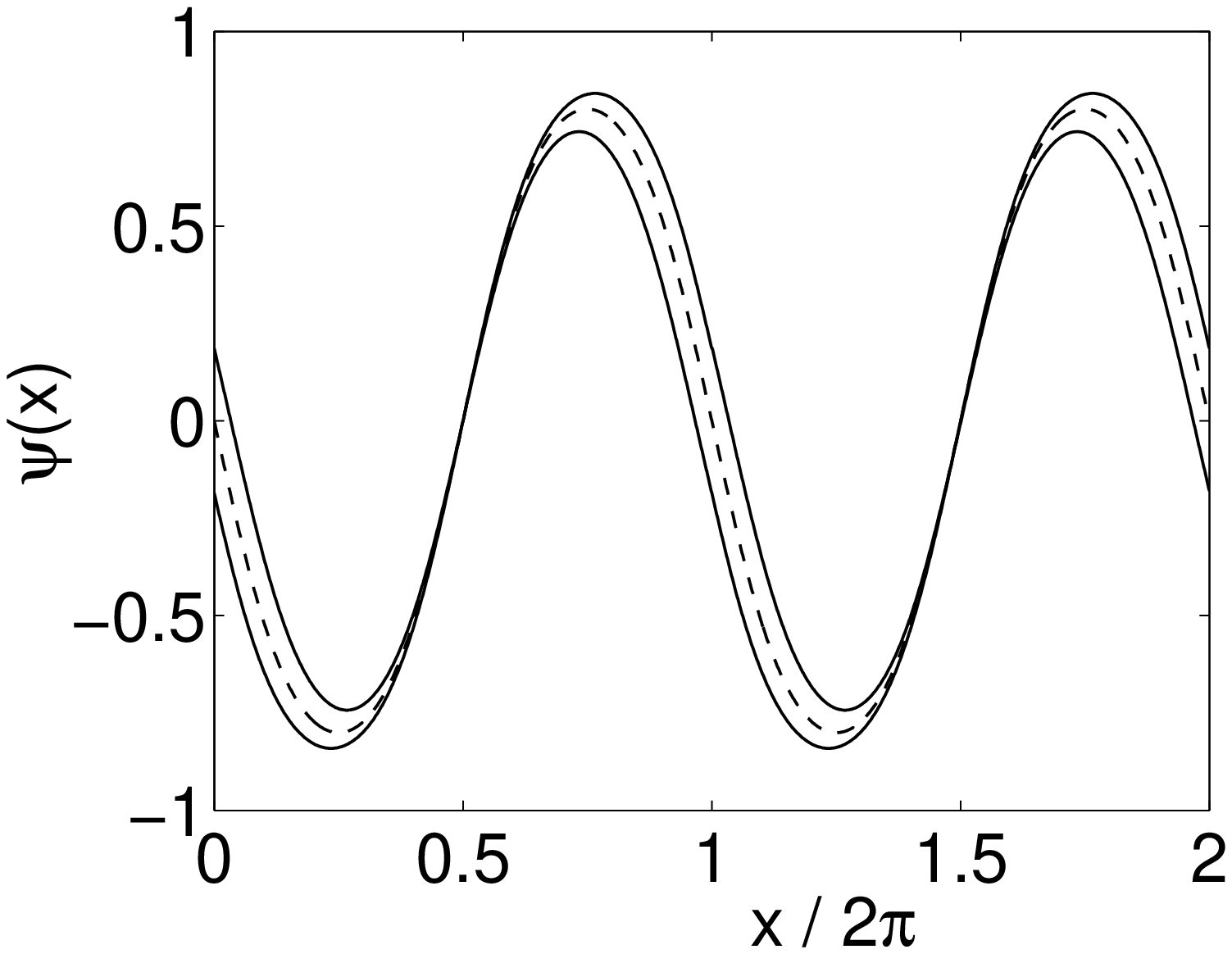}
\caption{\label{fig-gauss-bifur1}
Upper panel: Bifurcation of the trivial periodic state in a repulsive Gaussian comb
for $\mu = 1$.
Lower panel: The unstable $2 \pi$-periodic wave function (dashed line) in comparison
with the stable $4\pi$-periodic wave functions (solid lines) in a Gaussian comb
with $\lambda = 0.18$ and $\mu = 1$.}
\end{figure}

Again we find trivial and symmetric periodic solutions,
whose spatial stability properties are very similar to the case
of a delta-comb.
A pitchfork bifurcation of these periodic solutions occurs when a
control-parameter is varied. The bifurcation of the trivial
periodic state for $\lambda > 0$ is illustrated in the upper panel of
Fig.~\ref{fig-gauss-bifur1}. The lower panel shows a $4\pi$-periodic
wavefunction that emerges in the bifurcation.

\subsection{Dynamical (in)stability}

Up to now only the spatial stability of periodic solutions has been discussed,
not considering the {\it temporal} behavior, induced by the time-dependent
nonlinear Schr\"odinger equation
\be
  \left( - \frac{1}{2} \frac{\partial^2}{\partial x^2} + V(x) + g |\psi(x,t)|^2 \right)
  \psi(x,t) = \ri \frac{\partial \psi}{\partial t} \, .
\ee
In the following we analyze the temporal stability of the periodic solutions
in the Gaussian comb discussed above.
The time-dependent NLSE can lead to classical chaotic dynamics
\cite{Thom03} and dynamical instability, which usually leads to
a depletion of the Bose-condensed phase \cite{Cast97}.
Related work on the dynamical stability of the NLSE in a periodic potentials
can be found in \cite{Bron01a,Bron01b,Carr00c}.

The energetical stability of a Bloch state $u_\kappa(x)$ can be calculated from
perturbation theory \cite{Wu01,Wu03}. We make the ansatz
\be
  \psi(x) = \rme^{\rmi \kappa x} \left[ u_\kappa(x) + \delta u_\kappa(x) \right]
  \label{eqn-perturbation-ansatz1}
\ee
with a small perturbation $\delta u_\kappa(x)$, that can be decomposed into
different modes $\rme^{\pm \rmi q x}$.
A Bloch state $u_\kappa(x)$ is energetically stable and hence represents a
superflow, if it is a local minimum of the Hamiltonian
\be
  H = \int \psi^* \left( - \frac{1}{2} \frac{\partial^2}{\partial x^2}
  + V(x) \right) \psi + \frac{g}{2} |\psi|^4 - \mu |\psi|^2 \rd x .
\ee
Otherwise a perturbation can lower the energy of the solution and the system
becomes energetically unstable and superfluidity is lost, which is also called
Landau-instability.
Inserting the ansatz (\ref{eqn-perturbation-ansatz1}) into the Hamiltonian and
neglecting higher order terms reveals that the Bloch state is energetically stable
if the operator
\be
  M(q) = \left( \begin{array}{c c}
  L(\kappa+q) & g \psi^2 \\
  g \psi^{*2} & L(-\kappa+q) \\
  \end{array} \right)
\ee
with
\be
  L(q) = - \frac{1}{2} \left( \frac{\partial}{\partial x} + \ri q \right)^2
  + V(x) + 2 g \abs{\psi}^2 - \mu \, .
\ee
is positive definite for all $-1/2 \le q \le 1/2$ \cite{Wu03}.

Similarly one can deduce the dynamical stability by assuming a time-dependent
perturbation
\be
  \psi(x,t) = \rme^{\rmi k x- \rmi \mu t} \left[ u_k(x) + \delta u_k(x,t) \right].
\ee
The pertubation $\delta u_\kappa(x,t)$ grows exponentially in time if the operator
$\sigma_z M(q)$ with $\sigma_z ={\rm diag}(1,-1)$ has a complex eigenvalue for some $q$. Hence the Bloch state
$u_\kappa(x)$ is dynamically stable only if the spectrum of $\sigma_z M(q)$ is purely
real for all $-1/2 \le q \le 1/2$. Note that dynamical instability always
implies energetical instability, as shown in \cite{Wu03}.

\begin{figure}[htb]
\centering
\includegraphics[width=7cm,  angle=0]{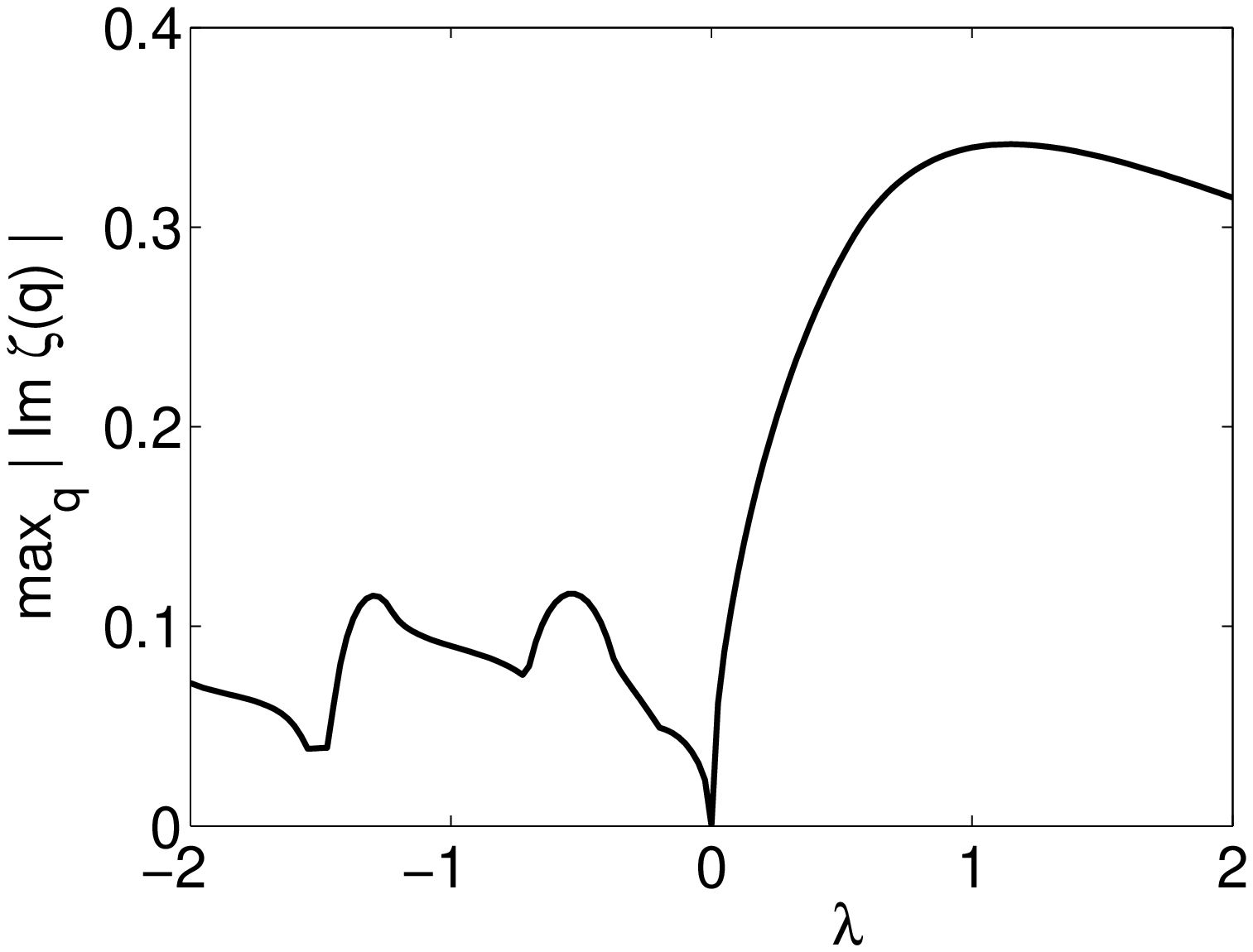}
\hspace{5mm}
\includegraphics[width=7cm,  angle=0]{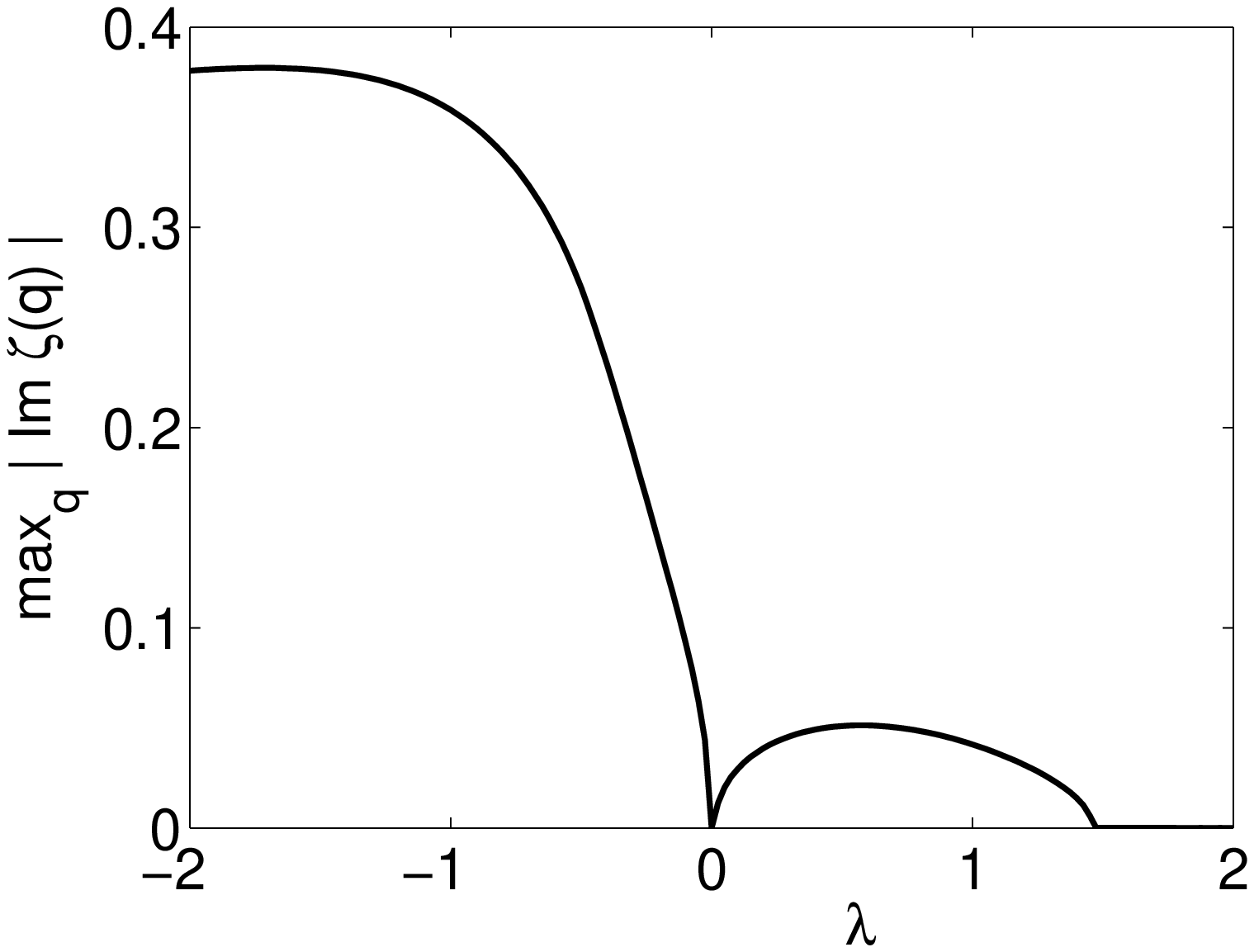}
\caption{\label{fig-gauss-zeta}
Eigenvalues of $\sigma_z M(q)$, indicating the dynamical (in)stability,
for the anti-symmetric (left) and symmetric (right) periodic solutions
in dependence of the potential strength $\lambda$.
Shown is the maximum modulus of the imaginary parts of the eigenvalues
of $\sigma_z M(q)$, where the maximum is taken over all eigenvalues and
all $q$ in the first Brillouin zone $q \in (-0.5,0.5)$.}
\end{figure}

Here we only consider strictly periodic states, which are Bloch states with $\kappa=0$.
The eigenvalues $\zeta(q)$ of $\sigma_z M(q)$ are calculated for the periodic
states discussed above in dependence of the potential strength $\lambda$.
In Fig.~\ref{fig-gauss-zeta} the maximum imaginary parts of the eigenvalues
$\zeta(q)$ are plotted, where the maximum is taken over all $q \in [-1/2,1/2]$.
One observes that the spectrum of $\sigma_z M(q)$ is never purely real
for the antisymmetric periodic states (except for $\lambda = 0$). Hence the
these states are always dynamically and energetically unstable.
The symmetric periodic state is dynamically unstable for attractive potentials
and not too strong repulsive potentials. It is stabilized if the potential strength
$\lambda$ exceeds a critical value $\lambda_d > 0$. However it still is energetically
unstable for all values of $\lambda$.
Similar results are obtained for the delta-comb discussed in the previous sections.

\begin{figure}[htb]
\centering
\includegraphics[width=7cm,  angle=0]{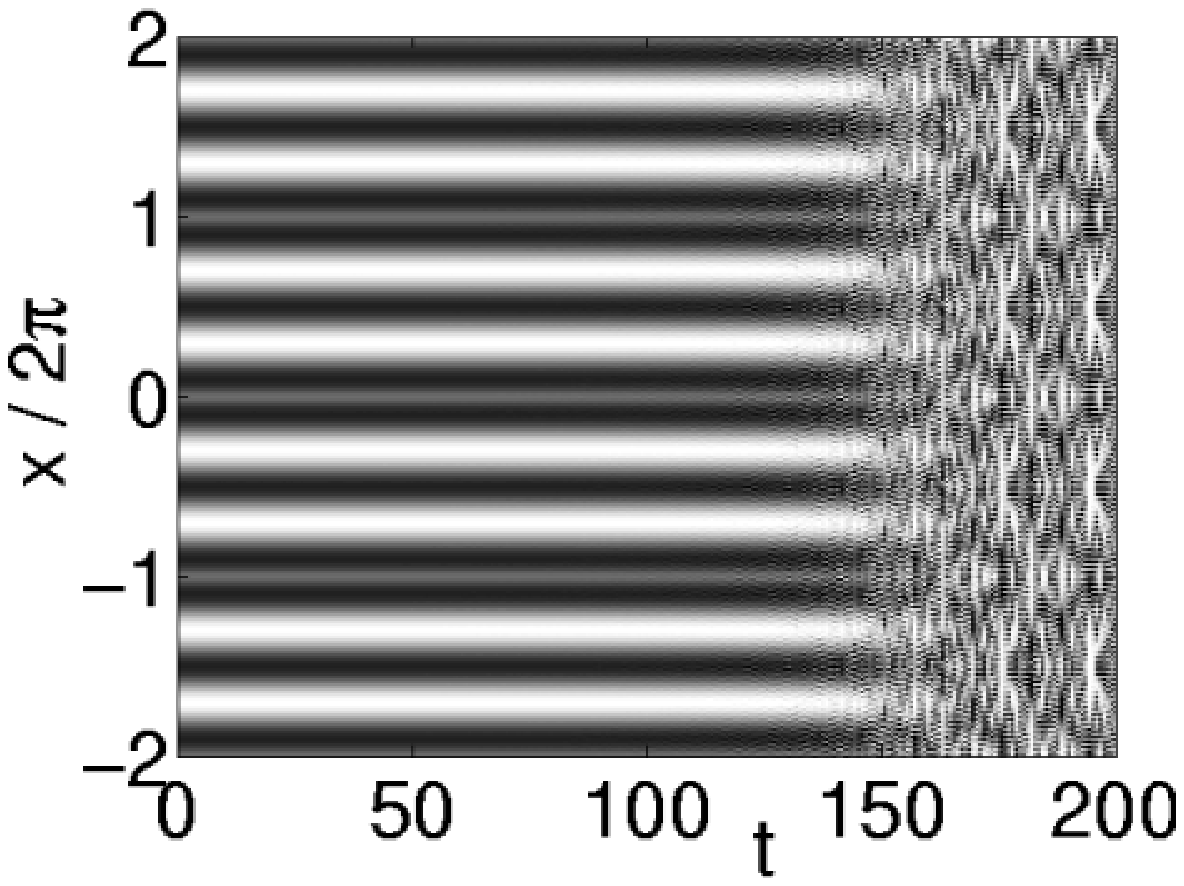}
\hspace{5mm}
\includegraphics[width=7cm,  angle=0]{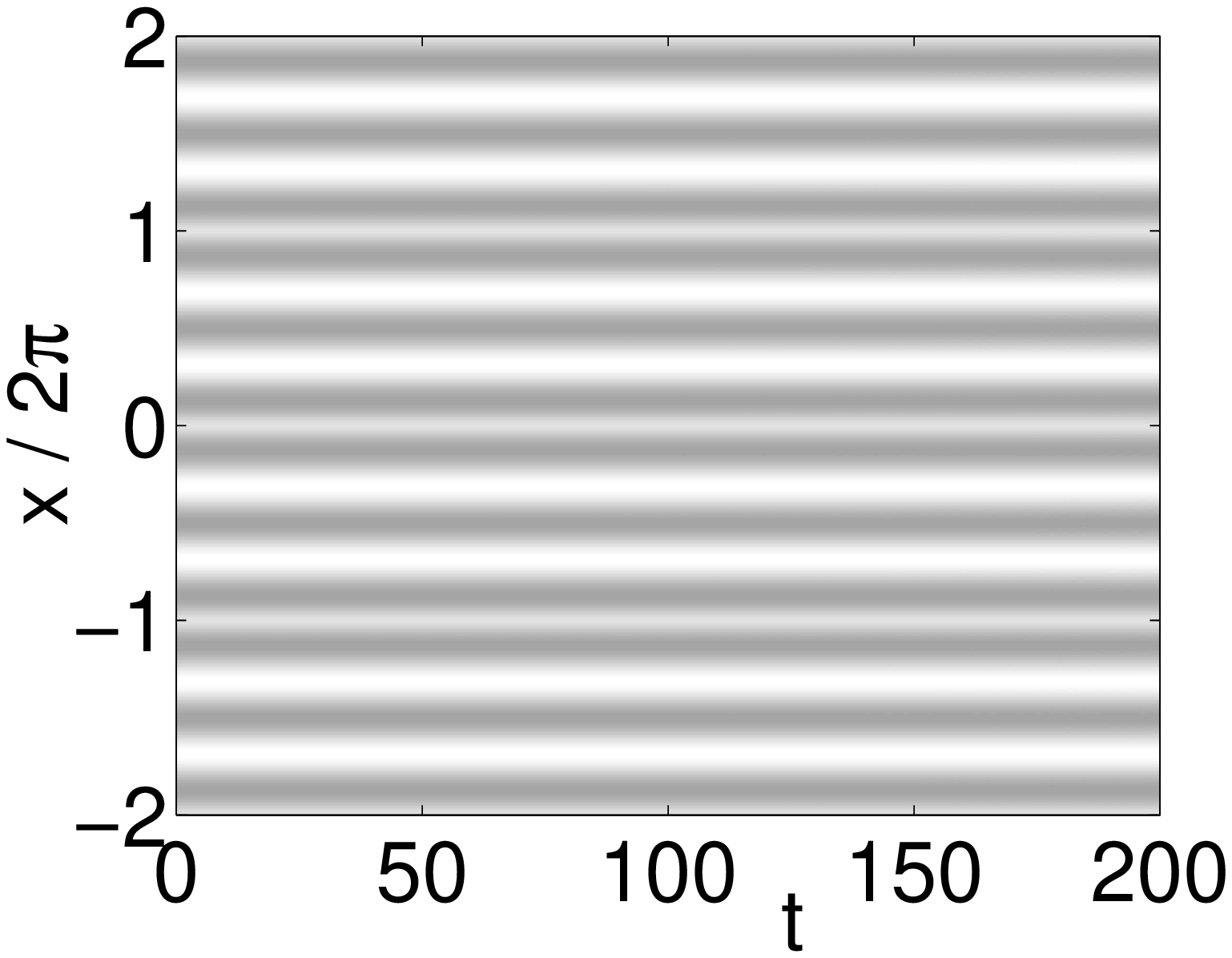}
\caption{\label{fig-gauss-wavefun}
Time evolution of the squared modulus $|\psi(x,t)|^2$ of a wave function
in a Gaussian comb of strength $\lambda = 1$ (left) resp. $\lambda = 2$ (right)
The initial state $\psi(x,t=0)$ was chosen as the symmetric periodic solution
for $\mu = 1$ plus a small random perturbation. }
\end{figure}

The dynamical (in)stability is illustrated in Fig.~\ref{fig-gauss-wavefun},
where the time evolution of the squared modulus $|\psi(x,t)|^2$ of a wavefunction
is plotted. The initial state $\psi(x,t=0)$ is chosen as a symmetric periodic
solution for $\lambda = 1$ resp. $\lambda = 2$ and $\mu = 1$ plus a small random
perturbation.
According to the results from perturbation theory illustrated in Fig.
\ref{fig-gauss-zeta}, the system is dynamically unstable for $\lambda = 1$
and dynamically stable for $\lambda = 2$. This is well confirmed by the results
of the wavepacket propagation.
For $\lambda =1$ the onset of dynamical instability is clearly visible at
$t \approx 150$, whereas no instability can be observed for $\lambda =2$.

Note that the temporal stability of the states is not affected at all by
the spatial stability properties. The onset of dynamical instability is
solely determined by the spectrum of $\sigma_z M(q)$ and the strength of
the initial random perturbation.
On the contrary one finds that the parameter values for which a periodic
state is spatially or dynamically stable are quite different.
The lowest symmetric state is spatially stable only for $\lambda_c < \lambda < 0$
(cf.~Fig.~\ref{fig-rep_fixp_gamma2}), whereas it is dynamically stable only for
$\lambda > \lambda_{d} > 0$. (cf.~Fig.~\ref{fig-gauss-zeta}).
One can even increase the potential strength $\lambda$ slowly through the
critical value $\lambda_c$ for a bifurcation without affecting the onset
of dynamical instability.

\section{Summary}
\label{sec-summary}

Summing up, the stationary solutions of the nonlinear Schr\"odinger
equation for a delta-comb potential are analyzed.
This somewhat artificial model system is of considerable interest since
it allows analytical solutions in terms of Jacobi elliptic functions.

The analogy to a classical nonlinear oscillator problem, the kicked nonlinear
Hill equation is pointed out and the dynamics in phase space is discussed.
It is shown that a repulsive nonlinearity typically leads
to spatially chaotic dynamics and finally a divergence of the wave function,
whereas the wave function always remains bounded for an attractive nonlinearity.
Periodic solutions, which are fixed points of the phase space dynamics,
are analyzed in detail. The fixed points are spatially stable only for
certain parameter values, for which a stability map is calculated.
The stability is lost in a Hamiltonian bifurcation scenario, giving rise
to new period doubled fixed points.

Nonlinear Bloch bands are calculated for the delta-comb showing the
celebrated loop structures.
It is shown that the Bloch states with zero quasi momentum are just the
periodic solutions mentioned above, which are given analytically in terms
of Jacobi elliptic functions. The critical nonlinearity parameter for the
emergence of loops is calculated analytically.
Furthermore the emergence of $4\pi$-periodic Bloch bands is discussed and
linked to the period doubling bifurcation of periodic solutions mentioned
above.

Finally an extension to a more realistic potential, a periodic arrangement
of narrow Gaussian potential barriers, shows that the properties of the
stationary solutions remain qualitatively the same as for the delta-comb.
In addition the {\it temporal} stability of the periodic solutions is
analyzed. Antisymmetric periodic solutions are always dynamically unstable,
whereas the symmetric periodic solutions are stabilized if the potential
strength exceeds a critical value $\lambda > \lambda_d > 0$.

The properties of stationary states of the nonlinear Schr\"odinger equation
are of fundamental interest for the study of BECs in optical lattices.
For example the emergence of loops leads to dynamical instability of BECs
in tilted optical lattices because an adiabatic evolution is no longer
possible here \cite{Liu03,Fall04}.
But not all solutions in periodic potential must take the form of Bloch states,
as it was shown in the present paper. In contrast, most solutions of the nonlinear
Schr\"odinger equation with a repulsive potential will be spatially chaotic
or even divergent. This deserves further studies.


\section*{Acknowledgments}
Support from the Deutsche Forschungsgemeinschaft
via the Graduiertenkolleg  ``Nichtlineare Optik und Ultrakurzzeitphysik'' and the 
research fellowship programme (grant number WI 3415/1) is gratefully acknowledged.

\label{lastpage}


\begin{thebibliography}{10}

\bibitem{Abra72}
{\sc Abramowitz, M., and Stegun, I.~A.}
\newblock {\em Handbook of Mathematical Functions}.
\newblock Dover Publications, Inc., New York, 1972.

\bibitem{Adam05}
{\sc Adami, R., and Saccetti, A.}
\newblock The transition from diffusion to blow-up for a nonlinear
  {Schr\"odinger} equation in dimension 1.
\newblock {\em J. Phys. A 38\/} (2005), 8379.

\bibitem{Agui87}
{\sc {Aguiar, C. P. Malta, M. Baranger and K. T. R. Davis}, M.}
\newblock Bifurcations of periodic trajectories in non-integrable hamiltonian
  systems with two degrees of freedom: Numerical and analytical results.
\newblock {\em Ann. Phys. (N.Y.) 180\/} (1987), 167.

\bibitem{Albi05}
{\sc Albiez, M., Gati, R., F\"olling, J., Hunsmann, S., Cristiani, M., and
  Oberthaler, M.~K.}
\newblock Direct observation of tunneling and nonlinear self-trapping in a
  single bosonic {Josephson} junction.
\newblock {\em Phys. Rev. Lett. 95\/} (2005), 010402.

\bibitem{Alfi07}
{\sc Alfimov, G.~L., and Zezyulin, D.~A.}
\newblock Nonlinear modes for the {Gross-Pitaevskii} equation -- a
  demonstrative computation approach.
\newblock {\em Nonlinearity 20\/} (2007), 2075.

\bibitem{Bloc08}
{\sc Bloch, I., Dalibard, J., and Zwerger, W.}
\newblock Many-body physics with ultracold gases.
\newblock {\em Rev. Mod. Phys. 80\/} (2008).

\bibitem{Bowm81}
{\sc Bowman, T. T.~M.}
\newblock Bifurcation of solutions of nonlinear boundary value problem.
\newblock {\em Nonlin. Anal. Theory, Methods Appl. 5\/} (1981), 655.

\bibitem{Bron01b}
{\sc Bronski, J.~C., Carr, L.~D., Carretero-Gonz\'alez, R., Deconinck, B.,
  Kutz, J.~N., and Promislow, K.}
\newblock Stability of attractive {Bose-Einstein} condensates in a periodic
  potential.
\newblock {\em Phys. Rev. E 64\/} (2001), 056615.

\bibitem{Bron01a}
{\sc Bronski, J.~C., Carr, L.~D., Deconinck, B., and Kutz, J.~N.}
\newblock {Bose-Einstein} condensates in standing waves: The cubic nonlinear
  {Schr\"odinger} equation with a periodic potential.
\newblock {\em Phys. Rev. Lett. 86\/} (2001), 1402.

\bibitem{Burg99a}
{\sc Burger, S., Bongs, K., Dettmer, S., Ertmer, W., Sengstock, K., Sanpera,
  A., Shlyapnikov, G., and Lewenstein, M.}
\newblock Dark solitons in {Bose-Einstein} condensates.
\newblock {\em Phys. Rev. Lett. 83\/} (1999), 5198.

\bibitem{Carr00}
{\sc Carr, L.~D., Clark, C.~W., and Reinhardt, W.~P.}
\newblock Stationary solutions of the one-dimensional nonlinear {Schr\"odinger}
  equation.
\newblock {\em Phys. Rev. A 62\/} (2000), 063610, 063611.

\bibitem{Carr00c}
{\sc Carr, L.~D., Kutz, J.~N., and Reinhardt, W.~P.}
\newblock Stability of stationary states in the cubic nonlinear {Schr\"odinger}
  equation: Applications to {Bose-Einstein} condensates.
\newblock {\em Phys. Rev. E 63\/} (2001), 066604.

\bibitem{Carr08}
{\sc Carretero-Gonz\'alez, R., Frantzeskakis, D.~J., and Kevrekidis, P.~G.}
\newblock Nonlinear waves in {Bose-Einstein} condensates: physical relevance
  and mathematical techniques.
\newblock {\em Nonlinearity 21\/} (2008), R139.

\bibitem{Cart08}
{\sc Cartarius, H., Main, J., and Wunner, G.}
\newblock Discovery of exceptional points in a {Bose-Einstein} condensation of
  gases with attractive 1/r-interaction.
\newblock {\em Phys. Rev. A 77\/} (2008), 013618.

\bibitem{Cast97}
{\sc Castin, Y., and Dum, R.}
\newblock Instability and depletion of an excited {Bose-Einstein} condensate in
  a trap.
\newblock {\em Phys. Rev. Lett. 79\/} (1997), 3553.

\bibitem{Cast98}
{\sc Castin, Y., and Dum, R.}
\newblock Low-temperature {Bose-Einstein} condensates in time-dependent traps:
  Beyond the $u(1)$ symmetry-breaking approach.
\newblock {\em Phys. Rev. A 57\/} (1998), 3008.

\bibitem{Cerv03}
{\sc Cervero, J.~M.}
\newblock Pt-symmetry in one-dimensional quantum periodic potentials.
\newblock {\em Phys. Lett. A 317\/} (2003), 26.

\bibitem{Cerv04}
{\sc Cervero, J.~M., and Rodriguez, A.}
\newblock The band spectrum of periodic potentials with pt-symmetry.
\newblock {\em J. Phys. A 37\/} (2004), 10167.

\bibitem{Chir79}
{\sc Chirikov, B.~V.}
\newblock Universal instability of many-dimensional oscillator systems.
\newblock {\em Phys. Rep. 52\/} (1979), 263.

\bibitem{LeCo08}
{\sc Coz, S.~L., Fukuizumi, R., G, F., Ksherim, B., and Y.Sivan}.
\newblock Instability of bound states of a nonlinear {Schr\"odinger} equation
  with a {Dirac} potential.
\newblock {\em Physica D 38\/} (2008), 8379.

\bibitem{Dago00}
{\sc D'Agosta, R., Malomed, B.~A., and Presilla, C.}
\newblock Stationary solutions of the {Gross--Pitaevskii} equation with a
  linear counterpart.
\newblock {\em Phys. Lett. A 275\/} (2000), 424.

\bibitem{Dans07a}
{\sc Danshita, I., and Tsuchiya, S.}
\newblock Stability of {Bose-Einstein} condensates in a {Kronig-Penney}
  potential.
\newblock {\em Physical Review A 75\/} (2007), 033612.

\bibitem{Dodd82}
{\sc Dodd, R.~K., Eilbeck, J.~C., Gibbon, J.~D., and Morris, H.~C.}
\newblock {\em Solitons and nonlinear wave equations}.
\newblock Academic Press, London, 1982.

\bibitem{Eier04a}
{\sc Eiermann, B., Anker, T., Albie, M., Taglieber, M., Treutlein, P., Marzlin,
  K.-P., and Oberthaler, M.~K.}
\newblock Bright {Bose-Einstein} gap solitons of atoms with repulsive
  interaction.
\newblock {\em Phys. Rev. Lett. 92\/} (2004).

\bibitem{Fall04}
{\sc Fallani, L., Sarlo, L.~D., Lye, J.~E., Modugno, M., Sears, R., Fort, C.,
  and Inguscio, M.}
\newblock Observation of dynamical instability for a {Bose-Einstein} condensate
  in a moving 1d optical lattice.
\newblock {\em Phys. Rev. Lett. 93\/} (2004), 140406.

\bibitem{Fibi06}
{\sc Fibich, G., Sivan, Y., and Weinstein, M.~I.}
\newblock Bound states of nonlinear {Schr\"odinger} equations with a periodic
  nonlinear microstructure.
\newblock {\em Physica D 217\/} (2006), 31.

\bibitem{Fort07}
{\sc Fortagh, J., and Zimmermann, C.}
\newblock Magnetic microtraps for ultracold atoms.
\newblock {\em Rev. Mod. Phys. 79\/} (2007), 235.

\bibitem{06nlnh}
{\sc Graefe, E.~M., and Korsch, H.~J.}
\newblock Crossing scenario for a nonlinear non-hermitian two-level system.
\newblock {\em Czech. J. Phys. 56\/} (2006), 1007.

\bibitem{Grei01}
{\sc Greiner, M., Bloch, I., Mandel, O., H\"ansch, T.~W., and Esslinger, T.}
\newblock {Bose-Einstein} condensates in 1d- and 2d- optical lattices.
\newblock {\em Appl. Phys. B 73\/} (2001), 769.

\bibitem{Grif92}
{\sc Griffiths, D.~J., and Taussig, N.~F.}
\newblock Scattering from a locally periodic potential.
\newblock {\em Am. J. Phys. 60\/} (1992), 883.

\bibitem{Hsu74}
{\sc Hsu, C.~S.}
\newblock Some simple exact periodic responses for a nonlinear system under
  parametric excitation.
\newblock {\em J.~Appl.~Mech. 42\/} (1974), 176.

\bibitem{Jack04}
{\sc Jackson, R.~K., and Weinstein, M.~I.}
\newblock Geometric analysis of bifurcation and symmetry breaking in a
  {Gross-Pitaevskii} equation.
\newblock {\em J. Stat. Phys. 116\/} (2004), 881.

\bibitem{Jona03}
{\sc Jona-Lasinio, M., Morsch, O., Cristiani, M., Malossi, N., M\"uller, J.~H.,
  Courtade, E., Anderlini, M., and Arimondo, E.}
\newblock Asymmetric {Landau-Zener} tunneling in a periodic potential.
\newblock {\em Phys. Rev. Lett. 91\/} (2003), 230406.

\bibitem{Khay02}
{\sc Khaykovich, L., Schreck, F., Ferrari, G., Bourdel, T., Cubizolles, J.,
  Carr, L.~D., Castin, Y., and Salomon, C.}
\newblock Formation of a matter-wave bright soliton.
\newblock {\em Science 296\/} (2002).

\bibitem{Komi06}
{\sc Kominis, Y.}
\newblock Analytical solitary wave solutions of the nonlinear {Kronig-Penney}
  model in photonic structures.
\newblock {\em Phys. Rev. E 73\/} (2006), 066619.

\bibitem{08chaos}
{\sc Korsch, H.~J., Jodl, H.-J., and Hartmann, T.}
\newblock {\em Chaos -- A Program Collection for the {PC}, 3nd {Ed.}}
\newblock Springer--Verlag, Heidelberg, New-York, 2008.

\bibitem{Lawd89}
{\sc Lawden, D.~F.}
\newblock {\em Elliptic Functions and Applications}.
\newblock Springer, New York, 1989.

\bibitem{Li04}
{\sc Li, W.~D., and Smerzi, A.}
\newblock Nonlinear {Kronig-Penney} model.
\newblock {\em Phys. Rev. E 70\/} (2004), 016605.

\bibitem{Liu03}
{\sc Liu, J., Wu, B., and Niu, Q.}
\newblock Nonlinear evolution of quantum states in the adiabatic regime.
\newblock {\em Phys. Rev. Lett. 90\/} (2003), 170404.

\bibitem{Mach04}
{\sc Machholm, M., Nicolin, A., Pethick, C.~J., and Smith, H.}
\newblock Spatial period doubling in {Bose-Einstein} condensates in an optical
  lattice.
\newblock {\em Phys. Rev. A 69\/} (2004), 043604.

\bibitem{Makr08}
{\sc Makris, K.~G., El-Ganainy, R., Christodoulides, D.~N., and Musslimani,
  Z.~H.}
\newblock Beam dynamics in pt symmetric optical lattices.
\newblock {\em Phys. Rev. Lett. 100\/} (2008), 103904.

\bibitem{Meye70}
{\sc Meyer, K.~R.}
\newblock Generic bifurcations of periodic points.
\newblock {\em Trans. AMS 149\/} (1970), 95.

\bibitem{Mont82}
{\sc Month, L.~A., and Rand, R.~H.}
\newblock Bifurcation of 4:1 subharmonics in the nonlinear {Mathieu} equation.
\newblock {\em Mechanics Research Comm. 9\/} (1982), 233.

\bibitem{Moon87}
{\sc Moon, F.~C.}
\newblock {\em Chaotic {Vibrations}}.
\newblock John Wiley, New York, 1987.

\bibitem{Mors06}
{\sc Morsch, O., and Oberthaler, M.}
\newblock Dynamics of {Bose-Einstein} condensates in optical lattices.
\newblock {\em Rev. Mod. Phys. 78\/} (2006), 179.

\bibitem{Muss08}
{\sc Musslimani, Z.~H., Makris, K.~G., El-Ganainy, R., and Christodoulides,
  D.~N.}
\newblock Optical solitons in $pt$ periodic potentials.
\newblock {\em Phys. Rev. Lett. 100\/} (2008), 030402.

\bibitem{Naeg08}
{\sc N\"agerl, H.-C.}
\newblock private communication.

\bibitem{Nayf79}
{\sc Nayfeh, A.~H., and Mook, D.~T.}
\newblock {\em Nonlinear Oscillations}.
\newblock John Wiley, New York, 1979.

\bibitem{Paul05}
{\sc Paul, T., Richter, K., and Schlagheck, P.}
\newblock Nonlinear resonant transport of {Bose-Einstein} condensates.
\newblock {\em Phys. Rev. Lett. 94\/} (2005), 020404.

\bibitem{Paul07c}
{\sc Paul, T., Schlagheck, P., Leboeuf, P., and Pavloff, N.}
\newblock Superfluidity versus {Anderson} localization in a dilute {Bose} gas.
\newblock {\em Phys. Rev. Lett. 98\/} (2007), 210602.

\bibitem{Peth02}
{\sc Pethick, C.~J., and Smith, H.}
\newblock {\em {Bose-Einstein} Condensation in Dilute Gases}.
\newblock Cambridge University Press, Cambridge, 2002.

\bibitem{Pita03}
{\sc Pitaevskii, L., and Stringari, S.}
\newblock {\em Bose-Einstein Condensation}.
\newblock Oxford University Press, Oxford, 2003.

\bibitem{Port04}
{\sc Porter, M.~A., and Cvitanovic, P.}
\newblock A perturbative analysis of modulated amplitude waves in
  {Bose-Einstein} condensates.
\newblock {\em Chaos 14\/} (2004), 739.

\bibitem{09ddshell}
{\sc Rapedius, K., and Korsch, H.~J.}
\newblock Resonance solutions of the nonlinear schr\"odinger equation in an
  open double-well potential.
\newblock {\em J. Phys. B \phantom{0}\/} (2009), in press, e--print arxiv:
  0809.2748.

\bibitem{06nl_transport}
{\sc Rapedius, K., Witthaut, D., and Korsch, H.~J.}
\newblock Analytical study of resonant transport of {Bose-Einstein}
  condensates.
\newblock {\em Phys. Rev. A 73\/} (2006), 033608.

\bibitem{Sacc07}
{\sc Saccetti, A.}
\newblock Spectral splitting method for nonlinear {Schr\"odinger} equations
  with singular potential.
\newblock {\em J. Chem. Phys. 227\/} (2007), 1483.

\bibitem{Seam05}
{\sc Seaman, B.~T., Carr, L.~D., and Holland, M.~J.}
\newblock Nonlinear band structure in {Bose-Einstein} condensates: Nonlinear
  {Schr{\"o}dinger} equation with a {Kronig-Penney} potential.
\newblock {\em Phys. Rev. A 71\/} (2005), 033622.

\bibitem{Seam05b}
{\sc Seaman, B.~T., Carr, L.~D., and Holland, M.~J.}
\newblock Period doubling, two-color lattices, and the growth of swallowtails
  in {Bose-Einstein} condensates.
\newblock {\em Phys. Rev. A 72\/} (2005), 033602.

\bibitem{Tara99}
{\sc Taras-Semchuk, D., and Gunn, J. M.~F.}
\newblock Superfluid flow past an array of scatterers.
\newblock {\em Phys. Rev. B 60\/} (1999), 13139.

\bibitem{Theo97}
{\sc Theodorakis, S., and Leontidis, E.}
\newblock Bound states in a nonlinear {Kronig-Penney} model.
\newblock {\em J. Phys. A 30\/} (1997), 4835.

\bibitem{Thom03}
{\sc Thommen, Q., Garreau, J.~C., and Zehnl\'e, V.}
\newblock Classical chaos with {Bose-Einstein} condensates in tilted optical
  lattices.
\newblock {\em Phys. Rev. Lett. 91\/} (2003), 210405.

\bibitem{09phaseappl}
{\sc Trimborn, F., Witthaut, D., and Korsch, H.~J.}
\newblock Beyond mean-field dynamics of small {Bose-Hubbard} systems based on
  the number-conserving phase space approach.
\newblock {\em Phys. Rev. A 79\/} (2009), 013608.

\bibitem{Trim08}
{\sc Trimborn, F., Witthaut, D., and Wimberger, S.}
\newblock Mean-field dynamics of a two-mode {Bose-Einstein} condensate subject
  to noise and dissipation.
\newblock {\em J. Phys. B 41\/} (2008), 171001.

\bibitem{Wan90}
{\sc Wan, Y., and Soukoulis, C.~M.}
\newblock One-dimensional nonlinear {Schr\"odinger} equation: {A} nonlinear
  dynamical approach.
\newblock {\em Phys. Rev. A 41\/} (1990), 800.

\bibitem{06zener_bec}
{\sc Witthaut, D., Graefe, E.~M., and Korsch, H.~J.}
\newblock Towards a {Landau-Zener} formula for an interacting {Bose-Einstein}
  condensate in a two-level system.
\newblock {\em Phys. Rev. A 73\/} (2006), 063609.

\bibitem{07nlres}
{\sc Witthaut, D., Graefe, E.~M., Wimberger, S., and Korsch, H.~J.}
\newblock Bose-{Einstein} condensates in accelerated double-periodic optical
  lattices: coupling and crossing of resonances.
\newblock {\em Phys. Rev. A 75\/} (2007), 013617.

\bibitem{04nls_delta}
{\sc Witthaut, D., Mossmann, S., and Korsch, H.~J.}
\newblock Bound and resonance states of the nonlinear {Schr\"odinger} equation
  in simple model systems.
\newblock {\em J. Phys. A 38\/} (2005), 1777.

\bibitem{04bloch_bec}
{\sc Witthaut, D., Werder, M., Mossmann, S., and Korsch, H.~J.}
\newblock Bloch-oscillations of {Bose-Einstein} condensates: Breakdown and
  revival.
\newblock {\em Phys. Rev. E 71\/} (2005), 036625.

\bibitem{Wu00}
{\sc Wu, B., and Niu, Q.}
\newblock Nonlinear {Landau-Zener} tunneling.
\newblock {\em Phys. Rev. A 61\/} (2000), 023402.

\bibitem{Wu01}
{\sc Wu, B., and Niu, Q.}
\newblock Landau and dynamical instabilities of the superflow of
  {Bose-Einstein} condensates in optical lattices.
\newblock {\em Phys. Rev. A 64\/} (2001), 061603.

\bibitem{Wu03}
{\sc Wu, B., and Niu, Q.}
\newblock Superfluidity of {Bose-Einstein} condensates in an optical lattice:
  {Landau-Zener} tunneling and dynamical instability.
\newblock {\em New J. Phys. 5\/} (2003), 104.

\end{thebibliography}
\end{document}